\documentclass{article}
\usepackage{amssymb}
\usepackage{amsmath}

\input{tcilatex}
\begin{document}

\begin{center}
{\huge The STIRAP-based unitary decelerating and accelerating processes
\smallskip of a single free atom \bigskip }

{\large Xijia Miao\smallskip }

{\large Somerville, Massachusetts \smallskip }

{\large Date: May 2007\bigskip }

{\large Abstract\\[0pt]
}
\end{center}

The STIRAP-based unitary decelerating and accelerating processes have been
proposed to realize the time- and space-compressing processes in the quantum
control process to simulate the reversible and unitary state-insensitive
halting protocol (Arxiv: quant-ph/0607144). A standard three-state STIRAP
pulse sequence may act as a basic unitary decelerating sequence or a basic
unitary accelerating sequence. A STIRAP-based unitary decelerating
(accelerating) process then consists of a train of these basic STIRAP
unitary decelerating (accelerating) sequences. The present work is focused
on investigating analytically and quantitatively how the momentum
distribution of a momentum superposition state of a pure-state quantum
system such as a momentum Gaussian wave-packet state of a single freely
moving atom affects the STIRAP state transfer in these decelerating and
accelerating processes. The complete STIRAP state transfer and the unitarity
of these processes are stressed highly in the investigation. It has been
shown that the momentum distribution has an important influence upon the
STIRAP state-transfer efficiency. In the ideal adiabatic condition these
unitary decelerating and accelerating processes for a freely moving atom are
studied in detail, and it is shown that they can be used to manipulate and
control in time and space the center-of-mass position and momentum of a
Gaussian wave-packet motional state of a free atom. Two general adiabatic
conditions for the basic STIRAP decelerating and accelerating processes are
derived analytically. They are strict and accurate. They can be used to set
up a conventional STIRAP state-transfer experiment and also the basic STIRAP
decelerating and accelerating processes. With the help of the STIRAP theory
and the unitary quantum dynamics it confirms theoretically that the time-
and space-compressing processes of the quantum control process (Arxiv:
quant-ph/0607144) can be realized almost perfectly by the STIRAP-based
unitary decelerating and accelerating processes in the ideal or nearly ideal
adiabatic condition. \newline
\newline
\newline
{\Large 1. Introduction }

The stimulated Raman adiabatic passage (STIRAP) processes are very important
coherent double-photon processes [1, 2]. The STIRAP method has been
extensively applied to the complete population transfer in energy levels of
atoms and molecules in the laser spectroscopy [3, 4], the laser cooling in
neutral atomic ensembles [5, 6, 7, 8, 9] (here may include the conventional
Raman adiabatic processes), and the atomic quantum interference experiments
[10, 11, 12, 13, 14] as well as other science research fields. A standard
three-state STIRAP pulse sequence [3, 4, 15, 17, 18] consists of a pair of
Raman laser light beams which are selectively applied to three chosen energy
levels of an atomic or molecular system. The largest advantage of the
STIRAP\ method is that the complete population transfer or state transfer in
an atomic or molecular system may be achieved by the STIRAP pulse sequence
with delayed and overlapping Raman laser light beams [15, 16, 17], and the
STIRAP method is tolerant to the experimental imperfections. The basic
theory of the STIRAP method has been well set up [16, 17, 18]. The
experimental confirmation for the STIRAP method has been carried out first
in the atomic and molecular laser spectroscopy [3, 15] and then in diverse
other science research fields. The basic STIRAP theoretical and experimental
methods have been first extended to study and design the quantum
interference experiments in cold atomic ensembles [10, 11, 12, 13, 14],
which involve the atomic motional momentum transfer. The theory also has
been developed to study the Raman laser cooling processes in a neutral atom
ensemble [20, 21] by combining the velocity-selective coherent population
trapping [5, 19]. In these two types of experiments [5, 10, 11, 12, 13, 14,
19, 20] the Raman laser pulse sequence such as the STIRAP pulse sequence
usually consists of a pair of counterpropagating Raman laser light beams.

It has been shown that the dynamical state-locking pulse field plays a key
role in constructing the reversible and unitary state-insensitive halting
protocol [22] and solving efficiently the quantum search problem [22, 36]. A
general state-locking pulse field [22] consists of a sequence of the time-
and space-dependent electromagnetic pulse fields and could also contain the
time- and space-dependent potential fields which could be generated by the
external electric and / or magnetic field. A unitary decelerating (or
accelerating) laser light pulse sequence that is used to decelerate (or
accelerate) a moving free atom could be thought of as the component of a
dynamical state-locking pulse field. The standard three-state STIRAP pulse
sequence may act as either a basic unitary decelerating sequence or a basic
unitary accelerating sequence, which is dependent upon the parameter
settings of the two counterpropagating Raman adiabatic laser light beams of
the STIRAP pulse sequence. The STIRAP-based decelerating (accelerating)
process then consists of a train of the basic STIRAP unitary decelerating
(accelerating) sequences which are applied to a moving atom consecutively.
It has been proposed that the STIRAP-based unitary decelerating and / or
accelerating processes may be used to coherently manipulate the
halting-qubit atom in the quantum control process to simulate the reversible
and unitary state-insensitive halting protocol [22]. Thus, the STIRAP-based
unitary decelerating and accelerating processes are the important building
blocks of the reversible and unitary state-insensitive halting protocol and
the efficient quantum search process based on the unitary quantum dynamics
in time and space. A unitary decelerating process is much like the
conventional laser cooling process in an atomic ensemble. The essential
difference between the two processes is that the unitary decelerating
process is reversible and unitary, while the laser cooling process usually
is irreversible. A unitary accelerating process could be more like the
momentum transfer process in the quantum interference experiments of an
atomic ensemble. The atomic momentum transfer process in these quantum
interference experiments is generally transverse with respect to the initial
atomic moving direction, while here the unitary accelerating process is
longitudinal. The unitary decelerating and accelerating processes stress
their unitarity and complete state transfer in the quantum control process.

A quantum computation tends to avoid using a space-dependent unitary
operation as its basic building block, since such a unitary operation
usually is more complicated and inconvenient to manipulate and control in a
quantum system with respect to the conventional quantum-gate operations. It
is necessary to manipulate and control at will the internal motion of an
atom (or atomic ion) in quantum computation when the specific internal
states of the atom such as the hyperfine ground electronic states or the
nuclear spin states are taken as a quantum bit, but at the same time the
atomic center-of-mass motion tends to be kept unchanged simply (or to be
constrained simply) so that it does not affect these quantum-gate operations
and hence in quantum computational theory it is usually not considered
explicitly. However, as pointed out in the previous paper [22], it is of
crucial importance to unitarily manipulate and control at will in time and
space the center-of-mass motion, the internal motion, and the mutual
cooperation and coupling of the two motions of the halting-qubit atom in
order to realize the quantum control process to simulate the reversible and
unitary state-insensitive halting protocol. The unitary manipulation and
control in time and space for the atom is also a key step toward the
realization to solve efficiently the quantum search problem. An
electromagnetic wave field such as a laser light field can manipulate and
control in time and space not only the center-of-mass motion and the
internal electronic (or spin) motion of an atom separately, but also it can
create and control the coupling between the center-of-mass motion and the
internal motion of an atomic system. This is the theoretical fundament for
the laser cooling of a neutral atomic ensemble [21, 23, 24] and the unitary
decelerating and accelerating of a free atom [22]. One large advantage to
use a laser light field to manipulate an atom is that the space-selective
and / or the internal-state-selective unitary operations of the atom can be
realized easily. The STIRAP-based unitary decelerating and accelerating
processes could be a very useful double-photon method to coherently
manipulate and control in time and space the center-of-mass motion, the
internal motion, and the coupling of both the motions of a moving free atom.
It usually uses a pair of Raman adiabatic laser light beams to couple the
atomic center-of-mass motional state and internal electronic states (or spin
polarization states) to realize the coupling between the two motions. The
STIRAP-based unitary decelerating and accelerating processes have been
proposed to realize the unitary time- and space-compressing processes which
are necessary components of the quantum control process to simulate
efficiently the state-insensitive reversible and unitary halting protocol
[22]. On the other hand, as far as the Gaussian wave-packet state of a free
atom is concerned, the STIRAP-based unitary decelerating and accelerating
processes could generate a type of time- and space-dependent unitary
propagators which can manipulate and control the center-of-mass position and
momentum of the atomic Gaussian wave-packet state. This important result
will be shown in the paper.

The basic STIRAP\ theory generally does not consider explicitly the effect
of the center-of-mass motional momentum distribution of an atomic or
molecular system on the STIRAP population transfer in the laser spectroscopy
[16, 17, 18], although the Doppler effect in these atomic and molecular
systems has also been considered suitably in the STIRAP experiments [3, 15].
This could be due to that the center-of-mass motion of an atom or molecule
is generally much slower than the internal electronic motion of the atom or
molecule and could be neglected in the basic STIRAP theory, and the STIRAP
experimental settings [3, 15] are also favorable for the experiments to
minimize the center-of-mass motional effect, for example, the STIRAP
experiments may use a pair of copropagating Raman laser light beams.
However, a pure-state quantum system such as a single freely moving atom may
be in a superposition of the center-of-mass motional momentum states which
may has a broad momentum distribution. It is generally hard to realize the
complete state transfer in a quantum system with a broad momentum
distribution by the standard STIRAP method. One therefore must consider the
effect of the momentum distribution of a superposition state of the quantum
system on the STIRAP population or state transfer. In concept the momentum
distribution of a momentum superposition state of a pure-state quantum
system is essentially different from the conventional momentum distribution
of a quantum ensemble, the latter is a statistical distribution of momentum
of the particles which form the quantum ensemble. But the effect of the
momentum distribution on the STIRAP state transfer is similar for the two
types of quantum systems according to the unitary quantum dynamical
principle that both a closed quantum system and its ensemble obey the same
unitary quantum dynamics [22] (it seems that a quantum system in the
presence of an external electromagnetic field is not a closed quantum
system, but in theory such a quantum system may be treated conditionally as
if it is a closed quantum system, as shown in the section 11 in Ref. [40]).
It has been investigated in theory how the momentum distribution of an
atomic ensemble affects the\ population or state transfer in the atomic
laser cooling based on the velocity-selective coherent population trapping
[5, 19, 20] which also uses the Raman laser light beams, the STIRAP-based
momentum transfer in the cold atomic interference experiments [10, 12], and
the conventional Raman-laser-light-based cold atomic interference
experiments [11], but these investigations are usually either qualitative or
numerical. It is important to investigate analytically and quantitatively
how a superposition of the momentum states affects the STIRAP state transfer
in a pure-state quantum system when the STIRAP method is used to perform the
state transfer or the unitary operation in quantum computation. This paper
is devoted to such a theoretic investigation: how a superposition of the
momentum states affects the STIRAP state transfer in a single freely moving
atom. According to quantum mechanics a freely moving atom in the presence of
the STIRAP pulse sequence may be described by the complete set of the
product states of both the center-of-mass motional states and the internal
states (the electronic states or spin polarization states) of the atom. The
atomic center-of-mass motional state may be a superposition of the atomic
momentum eigenstates. When the atom is transferred from one internal state
to another by a STIRAP pulse sequence, the transfer efficiency is generally
dependent upon the atomic center-of-mass motional state. The purpose to
investigate this dependence is to understand how the momentum distribution
of the center-of-mass motional state affects the transfer efficiency and
then design a better STIRAP pulse sequence so as to achieve a complete
STIRAP state transfer over the whole effective momentum distribution of the
atomic center-of-mass motional state.

In order to manipulate and control the halting-qubit atom in time and space
in the quantum control process [22] it is necessary to investigate the time
evolution process of the halting-qubit atom in the STIRAP-based unitary
decelerating and accelerating processes. In order to investigate
quantitatively how a superposition of the momentum states affects the STIRAP
state transfer for a single freely moving atom in the STIRAP-based
decelerating or accelerating process it is also necessary to calculate the
time evolution process of the atom in the presence of the Raman laser light
field. For example, it needs to solve the unitary dynamical equation to set
up a general adiabatic condition for the basic STIRAP decelerating and
accelerating processes for the atom. It seems that a single freely moving
atom in the presence of an external electromagnetic field such as the Raman
laser light field is a simple quantum system, but the time evolution process
of the atomic system is not so simple, partly because the interaction
between the atom and the external electromagnetic field is usually
time-dependent, and on the other hand, because the atomic internal motion
and center-of-mass motion as well as the coupling of the two motions induced
by the external electromagnetic field need to be considered explicitly and
simultaneously in a theoretical treatment. The time evolution process of a
free atom in the presence of a laser light field becomes so complex that it
is generally difficult to solve exactly the unitary dynamical equation of
the atomic system. It is also quite inconvenient even to use an
approximation method to solve the unitary dynamical equation with a high
accuracy. The STIRAP-based unitary decelerating and accelerating processes
for a free atom are relatively simple, because these decelerating and
accelerating processes are adiabatic and the Raman laser light beams of the
STIRAP decelerating and accelerating pulse sequences affect only the three
chosen atomic internal states. These special points may simplify greatly the
investigation of the time evolution process of the atom in the STIRAP
decelerating and accelerating processes. Because the time evolution process
of the atom involves only the three chosen internal states of the atom, it
may be investigated conveniently in the atomic three-internal-state
subspace. On the other hand, it is well known that a unitary process of a
free atom absorbing or emitting a photon has to obey the energy, momentum,
and angular momentum conservation laws. The energy, momentum, and angular
momentum conservation laws put a restriction on the time evolution process
for the free atom during the STIRAP-based unitary decelerating and
accelerating processes. These laws lead to that the atomic motional momentum
cannot be changed arbitrarily by these Raman laser light beams in the STIRAP
unitary decelerating and accelerating processes, but rather it can be
changed only within the Raman-laser-light-induced momentum state subspace
[5, 19]. Then the time evolution process of a free atom may be investigated
conveniently in the Raman-laser-light-induced momentum state subspace in the
STIRAP unitary decelerating and accelerating processes. Here the unitarity
of these processes is emphasized again. This greatly simplifies the
evaluation for the time evolution process.

In this paper the basic STIRAP theory [15, 16, 17, 18] has been developed to
study and construct the STIRAP-based unitary decelerating and accelerating
processes for a free atom by combining the quantum superposition principle
[25] and the energy, momentum, and angular momentum conservation laws for
the atomic photon absorption and emission processes [5, 19]. This research
is focused on investigating analytically and quantitatively the effect of
the momentum distribution of a superposition of momentum states of the atom
on the STIRAP state transfer in these processes. Both the ideal and the real
adiabatic condition for the basic STIRAP decelerating and accelerating
processes are derived analytically. One important result of the paper is to
confirm theoretically the time- and space-compressing processes of the
quantum control process [22], which are realized by the STIRAP unitary
decelerating and accelerating processes, with the help of the STIRAP
state-transfer theory and the unitary quantum dynamics. \newline
\newline
\newline
{\Large 2. }{\large The Hamiltonian for a single atom in the presence of the
Raman laser light beams}

The STIRAP-based laser cooling processes and the unitary decelerating and
accelerating processes are generally involved in the center-of-mass motion,
the internal electronic motion, and the interaction between the two motions
in an atomic system. A complete theoretical description for these processes
need consider the atomic center-of-mass motion, the internal electronic
motion, and the coupling of both the motions of the atom. When an atom is
irradiated by an externally applied electromagnetic field such as the Raman
laser light beams in the STIRAP experiments, the total Hamiltonian for the
physical system consisting of the atom and the electromagnetic field may be
generally written as [25]%
\begin{equation}
H=H_{a}+H_{rad}+H_{int}  \tag{1}
\end{equation}%
where $H_{a}$ is the atomic Hamiltonian in the absence of the externally
applied electromagnetic field,%
\begin{equation}
H_{a}=\sum_{k}\frac{1}{2m_{k}}p_{k}^{2}+V_{a},  \tag{2}
\end{equation}%
$H_{rad}$ the Hamiltonian for the electromagnetic field which may be written
as 
\begin{equation*}
H_{rad}=\frac{1}{8\pi }\int d^{3}x(E.E^{\ast }+B.B^{\ast }),
\end{equation*}%
and $H_{int}$ the interaction between the atom and the externally applied
electromagnetic field,%
\begin{equation*}
H_{int}=-\sum_{k}\frac{e_{k}}{m_{k}c}p_{k}.A(r_{k},t)+\sum_{k}\frac{e_{k}^{2}%
}{2m_{k}c^{2}}A(r_{k},t)^{2}.
\end{equation*}%
The atomic Hamiltonian $H_{a}$ describes both the atomic center-of-mass
motion and the internal electronic (or spin) motion of the atom, while the
interaction $H_{int}$ induced by the external electromagnetic field creates
the coupling between the atomic center-of-mass motion and the internal
electronic motion. The external electromagnetic field usually is weak in the
conventional STIRAP experiments and the interaction $H_{int}$ between the
atom and the electromagnetic field could be considered as a perturbation.

It is usually inconvenient to use directly the total Hamiltonian of Eq. (1)
to describe the STIRAP-based decelerating and accelerating processes,
although it is the most exact to use the total Hamiltonian to treat these
processes theoretically. Without losing generality one may use a simpler
form of the total Hamiltonian of Eq. (1) to describe clearly the unitary
decelerating and accelerating processes. It is well known that the
semiclassical theory of electromagnetic radiation has been extensively used
to describe the laser spectroscopy in the atomic and molecular systems [1,
16, 17, 18], the atomic coherent laser-cooling processes [20, 21, 23], and
the atomic quantum interference experiments [10, 11, 12]. If the
semiclassical theory is also reasonable for the unitary decelerating and
accelerating processes, then the Hamiltonian $H_{rad}$ of the
electromagnetic field itself may be omitted from the total Hamiltonian of
Eq. (1). Though the semiclassical theory of electromagnetic radiation
generally can not describe exactly the spontaneous emission in an atomic
system [1, 25] and especially the spontaneous emission in the long-time
atomic laser cooling process [26], it may be suited to describe the
STIRAP-based unitary decelerating and accelerating processes as these
unitary STIRAP processes can avoid the atomic spontaneous emission by
setting the suitable experimental parameters. Furthermore, if the electric
dipole approximation for the atomic system is also reasonable, then one may
use conveniently the electric dipole interaction $H_{d}$ to replace the
interaction $H_{int}$ to describe the unitary decelerating and accelerating
processes. The electric dipole interaction $H_{d}$ may be written as 
\begin{equation}
H_{d}=-D.E(x,t)  \tag{3}
\end{equation}%
where $D$ is the atomic electric dipole moment and $E(x,t)$ is the electric
field of the externally applied electromagnetic field, and the coordinate $x$
in the electric field $E(x,t)$ is the center-of-mass position of the atom.
The electric dipole approximation is reasonable when the wave length of the
external electromagnetic field is much larger than the atomic dimension
under study so that the atom may be considered as a point particle --- a
point electric dipole --- in the electromagnetic field [1]. It has been
shown that in the electric dipole approximation the electric-dipole
Hamiltonian $H_{d}$ is really equivalent to the interaction $H_{int}$ up to
a gauge transformation which is also a unitary transformation [1, 27]. The
electric dipole approximation is very popular in the theoretical description
of a variety of laser light and matter interactions. For example, one
generally uses the electric-dipole approximation to deal with theoretically
the dynamical process of atomic laser cooling in an atomic ensemble [1, 4,
5, 10, 19, 20, 26]. Now in the semiclassical theory of electromagnetic
radiation and in the electric dipole approximation the total Hamiltonian
(1)\ of the atom in the external electromagnetic field is reduced to the
form 
\begin{equation}
H=P^{2}/(2M)+V(x)+H(r)+H_{d}  \tag{4}
\end{equation}%
where the sum of the first three terms is just the atomic Hamiltonian $H_{a}$
of Eq. (2).\ The center-of-mass Hamiltonian $H_{cm}=P^{2}/(2M)+V(x)$
describes the atomic center-of-mass motion, while the internal Hamiltonian $%
H(r)$ describes the internal electronic (or spin) motion of the atom. In the
center-of-mass Hamiltonian $H_{cm}$ the term $H_{K}=P^{2}/(2M)$ with the
atomic mass $M$ and momentum $P$ is the atomic kinetic energy and the term $%
V(x)$\ the atomic potential energy in an external potential field. In the
unitary decelerating and accelerating processes of a free atom the external
potential energy $V(x)$ of the atom is zero, i.e., $V(x)=0.$ If there is an
external electric (or magnetic) field during these processes, then the
external potential energy $V(x)$ could not be zero. For example, the
potential energy $V(x)\neq 0$ when an atom in a harmonic potential well is
applied by an external electromagnetic field. Generally, both the
center-of-mass kinetic energy operator $P^{2}/(2M)$ and the internal
Hamiltonian $H(r)$ are not commutable with the electric dipole interaction $%
H_{d}$. The Hamiltonian $H$ of Eq. (4) still needs to be further simplified
so that it can describe conveniently the standard three-state STIRAP
experiment. This simplification is based on the facts that the externally
applied electromagnetic field in the STIRAP experiments can only affect some
specific internal electronic energy levels of the atom and the internal
electronic states are discrete. In the following only one-dimensional
center-of-mass motion of the atom is considered.

Because there are the center-of-mass motion, the internal electronic motion,
and even the coupling between the two motions of an atom in the STIRAP
processes, one must use simultaneously both the atomic center-of-mass
motional states and the internal electronic states of the atom to describe
exactly the STIRAP processes. Suppose that the wave functions $|\psi
_{P}(x)\rangle $ and $|\psi _{j}(r)\rangle $ are the eigenstates of the
center-of-mass Hamiltonian $H_{cm}$ and the internal Hamiltonian $H(r)$,
respectively, 
\begin{equation*}
H_{cm}|\psi _{P}(x)\rangle =E_{P}|\psi _{P}(x)\rangle ,\text{ }H(r)|\psi
_{j}(r)\rangle =E_{j}|\psi _{j}(r)\rangle ,
\end{equation*}%
where $E_{P}$ and $E_{j}$ are the eigen-energy of the center-of-mass
motional state $|\psi _{P}(x)\rangle $ and the internal state $|\psi
_{j}(r)\rangle ,$ respectively. For the STIRAP processes of a free atom the
eigen-energy $E_{P}$ is equal to the atomic center-of-mass kinetic energy $%
P^{2}/2M$ since the potential energy $V(x)=0$ and $H_{cm}|\psi
_{P}(x)\rangle =H_{K}|\psi _{P}(x)\rangle =P^{2}/(2M)|\psi _{P}(x)\rangle .$
According to quantum mechanics [25] the complete set $\{|\psi _{P}(x)\rangle
|\psi _{j}(r)\rangle \}$ of the product states of the center-of-mass
motional states and the internal states of the atom\ can be used to describe
completely both the center-of-mass motion and the internal electronic motion
as well as the coupling of the two motions of the atom. Now the total wave
function $|\Psi (x,r,t)\rangle $ of the atom may be generally expressed as a
linear combination of these product states [25], 
\begin{equation*}
|\Psi (x,r,t)\rangle =\sum_{j,P}a(j,P,t)|\psi _{P}(x)\rangle |\psi
_{j}(r)\rangle
\end{equation*}%
where $a(j,P,t)$ is an expansion coefficient. In the decelerating and
accelerating processes an atom may be in a given internal state, while its
center-of-mass motional state is a superposition state. Then in this case
the total wave function of the atom may be expanded as%
\begin{equation}
|\Psi (x,r,t)\rangle =\sum_{j,p}a(j,p,t)|\psi _{p}(x,t)\rangle |\psi
_{j}(r)\rangle  \tag{5}
\end{equation}%
where the wave function $|\psi _{p}(x,t)\rangle $ is a superposition of the
atomic center-of-mass motional states. For example, the center-of-mass
motional state $|\psi _{p}(x,t)\rangle $ may be a wave-packet state of the
halting-qubit atom in the quantum control process, which was denoted as $%
|CM,R\rangle $ in the previous paper [22]. In the decelerating and
accelerating processes of a free atom any center-of-mass motional state $%
|\psi _{p}(x,t)\rangle $ of the atom may be expanded in terms of the
complete eigenstate set $\{|\psi _{P}(x)\rangle \}$ [25], 
\begin{equation}
|\psi _{p}(x,t)\rangle =\underset{P}{\sum }a(p,P)\exp (-\frac{i}{\hslash }%
\frac{P^{2}}{2M}t)|\psi _{P}(x)\rangle .  \tag{6}
\end{equation}%
Note that the state $|\psi _{P}(x)\rangle $ is also an eigenstate of the
atomic center-of-mass momentum operator $P$ and in one-dimensional case it
may be written as 
\begin{equation}
|\psi _{P}(x)\rangle \equiv |P\rangle =\frac{1}{\sqrt{2\pi }}\exp
(iPx/\hslash ).  \tag{7}
\end{equation}%
The momentum wave function $|P\rangle $ represents that the atom moves along
the direction $+x$ with the center-of-mass motional velocity $P/M$.

In what follows it is supposed that the halting-qubit atom has a three-level 
$\Lambda $ configuration for the three-state STIRAP experiments. In the
STIRAP experiment the external electromagnetic field can have a real effect
only on the specific three-state subspace $\{|\psi _{0}(r)\rangle ,$ $|\psi
_{1}(r)\rangle ,$ $|\psi _{2}(r)\rangle \}$ of the internal electronic
states of the atom. This three-state subspace will further simplify the
Hamiltonian of Eq. (4) as the time evolution process of the internal states
of the atom in the STIRAP experiment is confined in the three-state
subspace. Those internal states of the atom outside the three-state subspace
will not be affected by the external electromagnetic field and are not
considered in the STIRAP experiment. Thus, it is sufficient to use the
internal Hamiltonian $(H(r))$ projection onto the three-state subspace to
describe the three-state STIRAP experiment. Since the internal states $%
\{|\psi _{j}(r)\rangle \}$ are the eigenstates of the internal Hamiltonian $%
H(r)$, this projection Hamiltonian onto the three-state subspace may be
given by 
\begin{equation}
H(r)=E_{0}|\psi _{0}(r)\rangle \langle \psi _{0}(r)|+E_{1}|\psi
_{1}(r)\rangle \langle \psi _{1}(r)|+E_{2}|\psi _{2}(r)\rangle \langle \psi
_{2}(r)|.  \tag{8}
\end{equation}%
On the other hand, the electric dipole interaction $H_{d}$ of Eq. (3)\ can
also be simplified further in the three-state subspace. In the STIRAP-based
decelerating and accelerating processes the total external electromagnetic
field generally consists of a pair of counterpropagating electromagnetic
fields, which may be amplitude- and phase-modulation Raman laser light
beams. The total electric field for the two counterpropagating linearly
polarized Raman laser light beams may be expressed as 
\begin{equation*}
E(x,t)=\frac{1}{2}E_{L0}(t)\exp [i(-k_{L0}.x-\omega _{L0}t)]
\end{equation*}%
\begin{equation}
+\frac{1}{2}E_{L1}(t)\exp [i(k_{L1}.x-\omega _{L1}t)]+C.C.  \tag{9}
\end{equation}%
where $C.C.$ stands for the complex (or hermite) conjugate of the first two
terms. Notice that the three internal states $\{|\psi _{k}(r)\rangle \}$
have quite different energy eigenvalues. The first Raman laser light beam
with the electric field $E_{L0}(t)$ couples only the two internal states $%
|\psi _{0}(r)\rangle $ and $|\psi _{2}(r)\rangle $ of the atom, while the
second with the electric field $E_{L1}(t)$ connects only the two internal
states $|\psi _{1}(r)\rangle $ and $|\psi _{2}(r)\rangle $. The frequency
difference $|\omega _{L0}-\omega _{L1}|$ should be near the resonance
frequency of the two atomic internal energy levels $|\psi _{0}(r)\rangle $
and $|\psi _{1}(r)\rangle .$ The first Raman laser light beam usually is
named the pumping laser pulse and the second the Stokes laser pulse in the
laser spectroscopy [15]. If among the three internal states the two states $%
|\psi _{0}(r)\rangle $ and $|\psi _{1}(r)\rangle $ which are usually the
ground states have the same energy eigenvalues, then one may use a pair of $%
\sigma _{+}$ and $\sigma _{-}$ circularly polarized laser light beams [19]
to replace the present two Raman laser light beams, one circularly polarized
laser light beam coupling only the two internal states $|\psi _{0}(r)\rangle 
$ and $|\psi _{2}(r)\rangle $ and another connecting only the two internal
states $|\psi _{1}(r)\rangle $ and $|\psi _{2}(r)\rangle .$ Then in the
three-state (internal) subspace the electric dipole interaction $H_{d}$ of
Eq. (3) may be written as, in the rotating wave approximation [1],%
\begin{equation*}
H_{d}=\hslash \Omega _{02}(t)\exp \{i(-k_{L0}.x-\omega _{L0}t)\}|\psi
_{2}(r)\rangle \langle \psi _{0}(r)|
\end{equation*}%
\begin{equation}
+\hslash \Omega _{12}(t)\exp \{i(k_{L1}.x-\omega _{L1}t)\}|\psi
_{2}(r)\rangle \langle \psi _{1}(r)|+C.C.  \tag{10}
\end{equation}%
where the Rabi frequencies for the two Raman laser light beams are defined
as 
\begin{equation*}
\Omega _{02}(t)=-\frac{1}{2}\langle \psi _{2}(r)|D.E_{L0}(t)|\psi
_{0}(r)\rangle ,
\end{equation*}%
\begin{equation*}
\Omega _{12}(t)=-\frac{1}{2}\langle \psi _{2}(r)|D.E_{L1}(t)|\psi
_{1}(r)\rangle .
\end{equation*}%
\newline
The electromagnetic field of the Raman laser light beams is usually weak in
the conventional STIRAP experiments, so that the effect of the
electromagnetic field on the atom could be considered as a perturbation.
Hence the rotating wave approximation is reasonable. On the other hand, if
each Raman laser light beams in the STIRAP experiment is replaced with a
pair of the laser light beams with the orthogonal electric field vectors and
the suitable phases [38] or one circularly polarized laser light beam [19],
then the rotating-wave approximation may be eliminated and hence the
electric dipole interaction (10)\ may be constructed exactly. The total
Hamiltonian of Eq. (4) associated with the electric dipole interaction $%
H_{d} $ of Eq. (10) and the internal Hamiltonian $H(r)$\ of Eq. (8) and the
product basis set $\{|\psi _{P}(x)\rangle |\psi _{j}(r)\rangle \}$ may be
used conveniently to describe the STIRAP-based unitary decelerating and
accelerating processes of a free atom. The transition matrix elements of the
electric dipole interaction $H_{d}$ can be calculated in the product basis
set, 
\begin{equation*}
W(j^{\prime },P^{\prime };j,P)=\langle \psi _{j^{\prime }}(r)|\langle \psi
_{P^{\prime }}(x)|H_{d}|\psi _{P}(x)\rangle |\psi _{j}(r)\rangle .
\end{equation*}%
These matrix elements are not zero only when both the internal states $|\psi
_{j}(r)\rangle $ and $|\psi _{j^{\prime }}(r)\rangle $ are in the
three-state (internal) subspace. They are also subjected to the constraint
of the energy, momentum, and angular momentum conservation laws for the
atomic photon absorption and emission process in the STIRAP decelerating and
accelerating processes. This is an instance of the velocity-selective rules
which have been used in the atomic laser cooling processes [5, 19]. Below it
is shown how the energy and momentum conservative laws have a constraint on
the electric dipole transition matrix elements $\{W(j^{\prime },P^{\prime
};j,P)\}$ in the three-state STIRAP experiments of the STIRAP-based unitary
decelerating and accelerating processes.

In the STIRAP experiments the internal states $|\psi _{0}(r)\rangle $ and $%
|\psi _{1}(r)\rangle $ of the three-state subspace usually are taken as the
hyperfine ground electronic states $|g_{0}\rangle $ and $|g_{1}\rangle $ of
an atom, while the internal state $|\psi _{2}(r)\rangle $ may be taken as
some excited state $|e\rangle $ of the atom. For example, the two internal
states $|g_{0}\rangle $ and $|g_{1}\rangle $ may be the hyperfine ground
electronic states $3S_{1/2}$ ($F=1$) and $3S_{1/2}$ ($F=2$) of sodium atom ($%
Na$), respectively, while the excited state $|e\rangle $ may be the excited
electronic state $3P_{3/2}$ ($F=2$) of the sodium atom. Suppose that at the
initial time in the STIRAP experiment the atom is in the ground internal
state $|g_{0}\rangle $ and the center-of-mass momentum state $|P^{\prime
}\rangle =(\sqrt{2\pi })^{-1}\exp (iP^{\prime }x/\hslash ),$ which also
means that the atom is in the product state $|P^{\prime }\rangle
|g_{0}\rangle $ and it travels along the direction $+x$ with the velocity $%
v^{\prime }=P^{\prime }/M$. Now a laser light field propagating along the
direction $-x$ is applied to the atom. Then the moving atom may absorb a
photon from the laser light field if the frequency ($\omega =k_{0}c$) of the
laser light field is just equal to the transition frequency of the atom in
motion between the ground internal state $|g_{0}\rangle $ and the excited
state $|e\rangle $ after the Doppler effect is taken into account. After the
atom absorbs a photon from the laser light field, the atomic motional
momentum becomes $P^{\prime }-\hslash k_{0}$ according to the momentum
conservation law and hence the atom is decelerated by $\hslash k_{0}/M$.
Since the atomic internal energy levels are discrete, the momentum change of
the moving atom is also discrete after the atom absorbs a photon. This means
that when the atom is excited to the internal state $|e\rangle ,$ its
motional momentum can not take an arbitrary value but it has to be $%
P^{\prime }-\hslash k_{0}$ due to the fact that the atomic optical
absorption process obeys the energy and momentum conservation laws [19].
After the atom absorbs a photon it jumps to the excited state $|e\rangle $
from the ground state $|g_{0}\rangle $ and its initial motional state $%
|P^{\prime }\rangle $ is changed to $|P^{\prime }-\hslash k_{0}\rangle $ and
hence the atom is in the product excited state $|P^{\prime }-\hslash
k_{0}\rangle |e\rangle .$ This is just the atomic decelerating process based
on the optical absorption mechanism [23, 24]. The atom in the product
excited state $|P^{\prime }-\hslash k_{0}\rangle |e\rangle $ may be further
decelerated by another laser light field. This laser light field travels
along the same direction $+x$ as the atom and its frequency ($\omega
=k_{1}c, $ $k_{1}\neq k_{0}$) is equal to the transition frequency of the
moving atom between the ground internal state $|g_{1}\rangle $ and the
excited state $|e\rangle $ after the Doppler effect is taken into account.
Thus, this laser light field may stimulate the atom in the excited internal
state $|e\rangle $ to jump to the ground internal state $|g_{1}\rangle .$
Because the energy difference between the two internal states $|g_{1}\rangle 
$ and $|e\rangle $ is quite different from that one between the two internal
states $|g_{0}\rangle $ and $|e\rangle ,$ this laser light field will not
affect the transition between the two internal states $|g_{0}\rangle $ and $%
|e\rangle .$ Likewise, the first laser light field does not affect the
transition between the two internal states $|g_{1}\rangle $ and $|e\rangle .$
Different from the first decelerating process this atomic decelerating
process is based on the stimulated optical emission mechanism [23, 24]. An
atom in an excited state may jump to the ground state when it is stimulated
by an external laser light field [25]. When the atom in the excited state $%
|e\rangle $ jumps to the ground state $|g_{1}\rangle ,$ it may emit
coherently a photon to the laser light field. Since the atomic motion
direction is the same as the propagation direction of the laser light field,
the atom really sends part of its motional momentum to the laser light field
and hence is decelerated in the stimulated transition process from the
excited state $|e\rangle $ to the ground state $|g_{1}\rangle $. This part
of motional momentum is just $\hslash k_{1}$ according to the momentum
conservation law and accordingly the atom is decelerated by $\hslash
k_{1}/M. $ Thus, after the stimulated optical emission process the atom is
in the ground internal state $|g_{1}\rangle $ and has to be in the motional
state $|P^{\prime }-\hslash k_{0}-\hslash k_{1}\rangle ,$ that is, the atom
is in the product state $|P^{\prime }-\hslash k_{0}-\hslash k_{1}\rangle
|g_{1}\rangle .$ Both the atomic optical absorption and emission processes
are required to be unitary here, as pointed out before [22]. The unitary
decelerating process based on the three-state STIRAP process just consists
of the reversible optical absorption and emission processes mentioned above,
where the two Raman laser light beams are adiabatic and usually
counterpropagating. This basic STIRAP-based unitary decelerating process may
be expressed in an intuitive form 
\begin{equation}
|P+\hslash k_{0}\rangle |g_{0}\rangle \rightarrow |P\rangle |e\rangle
\rightarrow |P-\hslash k_{1}\rangle |g_{1}\rangle  \tag{11}
\end{equation}%
Here for convenience in the later discussion the atomic motion momentum $%
P^{\prime }$ is denoted as $P+\hslash k_{0}$. Thus, a conventional
three-state STIRAP pulse sequence may be really used as a basic unitary
decelerating sequence if the two Raman laser light beams of the STIRAP pulse
sequence are arranged suitably such that the moving atom is decelerated
consecutively by the two Raman laser light beams. Obviously, the inverse
process of the STIRAP-based unitary decelerating process (11) may be used to
accelerate the atom in motion. However, it is more convenient to use
directly the reversible optical absorption and emission processes to
accelerate an atom in motion and this may be achieved by setting suitably
the parameters of the two Raman laser light beams of the STIRAP\ pulse
sequence. In the STIRAP-based unitary accelerating process the first laser
light field that induces the optical absorption process travels along the
motional direction ($+x)$ of the atom, while the second laser light field
that stimulates the atomic optical emission process propagates along the
opposite direction $(-x)$ to the moving atom. When the atom in the initial
product state $|P^{\prime }\rangle |g_{0}\rangle $ is excited to the product
state $|P^{\prime }+\hslash k_{0}^{\prime }\rangle |e\rangle $ by the first
laser light field, it absorbs a photon from the laser light field and is
accelerated by $\hslash k_{0}^{\prime }/M.$ When the atom in the excited
state $|P^{\prime }+\hslash k_{0}^{\prime }\rangle |e\rangle $ jumps to the
ground state $|P^{\prime }+\hslash k_{0}^{\prime }+\hslash k_{1}^{\prime
}\rangle |g_{1}\rangle $ under the stimulation of the second laser light
field, it emits a photon to the laser light field and is accelerated further
by $\hslash k_{1}^{\prime }/M.$ Therefore, the basic STIRAP-based unitary
accelerating process may be expressed in an intuitive form $(P^{\prime
}=P-\hslash k_{0}^{\prime })$ 
\begin{equation}
|P-\hslash k_{0}^{\prime }\rangle |g_{0}\rangle \rightarrow |P\rangle
|e\rangle \rightarrow |P+\hslash k_{1}^{\prime }\rangle |g_{1}\rangle . 
\tag{11a}
\end{equation}%
In what follows only the basic STIRAP-based unitary decelerating process
(11) is treated explicitly. In an analogous way, one can also deal with the
basic STIRAP-based unitary accelerating process (11a).

It follows from the basic decelerating sequence (11) that the time evolution
process of the atom in the unitary decelerating process (11) is restricted
within the three-state (product state) subspace $\{|P+\hslash k_{0}\rangle
|g_{0}\rangle ,$ $|P\rangle |e\rangle ,$ $|P-\hslash k_{1}\rangle
|g_{1}\rangle \}$ for a given atomic motion momentum $P$. Then during the
STIRAP-based unitary decelerating process (11)\ the total wave function $%
|\Psi (x,r,t)\rangle $ of the atom at any instant of time $t$ can be
expanded in the three-state (product state) subspace [5, 10, 12, 19, 20],
according to the superposition principle in quantum mechanics [25], 
\begin{equation*}
|\Psi (x,r,t)\rangle =\sum_{P}\rho (P)\{A_{0}(P,t)|P+\hslash k_{0}\rangle
|g_{0}\rangle
\end{equation*}%
\begin{equation}
+A_{1}(P,t)|P\rangle |e\rangle +A_{2}(P,t)|P-\hslash k_{1}\rangle
|g_{1}\rangle \}.  \tag{12}
\end{equation}%
where the sum over the momentum $P$ is due to the fact that the atom may be
in a superposition of momentum states, as can be seen in Eq. (6), $\rho (P)$
is the time-independent amplitude which has the physical meaning that $|\rho
(P)|^{2}$ is the probability in the superposition to find the atom in the
three-state subspace $\{|P+\hslash k_{0}\rangle |g_{0}\rangle ,$ $|P\rangle
|e\rangle ,$ $|P-\hslash k_{1}\rangle |g_{1}\rangle \}$ labelled by the
momentum $P,$ and the time-dependent amplitudes $\{A_{k}(P,t)\}$ satisfies
the normalization condition: 
\begin{equation}
|A_{0}(P,t)|^{2}+|A_{1}(P,t)|^{2}+|A_{2}(P,t)|^{2}=1.  \tag{13}
\end{equation}%
The amplitude $\rho (P)$ is time-independent because the two Raman laser
light beams induce a change only within the three-state subspace $%
\{|P+\hslash k_{0}\rangle |g_{0}\rangle ,$ $|P\rangle |e\rangle ,$ $%
|P-\hslash k_{1}\rangle |g_{1}\rangle \}$ for each given momentum $P$ during
the unitary decelerating process [19]. If at the initial time $t_{0}$ the
atom is in the ground internal state $|g_{0}\rangle $ and in a wave-packet
motional state, then the initial wave packet state of the atom may be
expanded as 
\begin{equation}
|\Psi (x,r,t_{0})\rangle =\underset{P^{\prime }}{\sum }\rho (P^{\prime
})|P^{\prime }\rangle |g_{0}\rangle .  \tag{12a}
\end{equation}%
Obviously, here the coefficients $A_{0}(P,t_{0})=1$ and $%
A_{1}(P,t_{0})=A_{2}(P,t_{0})=0,$ which can be deduced from Eqs. (12) and
(12a) with $P^{\prime }=P+\hslash k_{0}$, while $|\rho (P^{\prime })|^{2}$
is the probability to find the atom in the three-state subspace $%
\{|P+\hslash k_{0}\rangle |g_{0}\rangle ,$ $|P\rangle |e\rangle ,$ $%
|P-\hslash k_{1}\rangle |g_{1}\rangle \}.$ As an example, the initial state $%
|\Psi (x,r,t_{0})\rangle $ may be taken as the Gaussian wave-packet motional
state of the halting-qubit atom in the right-hand potential well of the
double-well potential field in the quantum control process [22]. The
three-state (product state) subspace and the basic decelerating process
(11)\ show that only special dipole transition matrix elements $%
\{W(j^{\prime },P^{\prime };j,P)\}$ can take nonzero values, that is, for a
given momentum $P$ there are only four matrix elements to take nonzero
value: $W(e,P;g_{0},P+\hslash k_{0})$, $W(e,P;g_{1},P-\hslash k_{1}),$ $%
W(g_{0},P+\hslash k_{0};e,P)$, and $W(g_{1},P-\hslash k_{1};e,P).$ For the
basic STIRAP decelerating process (11) the total electric field of the two
Raman laser light beams can be explicitly obtained from equation (9) by
setting the parameter sets: $(E_{L0}(t),k_{L0},\omega
_{L0})=(E_{01}(t),k_{0},\omega _{0})$ and $(E_{L1}(t),k_{L1},\omega
_{L1})=(E_{12}(t),k_{1},\omega _{1})$, where the first Raman laser light
beam $(E_{01}(t),k_{0},\omega _{0})$ couples the two internal states $%
|g_{0}\rangle $ and $|e\rangle $ and its propagating direction is opposite
to the motional direction of the atom, while the the second beam $%
(E_{12}(t),k_{1},\omega _{1})$ connects the two internal states $%
|g_{1}\rangle $ and $|e\rangle $ and it travels along the motional direction 
$(+x)$ of the atom. Here suppose that the energy difference (measured in
frequency unit) between the two ground internal states $|g_{0}\rangle $ and $%
|g_{1}\rangle $ is much larger than the detunings of the two Raman laser
light beams. Now these four nonzero electric-dipole-transition matrix
elements for the basic decelerating process (11) can be obtained with the
help of the total electric field of Eq. (9) with these parameter settings,
the electric dipole interaction $H_{d}$ of Eq. (10), and the momentum
eigenstates of Eq. (7) as well as their orthogonalizations, 
\begin{equation*}
W_{02}(t)=W_{20}^{\ast }(t)=\langle e|\langle P|H_{d}|P+\hslash k_{0}\rangle
|g_{0}\rangle =\hslash \Omega _{02}(t)\exp (-i\omega _{0}t),
\end{equation*}%
\begin{equation*}
W_{20}(t)=\langle g_{0}|\langle P+\hslash k_{0}|H_{d}|P\rangle |e\rangle ,
\end{equation*}%
\begin{equation*}
W_{12}(t)=W_{21}^{\ast }(t)=\langle e|\langle P|H_{d}|P-\hslash k_{1}\rangle
|g_{1}\rangle =\hslash \Omega _{12}(t)\exp (-i\omega _{1}t),
\end{equation*}%
\begin{equation*}
W_{21}(t)=\langle g_{1}|\langle P-\hslash k_{1}|H_{d}|P\rangle |e\rangle .
\end{equation*}%
Here the star $\star $ stands for the complex conjugate and the Rabi
frequencies $\Omega _{02}(t)$ and $\Omega _{12}(t)$ are defined in Eq. (10)
with the states: $|\psi _{0}(r)\rangle =|g_{0}\rangle ,$ $|\psi
_{1}(r)\rangle =|g_{1}\rangle ,$ and $|\psi _{2}(r)\rangle =|e\rangle $.

Now the time evolution process of the atom in the presence of the Raman
laser light beams in the basic decelerating process (11)\ is described by
the time-dependent Schr\"{o}dinger equation:%
\begin{equation}
i\hslash \frac{\partial }{\partial t}\Psi (x,r,t)=H(t)\Psi (x,r,t).  \tag{14}
\end{equation}%
Here the total Hamiltonian $H(t)$ is given by Eq. (4), in which $V(x)=0$ and 
$H(r)$ and $H_{d}$ are given by Eq. (8) and (10), respectively, while the
wave function $|\Psi (x,r,t)\rangle $ is given by Eq. (12). By using the
four nonzero electric-dipole-transition matrix elements and the
orthonormalization of the momentum eigenstates of Eq. (7) the Schr\"{o}%
dinger equation (14) can be reduced to a three-state Schr\"{o}dinger
equation for a given momentum $P,$ which may be written in the matrix form%
\begin{equation}
i\hslash \frac{\partial }{\partial t}\left( 
\begin{array}{c}
A_{0}(P,t) \\ 
A_{1}(P,t) \\ 
A_{2}(P,t)%
\end{array}%
\right) =\hat{H}(P,t)\left( 
\begin{array}{c}
A_{0}(P,t) \\ 
A_{1}(P,t) \\ 
A_{2}(P,t)%
\end{array}%
\right)  \tag{14a}
\end{equation}%
where the three-state vector $(A_{0}(P,t),A_{1}(P,t),A_{2}(P,t))^{T}$ (here $%
T$ stands for the vector transpose) satisfies the normalization of Eq. (13)
and the reduced Hamiltonian $\hat{H}(P,t)$ is a $3\times 3-$dimensional
Hermitian matrix, 
\begin{equation*}
\hat{H}(P,t)=\left[ 
\begin{array}{ccc}
\frac{(P+\hslash k_{0})^{2}}{2M}+E_{0} & W_{02}^{\ast }(t) & 0 \\ 
W_{02}(t) & \frac{P^{2}}{2M}+E_{2} & W_{12}(t) \\ 
0 & W_{12}^{\ast }(t) & \frac{(P-\hslash k_{1})^{2}}{2M}+E_{1}%
\end{array}%
\right] .
\end{equation*}%
Here the three basis vectors $|1\rangle =(1,0,0)^{T},$ $|2\rangle
=(0,1,0)^{T},$ and $|3\rangle =(0,0,1)^{T}$ of the three-state vector space $%
\{(A_{0}(P,t),A_{1}(P,t),A_{2}(P,t))^{T}\}$ stand for the three basis
product states $|P+\hslash k_{0}\rangle |g_{0}\rangle ,$ $|P\rangle
|e\rangle ,$ and $|P-\hslash k_{1}\rangle |g_{1}\rangle $ of the original
three-state subspace $\{|P+\hslash k_{0}\rangle |g_{0}\rangle ,$ $|P\rangle
|e\rangle ,$ $|P-\hslash k_{1}\rangle |g_{1}\rangle \}$, respectively. This
reduced three-state Schr\"{o}dinger equation can describe completely the
three-state $STIRAP$ experiments just like the original Schr\"{o}dinger
equation (14). By making a unitary transformation on the three-state vector
in Eq. (14a) [20, 12, 18, 25]: 
\begin{equation}
\bar{A}_{0}(P,t)=\exp [\frac{i}{\hslash }(\frac{(P+\hslash k_{0})^{2}}{2M}%
+E_{0})t]A_{0}(P,t),  \tag{15a}
\end{equation}%
\begin{equation}
\bar{A}_{1}(P,t)=\exp [\frac{i}{\hslash }(\frac{P^{2}}{2M}%
+E_{2})t]A_{1}(P,t),  \tag{15b}
\end{equation}%
\begin{equation}
\bar{A}_{2}(P,t)=\exp [\frac{i}{\hslash }(\frac{(P-\hslash k_{1})^{2}}{2M}%
+E_{1})t]A_{2}(P,t),  \tag{15c}
\end{equation}%
the Schr\"{o}dinger equation (14a) is further reduced to the form%
\begin{equation}
i\hslash \frac{\partial }{\partial t}\left( 
\begin{array}{c}
\bar{A}_{0}(P,t) \\ 
\bar{A}_{1}(P,t) \\ 
\bar{A}_{2}(P,t)%
\end{array}%
\right) =H(P,t)\left( 
\begin{array}{c}
\bar{A}_{0}(P,t) \\ 
\bar{A}_{1}(P,t) \\ 
\bar{A}_{2}(P,t)%
\end{array}%
\right) .  \tag{16}
\end{equation}%
Now the Hamiltonian $H(P,t)$ is a traceless Hermitian matrix, 
\begin{equation*}
H(P,t)=\left[ 
\begin{array}{ccc}
0 & \bar{W}_{02}^{\ast }(P,t) & 0 \\ 
\bar{W}_{02}(P,t) & 0 & \bar{W}_{12}(P,t) \\ 
0 & \bar{W}_{12}^{\ast }(P,t) & 0%
\end{array}%
\right] ,
\end{equation*}%
where the time- and momentum-dependent complex parameters $\bar{W}_{02}(P,t)$
and $\bar{W}_{12}(P,t)$ are given respectively by%
\begin{equation*}
\bar{W}_{02}(P,t)=\hslash \Omega _{02}(t)\exp \{-i[\frac{2Pk_{0}+\hslash
k_{0}^{2}}{2M}-(\omega _{02}-\omega _{0})]t\},
\end{equation*}%
\begin{equation*}
\bar{W}_{12}(P,t)=\hslash \Omega _{12}(t)\exp \{-i[\frac{-2Pk_{1}+\hslash
k_{1}^{2}}{2M}-(\omega _{12}-\omega _{1})]t\},
\end{equation*}%
and the transition frequencies for the atomic internal states are defined by 
$\hslash \omega _{02}=E_{2}-E_{0}$ and $\hslash \omega _{12}=E_{2}-E_{1}.$
Suppose that the Raman laser light beams are amplitude- and
phase-modulating. Then the Rabi frequencies for the two Raman laser light
beams (the pumping pulse ($\Omega _{p}(t)$) and the Stokes pulse ($\Omega
_{s}(t)$)) are written as 
\begin{equation*}
\Omega _{02}(t)=\Omega _{p}(t)\exp [-i\phi _{0}(t)],\text{ }\Omega
_{12}(t)=\Omega _{s}(t)\exp [-i\phi _{1}(t)],
\end{equation*}%
and the parameters in the Hamiltonian $H(P,t)$ therefore are given by 
\begin{equation*}
\bar{W}_{02}(P,t)=\hslash \Omega _{p}(t)\exp [-i\alpha _{p}(P,t)],\text{ }%
\bar{W}_{12}(P,t)=\hslash \Omega _{s}(t)\exp [-i\alpha _{s}(P,t)],
\end{equation*}%
where the phases $\alpha _{p}(P,t)$ and $\alpha _{s}(P,t)$ are dependent
upon the momentum $P$,%
\begin{equation*}
\alpha _{p}(P,t)=[\frac{2Pk_{0}+\hslash k_{0}^{2}}{2M}-(\omega _{02}-\omega
_{0})]t+\phi _{0}(t),
\end{equation*}%
\begin{equation*}
\alpha _{s}(P,t)=[\frac{-2Pk_{1}+\hslash k_{1}^{2}}{2M}-(\omega _{12}-\omega
_{1})]t+\phi _{1}(t).
\end{equation*}%
Now the Hamiltonian $H(P,t)$ can be rewritten in the explicit form 
\begin{equation}
H(P,t)=\hslash \left[ 
\begin{array}{ccc}
0 & \Omega _{p}(t)e^{i\alpha _{p}(P,t)} & 0 \\ 
\Omega _{p}(t)e^{-i\alpha _{p}(P,t)} & 0 & \Omega _{s}(t)e^{-i\alpha
_{s}(P,t)} \\ 
0 & \Omega _{s}(t)e^{i\alpha _{s}(P,t)} & 0%
\end{array}%
\right]  \tag{17}
\end{equation}%
This type of Hamiltonians often have been met in the three-state STIRAP
experiments in the laser spectroscopy [4, 15, 17, 18] and in the atomic
interference experiments [12]. The three-state Schr\"{o}dinger equation (16)
and the traceless Hamiltonian (17) are the theoretical basis to design the
Raman adiabatic pulses of the STIRAP-based decelerating and accelerating
processes. There is a special point in the STIRAP-based unitary decelerating
and accelerating processes that the Hamiltonian (17) is dependent upon the
center-of-mass momentum of the atom besides the frequency offsets $\{\omega
_{k2}-\omega _{k}\}$ ($k=0$ and $1$). This is similar to the situations of
the STIRAP-based atomic laser cooling [20] and quantum interference
experiments [10, 12]. The effect of the frequency offsets on the STIRAP
population transfer has been examined in detail in the laser spectroscopy
[18b]. Though here considers only the pure-state quantum system of a single
atom instead of an atomic ensemble, the atomic momentum distribution could
have a great effect on the population transfer of the atom from an internal
state to another in the STIRAP decelerating and accelerating processes. This
is because the atom may be in a superposition of the momentum eigenstates
and hence has a momentum distribution. For example, though a freely moving
atom is in a Gaussian wave-packet state in coordinate space, it is also in a
superposition of the momentum eigenstates of the atom in momentum space.
Here the position of the momentum $P$ in the unitary decelerating and
accelerating processes is similar to that one of the frequency offsets in
the STIRAP\ experiments of the conventional laser spectroscopy [18]. The
frequency offsets (with respect to the transition frequencies of given
atomic internal energy levels) could be set at will for a single atom, since
they are the parameters of the Raman laser light beams, while the
superposition of momentum eigenstates of the atom (i.e. the atomic momentum
distribution) is an inherent property of the atomic motional state. Whether
or not the Raman adiabatic pulses obtained from the Schr\"{o}dinger equation
(16) and the Hamiltonian (17) are suitable for the decelerating and
accelerating processes are dependent upon the atomic momentum distribution.
The excitation bandwidth for a good STIRAP pulse sequence must be much
larger than the effective spreading of the atomic momentum distribution.

In the quantum control process [22] to simulate the reversible and unitary
halting protocol and the quantum search process the halting-qubit atom needs
to be decelerated and accelerated by the STIRAP-based unitary decelerating
and accelerating processes, respectively. As required by the quantum control
process, if the halting-qubit atom is completely in the ground internal
state $|g_{0}\rangle $ ($|g_{1}\rangle $) at the initial time in the
decelerating or accelerating process, then it must be converted completely
into another ground internal state $|g_{1}\rangle $ ($|g_{0}\rangle $) at
the end of the process. Because the conversion efficiency from the initial
state to the end state in these processes has a great effect on the
performance of the quantum control process, it is of crucial importance to
achieve a high enough conversion efficiency in these processes. This is the
first guidance to design the STIRAP pulse sequences for the unitary
decelerating and accelerating processes. On the other hand, in order that a
high conversion efficiency is achieved in the STIRAP\ experiments one must
also consider the decoherence effect due to the atomic spontaneous emission
when the atom is in the excited internal state. The atomic spontaneous
emission becomes an important factor to cause the decoherence effect when
the atom is in a short-lifetime excited internal state in the STIRAP
experiments. Therefore, the atom should avoid being in the excited internal
state during the STIRAP experiments. This may be realized by setting the
favorable detunings for the two Raman laser light beams. Since the three
product states $|P+\hslash k_{0}\rangle |g_{0}\rangle ,$ $|P\rangle
|e\rangle ,$ and $|P-\hslash k_{1}\rangle |g_{1}\rangle $ are the
eigenstates of the total atomic Hamiltonian $H_{a}$ of Eq. (2)\ and have
eigenenergy: $(P+\hslash k_{0})^{2}/(2M)+E_{0},$ $P^{2}/(2M)+E_{2},$ and $%
(P-\hslash k_{1})^{2}/(2M)+E_{1},$ respectively. Then the energy difference
between the ground state $|P+\hslash k_{0}\rangle |g_{0}\rangle $ and the
excited state $|P\rangle |e\rangle $ is given by $(E_{2}-E_{0})-P\hslash
k_{0}/M-\hslash ^{2}k_{0}^{2}/(2M)$ and the energy difference between the
ground state $|P-\hslash k_{1}\rangle |g_{1}\rangle $ and the excited state $%
|P\rangle |e\rangle $ is $(E_{2}-E_{1})+P\hslash k_{1}/M-\hslash
^{2}k_{1}^{2}/(2M).$ Since the Raman laser light beam with the carrier
frequency $\omega _{0}$ couples the ground state $|P+\hslash k_{0}\rangle
|g_{0}\rangle $ and the excited state $|P\rangle |e\rangle ,$ the detuning $%
\Delta _{p}$ of the beam is given by 
\begin{equation*}
\Delta _{p}(P)=(\omega _{02}-\omega _{0})-(\frac{k_{0}}{M})P-\frac{\hslash
k_{0}^{2}}{2M},
\end{equation*}%
while the detuning $\Delta _{s}$ of another Raman laser light beam with the
carrier frequency $\omega _{1}$ is 
\begin{equation*}
\Delta _{s}(P)=(\omega _{12}-\omega _{1})+(\frac{k_{1}}{M})P-\frac{\hslash
k_{1}^{2}}{2M}.
\end{equation*}%
By using the two formulae one may set conveniently the detunings $\Delta
_{p} $ and $\Delta _{s}$ for the STIRAP-based decelerating and accelerating
processes. Both the detunings are dependent upon the motional momentum of
the atom and also their frequency offsets, respectively. If the atom is in a
wave-packet state or generally in a superposition of momentum states, then
there is a distribution of momentum and the detuning settings should
consider the momentum distribution. In next sections the parameters of the
Raman laser light beams will be determined so as to arrive at the main goal
that the conversion efficiency from the initial state to the end state in
the decelerating and accelerating processes must be as close to 100\% as
possible. There the atomic momentum distribution must be taken into account.
Thus, this is involved in the adiabatic condition for the three-state
STIRAP\ state transfer of the decelerating and accelerating processes. 
\newline
\newline
{\large 3. The basic differential equations for the STIRAP-based
decelerating and accelerating processes}

The three-state STIRAP experiments have been studied extensively both in
theory and experiment in the laser spectroscopy [15, 16, 17, 18]. These
studies tell ones which conditions the complete population transfer can be
achieved by the STIRAP method among the energy levels of atomic and
molecular systems. From the point of view of theory an important feature of
the three-state STIRAP experiments is that there is the special eigenstate
of the Hamiltonian (17) that corresponds to the atomic trapping state [28].
This special eigenstate is independent of the intermediate state [17], which
is usually taken as the excited internal state of the atom in the STIRAP
experiments. Under the adiabatic condition the complete state or population
transfer through the special eigenstate can occur from the initial state
directly to the final state, while the intermediate state is bypassed in the
transfer process. Such an adiabatic state-transfer process is particularly
favorable for the STIRAP-based decelerating and accelerating processes. This
is because the atomic spontaneous emission could be avoided in the
decelerating and accelerating processes when the excited state of the atom
is bypassed in the STIRAP state-transfer processes. This also shows that the
semiclassical theory of\ electromagnetic radiation is suited to treat the
STIRAP experiments. On the other hand, from the experimental point of view
the three-state STIRAP experiment needs to set suitably the experimental
parameters for the Raman laser light beams. The important thing in
experiment is the settings for the Rabi frequencies of the two Raman laser
light beams and for the pulse delay between the two Raman laser light beams
[16]. The atomic system should first interact with the Stokes pulse and then
with the pumping pulse, and an appropriate overlapping between the two Raman
laser light beams is also required in experiment [3, 4, 15, 16, 17, 18].
These requirements are generally related to the adiabatic condition of the
three-state STIRAP experiment.

The special point for the STIRAP decelerating and accelerating processes is
that one must consider the atomic momentum distribution when a general
adiabatic condition is set up for these processes. According to the
adiabatic theorem and the adiabatic approximation method in quantum
mechanics [29, 30] one should first calculate the three eigenvectors and
eigenvalues of the instantaneous Hamiltonian $H(P,t)$ of Eq. (17), 
\begin{equation}
H(P,t)|g(P,t)\rangle =E(P,t)|g(P,t)\rangle .  \tag{18}
\end{equation}%
The three eigenvectors $\{|g(P,t)\rangle \}$ and their eigenvalues $%
\{E(P,t)\}$ of the Hamiltonian $H(P,t)$ are usually named the adiabatic
eigenvectors and eigenvalues, respectively. Notice that the three basis
vectors of the three-state vector space $%
\{(A_{0}(P,t),A_{1}(P,t),A_{2}(P,t))^{T}\}$ are $|1\rangle =(1,0,0)^{T},$ $%
|2\rangle =(0,1,0)^{T},$ and $|3\rangle =(0,0,1)^{T},$ respectively. The
three basis vectors really stand for the three basis product states $%
|P+\hslash k_{0}\rangle |g_{0}\rangle ,$ $|P\rangle |e\rangle ,$ and $%
|P-\hslash k_{1}\rangle |g_{1}\rangle $ of the original three-state subspace
for each given momentum $P$, respectively. Any one of the three eigenvectors
of the Hamiltonian $H(P,t)$ may be expanded in terms of the three basis
vectors. By the explicit Hamiltonian of Eq. (17) one can obtain one of the
three eigenvectors [15, 17, 18], 
\begin{equation}
|g^{0}(P,t)\rangle =\exp [-i\gamma (t)]\{\cos \theta (t)|1\rangle -\sin
\theta (t)\exp [-i(\alpha _{p}(P,t)-\alpha _{s}(P,t))]|3\rangle \}  \tag{19a}
\end{equation}%
where the phase $\gamma (t)\ $is a global phase, the mixing angle $\theta
(t) $ and the Rabi frequency $\Omega (t)$ are defined respectively by%
\begin{equation*}
\cos \theta (t)=\frac{\Omega _{s}(t)}{\Omega (t)},\text{ }\sin \theta (t)=%
\frac{\Omega _{p}(t)}{\Omega (t)},\text{ and }\Omega (t)=\sqrt{\Omega
_{p}(t)^{2}+\Omega _{s}(t)^{2}}.
\end{equation*}%
The eigenvalue corresponding to the eigenvector $|g^{0}(P,t)\rangle $ is $%
E^{0}\equiv \hslash \omega ^{0}=0.$ The eigenvector $|g^{0}(P,t)\rangle $ is
special in that it does not contain the intermediate eigenvector $|2\rangle $
which contains the excited internal state $|e\rangle $ of the atom. It is
the so-called atomic trapping state [28]. The adiabatic population transfer
is achieved through the eigenvector $|g^{0}(P,t)\rangle $ in all the
three-state STIRAP experiments [15, 17]. Therefore, when the initial state $%
|1\rangle $ is transferred to the final state $|3\rangle $ through the
eigenvector $|g^{0}(P,t)\rangle $ in the STIRAP state-transfer process, the
intermediate state $|2\rangle $ is not involved and hence is bypassed. Since
the phase difference $\alpha _{p}(P,t)-\alpha _{s}(P,t))$ in the eigenvector 
$|g^{0}(P,t)\rangle $ is dependent upon the atomic motional momentum $P$ the
adiabatic state transfer in the STIRAP experiments is really affected by the
atomic momentum distribution. In the laser spectroscopy the similar trapping
state is obtained [4, 15, 17, 18] but that state is not dependent upon the
momentum $P$. Thus, there is not any theoretical problem involved in the
effect of the momentum distribution on the adiabatic state transfer in the
laser spectroscopy. Of course, in the laser spectroscopy the Doppler effect
usually is considered in the STIRAP experiments of those quantum systems
such as an atomic or molecular beam [15]. However, in the atomic laser
cooling [5, 19, 20, 21], quantum interference experiments [10, 11, 12, 13,
14], and the atomic decelerating and accelerating processes the atomic
momentum distribution generally needs to be considered explicitly. The other
two adiabatic eigenstates of the Hamiltonian of Eq. (17) are given by [15,
17, 18]%
\begin{equation*}
|g^{\pm }(P,t)\rangle =\frac{1}{\sqrt{2}}\exp [-i\delta (t)]\{\sin \theta
(t)|1\rangle \mp \exp [-i\alpha _{p}(P,t)]|2\rangle
\end{equation*}%
\begin{equation}
+\cos \theta (t)\exp [-i(\alpha _{p}(P,t)-\alpha _{s}(P,t))]|3\rangle \} 
\tag{19b}
\end{equation}%
and their corresponding eigenvalues are $E^{\pm }\equiv \hslash \omega ^{\pm
}=\mp \hslash \Omega (t).$ Here the phase $\delta (t)$ is also a global
phase. The phase difference $\alpha _{p}(P,t)-\alpha _{s}(P,t)$ in the
adiabatic eigenvectors is given by%
\begin{equation*}
\alpha _{p}(P,t)-\alpha _{s}(P,t)=\phi _{0}(t)-\phi _{1}(t)
\end{equation*}%
\begin{equation}
+\{-[\omega _{02}-\omega _{0}]+[\omega _{12}-\omega _{1}]+\frac{%
P(k_{0}+k_{1})}{M}+\frac{\hslash k_{0}^{2}-\hslash k_{1}^{2}}{2M}\}t, 
\tag{20}
\end{equation}%
and there is a relation between the phase difference and the detunings: 
\begin{equation*}
\alpha _{p}(P,t)-\alpha _{s}(P,t)=[\phi _{0}(t)-\phi _{1}(t)]+[\Delta
_{s}(P)-\Delta _{p}(P)]t.
\end{equation*}%
The phase difference is dependent upon both the frequency offsets $(\omega
_{02}-\omega _{0})$ and $(\omega _{12}-\omega _{1})$ and also the momentum $%
P.$ It could be considered that the term $(k_{0}+k_{1})P/M$ in the phase
difference is generated by the Doppler effect. A large Doppler effect is not
favorable for the decelerating and accelerating processes. Assume that the
atomic motional state is a wave-packet state with a finite wave-packet
spreading. Then in order to minimize the effect of the Doppler-effect term $%
(k_{0}+k_{1})P/M$ on the STIRAP state transfer in the decelerating and
accelerating processes one must choose suitably the frequency offsets $%
(\omega _{02}-\omega _{0})$ and $(\omega _{12}-\omega _{1})$ for the two
Raman laser light beams. Now $P_{0}$ and $\Delta P_{M}$ are denoted as the
central position and the effective momentum bandwidth of the atomic momentum
wave-packet state$,$ respectively, and $\Delta P=P-P_{0}$ as the deviation
of the position $P$ of the momentum wave-packet state from the central
position $P_{0}.$ For simplicity, hereafter assume that the momentum $P_{0}$
is always much greater than $\Delta P_{M}/2$ in the unitary decelerating
process, this means that the atom always moves along the same direction $+x$
in the decelerating process. Suppose that the absolute amplitude at the
position $P$ of the momentum wave-packet state decays exponentially as the
deviation $\Delta P$. A typical example of such wave-packet states is
Gaussian wave-packet states [25, 31]. Then the amplitude of the position $P$
outside the effective momentum region $[P]=[P_{0}-\Delta P_{M}/2,$ $%
P_{0}+\Delta P_{M}/2]$ in the momentum wave-packet state is almost zero and
the probability to find that the atom is not in the effective momentum
region $[P]$ is so small that it can be negligible. This is just the
definition of the effective momentum bandwidth $\Delta P_{M}$ of the atomic
momentum wave-packet state. Then it is sufficient to consider only the
effective momentum region $[P]$ of the momentum wave-packet state when the
adiabatic condition is investigated for the three-state STIRAP experiments
of the decelerating and accelerating processes. Here the carrier frequencies 
$\omega _{0}$ and $\omega _{1}$ of the two Raman laser light beams are not
determined until the adiabatic condition for the STIRAP experiments is
examined later.

The unitary decelerating and accelerating processes of the quantum control
process [22] require that the atom in the initial internal state $%
|g_{0}\rangle $ ($|g_{1}\rangle $) be completely transferred to another
internal state $|g_{1}\rangle $ ($|g_{0}\rangle $) and at the same time the
initial atomic wave-packet motional state be completely converted into
another wave-packet motional state. It is well known that in an ideal
condition the STIRAP population transfer process can achieve 100\% transfer
efficiency from one atomic internal state to another [3, 4, 15, 17, 18]. The
adiabatic state-transfer channel for the three-state STIRAP experiments is
formed generally through the special adiabatic eigenstate $%
|g^{0}(P,t)\rangle $ of the Hamiltonian $H(P,t)$ [17]$.$ Therefore, if the
unitary decelerating and accelerating processes want to achieve their main
goal, they had better make full use of this adiabatic state-transfer
channel. If now the atom is prepared in the adiabatic eigenstate $%
|g^{0}(P,t_{0})\rangle $ of the Hamiltonian $H(P,t_{0})$ at the initial time 
$t_{0},$ then according as the adiabatic theorem [25, 30] the atom will be
in the adiabatic eigenstate $|g^{0}(P,t)\rangle $ of the Hamiltonian $H(P,t)$
at any instant of time $t$ in the adiabatic process for $t_{0}\leq t\leq
t_{0}+T$ if the adiabatic condition is met, that is, if the time period $T$
is infinitely large or more generally if the eigenvectors of the Hamiltonian 
$H(P,t)$ vary infinitely slowly [30]. However, in practice the time period $%
T $ can not be infinitely large and rotating of the eigenvectors of the
Hamiltonian are not infinitely slow. Thus, the ideal adiabatic condition can
not met perfectly in practice. In fact, the quantum control process does not
allow the time period $T$ of the adiabatic process to take an infinitely
large value. Obviously, if the time interval $T$ is taken as a finite value
but large enough or the eigenvectors of the Hamiltonian $H(P,t)$ rotate
sufficiently slowly, then the adiabatic theorem still holds approximately
and the real adiabatic state at the end time $t_{0}+T$ of the adiabatic
process will be very close\ to the ideal adiabatic eigenstate $%
|g^{0}(P,t_{0}+T)\rangle $. Since the real adiabatic state at the end of the
adiabatic process is close to the ideal adiabatic eigenstate $%
|g^{0}(P,t_{0}+T)\rangle ,$ one may calculate the real wave-packet state of
the atom at the end of the STIRAP-based unitary decelerating or accelerating
process with the help of the ideal adiabatic eigenstate $|g^{0}(P,t_{0}+T)%
\rangle $. Then this simplifies greatly the calculation for the real
wave-packet state of the atom at the end of the decelerating or accelerating
process, although such a calculation could generate an error for the real
wave-packet state of the atom. Similarly, if the real transfer efficiency
for the $STIRAP$ process is calculated with the help of the ideal adiabatic
eigenstate $|g^{0}(P,t_{0}+T)\rangle $ instead of the real adiabatic state
at the final time of the decelerating or accelerating process, then there
exists certainly an error for the real transfer efficiency. However, these
errors may be estimated. In fact, if one finds the real adiabatic condition
for the $STIRAP$ process, one may estimate these errors, as shown below.
Therefore, for the STIRAP\ process satisfying the real adiabatic condition
there is a simple scheme to calculate the real transfer efficiency and the
real wave-packet state of the atom at the end of the STIRAP process: one may
use the ideal adiabatic eigenstate $|g^{0}(P,t_{0}+T)\rangle $ of the
Hamiltonian $H(P,t_{0}+T)$ to simplify the calculation of the real transfer
efficiency and the real wave-packet state of the atom, then evaluate the
generated errors and control these errors to be within the desired upper
bound by setting suitably the experimental parameters for the STIRAP
process. This is really the procedure to design the STIRAP pulse sequences
for the unitary decelerating and accelerating processes. Thus, the problem
to be solved below is how to design the STIRAP\ pulse sequence such that in
the real adiabatic condition the final adiabatic state for the real
adiabatic process is still very close to the ideal adiabatic eigenstate $%
|g^{0}(P,t_{0}+T)\rangle $ even if the time period $T$ takes a finite value.

In the STIRAP-based decelerating and accelerating processes the adiabatic
evolution process of the atom\ could occur simultaneously and in a parallel
form in these three-state subspaces $\{|P+\hslash k_{0}\rangle |g_{0}\rangle
,$ $|P\rangle |e\rangle ,$ $|P-\hslash k_{1}\rangle |g_{1}\rangle \}$ for
all the given momentum $P$ within the effective momentum region $[P]$ if the
adiabatic condition is met in the effective momentum region $[P].$ It is
sufficient to examine the adiabatic evolution process of the atom in a
three-state subspace $\{|P+\hslash k_{0}\rangle |g_{0}\rangle ,$ $|P\rangle
|e\rangle ,$ $|P-\hslash k_{1}\rangle |g_{1}\rangle \}$ with a given
momentum $P$ of the effective momentum region $[P]$ to set up the adiabatic
condition for the STIRAP experiments of the decelerating and accelerating
processes. Obviously, any atomic state in the three-state subspace during
the adiabatic evolution process can be expanded in terms of the three basis
vectors of the subspace. If now the three basis vectors of the three-state
subspace $\{|P+\hslash k_{0}\rangle |g_{0}\rangle ,$ $|P\rangle |e\rangle ,$ 
$|P-\hslash k_{1}\rangle |g_{1}\rangle \}$ are taken as the three
orthonormal adiabatic eigenvectors $\{|g^{0}(P,t)\rangle ,$ $|g^{\pm
}(P,t)\rangle \}$ of the Hamiltonian $H(P,t),$ then the atomic three-state
vector $|\Phi (P,t)\rangle =(\bar{A}_{0}(P,t),\bar{A}_{1}(P,t),\bar{A}%
_{2}(P,t))^{T}$ in the Schr\"{o}dinger equation (16) at any instant of time $%
t$ in the adiabatic evolution process may be expanded as [4, 5, 10, 12, 15,
17, 18, 19, 20, 25] 
\begin{equation*}
|\Phi (P,t)\rangle =a_{0}(P,t)|g^{0}(P,t)\rangle
+a_{+}(P,t)|g^{+}(P,t)\rangle \exp [i\int_{t_{0}}^{t}dt^{\prime }\Omega
(t^{\prime })]
\end{equation*}%
\begin{equation}
+a_{-}(P,t)|g^{-}(P,t)\rangle \exp [-i\int_{t_{0}}^{t}dt^{\prime }\Omega
(t^{\prime })].  \tag{21}
\end{equation}%
By substituting the adiabatic eigenstates of Eqs. (19a) and (19b) into the
state $|\Phi (P,t)\rangle $ of Eq. (21) one can find that the coefficients $%
\{a_{0}(P,t),$ $a_{\pm }(P,t)\}$ in Eq. (21) are related to those $\{\bar{A}%
_{l}(P,t),$ $l=0,1,2\}$ in Eq. (16) by 
\begin{equation*}
\bar{A}_{0}(P,t)=a_{0}(P,t)\cos \theta (t)\exp [-i\gamma (t)]
\end{equation*}%
\begin{equation*}
+\frac{1}{\sqrt{2}}a_{+}(P,t)\sin \theta (t)\exp [-i\delta (t)]\exp
[i\int_{t_{0}}^{t}dt^{\prime }\Omega (t^{\prime })]
\end{equation*}%
\begin{equation}
+\frac{1}{\sqrt{2}}a_{-}(P,t)\sin \theta (t)\exp [-i\delta (t)]\exp
[-i\int_{t_{0}}^{t}dt^{\prime }\Omega (t^{\prime })],  \tag{22a}
\end{equation}%
\begin{equation*}
\bar{A}_{1}(P,t)=\frac{1}{\sqrt{2}}\exp [-i\alpha _{p}(P,t)]\exp [-i\delta
(t)]\{-a_{+}(P,t)\exp [i\int_{t_{0}}^{t}dt^{\prime }\Omega (t^{\prime })]
\end{equation*}%
\begin{equation}
+a_{-}(P,t)\exp [-i\int_{t_{0}}^{t}dt^{\prime }\Omega (t^{\prime })]\} 
\tag{22b}
\end{equation}%
\begin{equation*}
\bar{A}_{2}(P,t)=\exp [-i(\alpha _{p}(P,t)-\alpha
_{s}(P,t))]\{-a_{0}(P,t)\sin \theta (t)\exp [-i\gamma (t)]
\end{equation*}%
\begin{equation*}
+\frac{1}{\sqrt{2}}a_{+}(P,t)\cos \theta (t)\exp [-i\delta (t)]\exp
[i\int_{t_{0}}^{t}dt^{\prime }\Omega (t^{\prime })]
\end{equation*}%
\begin{equation}
+\frac{1}{\sqrt{2}}a_{-}(P,t)\cos \theta (t)\exp [-i\delta (t)]\exp
[-i\int_{t_{0}}^{t}dt^{\prime }\Omega (t^{\prime })]\}.  \tag{22c}
\end{equation}%
Thus, the atomic three-state vector $|\Phi (P,t)\rangle $ may be determined
if one knows the coefficients $a_{0}(P,t)$ and $a_{\pm }(P,t).$ Assume that
at the initial time $t_{0}$ the atom is in the adiabatic eigenstate $%
|g^{0}(P,t_{0})\rangle .$ Then the adiabatic theorem shows that the atom is
kept in the adiabatic eigenstate $|g^{0}(P,t)\rangle $ of the Hamiltonian $%
H(P,t)$ at any instant of time $t$ during the adiabatic process if the ideal
adiabatic condition is met. Then it follows from the atomic state $|\Phi
(P,t)\rangle $ of Eq. (21)\ that if the ideal adiabatic condition is met,
the time-dependent coefficient $|a_{0}(P,t)|$ should be kept almost
unchanged and is very close to unity over the whole adiabatic process, while
two other coefficients $\{|a_{\pm }(P,t)|\}$ should be very close to zero.
Generally, these coefficients may be evaluated by solving the Schr\"{o}%
dinger equation (16) that the state $|\Phi (P,t)\rangle $ of Eq. (21)\
obeys. Inserting the state $|\Phi (P,t)\rangle $ of Eq. (21) into the Schr%
\"{o}dinger equation (16) one obtains a set of the three differential
equations with the variables $a_{0}(P,t)$ and $a_{\pm }(P,t),$%
\begin{equation*}
\frac{\partial }{\partial t}a_{0}(P,t)+i\omega
_{0}(P,t)a_{0}(P,t)+a_{+}(P,t)\omega _{0,+}(P,t)\exp
[i\int_{t_{0}}^{t}dt^{\prime }\Omega (t^{\prime })]
\end{equation*}%
\begin{equation}
+a_{-}(P,t)\omega _{0,-}(P,t)\exp [-i\int_{t_{0}}^{t}dt^{\prime }\Omega
(t^{\prime })]=0,  \tag{23a}
\end{equation}%
\begin{equation*}
\frac{\partial }{\partial t}a_{\pm }(P,t)+i\omega _{\pm }(P,t)a_{\pm
}(P,t)+a_{\mp }(P,t)\omega _{\pm ,\mp }(P,t)\exp [\mp
2i\int_{t_{0}}^{t}dt^{\prime }\Omega (t^{\prime })]
\end{equation*}%
\begin{equation}
+a_{0}(P,t)\omega _{\pm ,0}(P,t)\exp [\mp i\int_{t_{0}}^{t}dt^{\prime
}\Omega (t^{\prime })]=0,  \tag{23b}
\end{equation}%
where the coefficients $\{\omega _{k}(P,t),$ $\omega _{k,l}(P,t)\}$ are
defined as 
\begin{equation*}
i\omega _{k}(P,t)=\langle g^{k}(P,t)|\partial g^{k}(P,t)/\partial t\rangle 
\text{ }(k=0,\pm ),
\end{equation*}%
\begin{equation*}
\omega _{k,l}(P,t)=\langle g^{k}(P,t)|\frac{\partial }{\partial t}%
|g^{l}(P,t)\rangle \text{ }(k\neq l,\text{ }k,l=0,\pm ).
\end{equation*}%
These coefficients can be obtained explicitly from the adiabatic eigenstates
of Eqs. (19), 
\begin{equation}
\omega _{0}(P,t)=-\frac{\partial }{\partial t}\gamma (t)-\sin ^{2}\theta (t)%
\frac{\partial }{\partial t}[\alpha _{p}(P,t)-\alpha _{s}(P,t)],  \tag{24a}
\end{equation}%
\begin{equation*}
\omega _{+}(P,t)=\omega _{-}(P,t)=-\frac{\partial }{\partial t}\delta (t)-%
\frac{1}{2}\frac{\partial }{\partial t}\alpha _{p}(P,t)
\end{equation*}%
\begin{equation}
-\frac{1}{2}\cos ^{2}\theta (t)\frac{\partial }{\partial t}[\alpha
_{p}(P,t)-\alpha _{s}(P,t)]  \tag{24b}
\end{equation}%
and 
\begin{equation*}
\omega _{0,\pm }(P,t)=-\omega _{\pm ,0}(P,t)^{\ast }=\frac{1}{\sqrt{2}}\exp
[i(\gamma (t)-\delta (t))]
\end{equation*}%
\begin{equation}
\times \{\frac{\partial }{\partial t}\theta (t)+i\sin \theta (t)\cos \theta
(t)\frac{\partial }{\partial t}[\alpha _{p}(P,t)-\alpha _{s}(P,t)]\}, 
\tag{24c}
\end{equation}%
\begin{equation*}
\omega _{+,-}(P,t)=\omega _{-,+}(P,t)=i\frac{1}{2}\frac{\partial }{\partial t%
}\alpha _{p}(P,t)
\end{equation*}%
\begin{equation}
-i\frac{1}{2}\cos ^{2}\theta (t)\frac{\partial }{\partial t}[\alpha
_{p}(P,t)-\alpha _{s}(P,t)].  \tag{24d}
\end{equation}%
\newline
Though these coefficients contain the global phases $\gamma (t)$ and $\delta
(t)$, the final results of the adiabatic process obtained from these
coefficients will not be affected by these global phase factors [18b], as
can be seen later. In order to solve conveniently the equations (23) it
could be better to make variable transformations [25, 30]: 
\begin{equation}
a_{k}(P,t)=b_{k}(P,t)\exp [-i\int_{t_{0}}^{t}dt^{\prime }\omega
_{k}(P,t^{\prime })]\text{ }(k=0,\pm ).  \tag{25}
\end{equation}%
Then with the new variables $\{b_{k}(P,t)\}$ these equations (23) are
changed to%
\begin{equation*}
\frac{\partial }{\partial t}b_{0}(P,t)=\frac{\exp [i(\gamma (t_{0})-\delta
(t_{0}))]}{\sqrt{2}}\Theta (P,t)^{\ast }
\end{equation*}%
\begin{equation}
\times \{b_{+}(P,t)\exp [i\int_{t_{0}}^{t}dt^{\prime }\Omega
_{+}(P,t^{\prime })]+b_{-}(P,t)\exp [-i\int_{t_{0}}^{t}dt^{\prime }\Omega
_{-}(P,t^{\prime })]\},  \tag{26a}
\end{equation}%
\begin{equation*}
\frac{\partial }{\partial t}b_{\pm }(P,t)=-i\frac{1}{2}b_{\mp }(P,t)\Gamma
(P,t)\exp [\mp i\int_{t_{0}}^{t}dt^{\prime }2\Omega (t^{\prime })]
\end{equation*}%
\begin{equation}
-\frac{\exp [-i(\gamma (t_{0})-\delta (t_{0}))]}{\sqrt{2}}b_{0}(P,t)\Theta
(P,t)\exp [\mp i\int_{t_{0}}^{t}dt^{\prime }\Omega _{\pm }(P,t^{\prime })]. 
\tag{26b}
\end{equation}%
The coefficients in Eqs. (26) are obtained from those in Eq. (24): 
\begin{equation*}
\Omega _{\pm }(P,t)=\Omega (t)\pm \{[\dot{\alpha}_{s}(P,t)-\frac{1}{2}\dot{%
\alpha}_{p}(P,t)]\sin ^{2}\theta (t)
\end{equation*}%
\begin{equation}
+[\dot{\alpha}_{p}(P,t)-\frac{1}{2}\dot{\alpha}_{s}(P,t)]\cos ^{2}\theta
(t)\},  \tag{27a}
\end{equation}%
\begin{equation}
\Theta (P,t)=-\dot{\theta}(t)+i\frac{1}{2}\sin 2\theta (t)[\dot{\alpha}%
_{p}(P,t)-\dot{\alpha}_{s}(P,t)],  \tag{27b}
\end{equation}%
\begin{equation}
\Gamma (P,t)=\sin ^{2}\theta (t)\dot{\alpha}_{p}(P,t)+\cos ^{2}\theta (t)%
\dot{\alpha}_{s}(P,t),  \tag{27c}
\end{equation}%
where $\dot{\theta}(t)=\frac{d}{dt}\theta (t),$ $\dot{\alpha}_{s}(P,t)=\frac{%
\partial }{\partial t}\alpha _{s}(P,t),$ and so on. The equations (26) are
the basic differential equations to describe completely the STIRAP-based
decelerating and accelerating processes. The set of basic equations (26) is
a generalization of the basic equations to describe the conventional
three-state STIRAP experiments [17, 16, 18] when the effect of a momentum
distribution is taken into account on the STIRAP experiments. The basic
equations (26) may be used to set up a general adiabatic condition for the
three-state STIRAP-based decelerating and accelerating processes.

The basic differential equations (26) may be rewritten in the matrix form%
\begin{equation}
i\frac{\partial }{\partial t}B(P,t)=M(P,t)B(P,t).  \tag{28}
\end{equation}%
Here the normalization three-state vector $B(P,t)$ is defined as $%
(b_{0}(P,t),b_{+}(P,t),$ $b_{-}(P,t))^{T}$ and the $3\times 3-$dimensional
hermitian Hamiltonian $M(P,t)$ is given by%
\begin{equation*}
M(P,t)=\left[ 
\begin{array}{ccc}
0 & M_{12} & M_{13} \\ 
M_{12}^{\ast } & 0 & M_{23} \\ 
M_{13}^{\ast } & M_{23}^{\ast } & 0%
\end{array}%
\right] ,
\end{equation*}%
in which the matrix elements $\{M_{ij}\}$ are defined by 
\begin{equation*}
M_{12}=\frac{i}{\sqrt{2}}\exp [i(\gamma (t_{0})-\delta (t_{0}))]\Theta
(P,t)^{\ast }\exp [i\int_{t_{0}}^{t}dt^{\prime }\Omega _{+}(P,t^{\prime })],
\end{equation*}%
\begin{equation*}
M_{13}=\frac{i}{\sqrt{2}}\exp [i(\gamma (t_{0})-\delta (t_{0}))]\Theta
(P,t)^{\ast }\exp [-i\int_{t_{0}}^{t}dt^{\prime }\Omega _{-}(P,t^{\prime })],
\end{equation*}%
\begin{equation*}
M_{23}=\frac{1}{2}\Gamma (P,t)\exp [-i\int_{t_{0}}^{t}dt^{\prime }2\Omega
(t^{\prime })].
\end{equation*}%
Notice that the momentum $P$ in the hermitian Hamiltonian $M(P,t)$ is a
parameter instead of an operator. The three-state Schr\"{o}dinger equation
(28) may have the formal solution:%
\begin{equation*}
B(P,t)=T\exp \{-i\int_{t_{0}}^{t}dt^{\prime }M(P,t^{\prime })\}B(P,t_{0})
\end{equation*}%
\begin{equation}
=\{1+(\frac{1}{i})\int_{t_{0}}^{t}dt_{1}M(P,t_{1})+(\frac{1}{i}%
)^{2}\int_{t_{0}}^{t}%
\int_{t_{0}}^{t_{1}}dt_{1}dt_{2}M(P,t_{1})M(P,t_{2})+...\}B(P,t_{0}). 
\tag{29}
\end{equation}%
The Dyson series solution (29) may be useful to set up a general adiabatic
condition for the STIRAP-based decelerating and accelerating processes. The
detailed discussion will appear in the section seven of the paper. \newline
\newline
{\large 4. The STIRAP state-transfer process in the ideal adiabatic
condition }

Consider first the special case: the ideal adiabatic condition. The ideal
adiabatic condition usually means that the time interval $T$ of the
adiabatic process is infinitely large or the adiabatic eigenstates of the
Hamiltonian (17) rotate infinitely slowly. Here the ideal adiabatic
condition means that for any instant of time of the adiabatic process the
integrations of the basic differential equations (26) approach zero for any
given momentum $P$ within the effective momentum region $[P]$ of the atomic
wave-packet motional state. The ideal adiabatic condition may be expressed
as 
\begin{equation}
\int_{t_{0}}^{t}dt^{\prime }[\frac{\partial }{\partial t^{\prime }}%
b_{l}(P,t^{\prime })]\rightarrow 0,\text{ }l=0,\pm ;\text{ }t_{0}\leq t\leq
t_{0}+T;\text{ }P\in \lbrack P].  \tag{30}
\end{equation}%
The ideal adiabatic condition (30)\ shows that the basic equations (26)\
have the solution $b_{l}(P,t)\rightarrow b_{l}(P,t_{0})$ ($l=0,\pm $) for
any instant of time $t$ of the adiabatic process and any given momentum $P$
within the effective momentum region $[P].$ If at the initial time $t_{0}$
the atom is prepared in the adiabatic eigenstate $|g^{0}(P,t_{0})\rangle ,$
which is the initial atomic state $|\Phi (P,t_{0})\rangle $ of Eq. (21)\
with $a_{0}(P,t_{0})=1$ and $a_{\pm }(P,t_{0})=0,$ then according to the
adiabatic theorem [25, 30]\ one should find that at any instant of time $t$
of the adiabatic process the atom is in the adiabatic eigenstate $%
|g^{0}(P,t)\rangle $ of the Hamiltonian $H(P,t).$ In fact, by the equations
(25) and (26) one can obtain $|a_{l}(P,t)|\rightarrow |a_{l}(P,t_{0})|$ ($%
l=0,\pm $) in the ideal adiabatic condition (30). Hence $|a_{0}(P,t)|%
\rightarrow |a_{0}(P,t_{0})|=1$ and $a_{\pm }(P,t)\rightarrow a_{\pm
}(P,t_{0})=0$ for the atomic state $|\Phi (P,t)\rangle $ at the instant of
time $t$ of the ideal adiabatic process, while this atomic state $|\Phi
(P,t)\rangle $ is just the adiabatic eigenstate $|g^{0}(P,t)\rangle $ of the
Hamiltonian $H(P,t),$ as can be seen from Eq. (21). The basic equations
(26)\ show that the ideal adiabatic condition (30) could be achieved if the
coefficients $\Theta (P,t)$ and $\Gamma (P,t)$ approach zero sufficiently
and the Rabi frequencies $\Omega (t)$ and $\Omega _{\pm }(P,t)$ are
sufficiently large. The Rabi frequencies $\Omega (t)$ and $\Omega _{\pm
}(P,t)$ can have a great effect on the adiabatic condition of the STIRAP
experiments. This point is very important for the quantum control process,
since a sufficiently long time period $T$ of the adiabatic process is not
accepted for the quantum control process. Then one may likely make the
adiabatic condition to be met by setting suitably the Rabi frequencies of
the Raman laser light beams, although the adiabatic condition is met usually
by making the time interval $T$ sufficiently large. A general adiabatic
condition will be discussed in detail later.

In order to optimize the adiabatic condition the carrier frequencies $\omega
_{0}$ and $\omega _{1}$ of the Raman laser light beams could be chosen
suitably such that 
\begin{equation}
\alpha _{p}(P,t)=[\frac{2Pk_{0}+\hslash k_{0}^{2}}{2M}-(\omega _{02}-\omega
_{0})]t+\phi _{0}(t)=\frac{\Delta P}{M}k_{0}t+\varphi _{0}(t)  \tag{31}
\end{equation}%
and%
\begin{equation}
\alpha _{s}(P,t)=[\frac{-2Pk_{1}+\hslash k_{1}^{2}}{2M}-(\omega _{12}-\omega
_{1})]t+\phi _{1}(t)=\frac{-\Delta P}{M}k_{1}t+\varphi _{1}(t)  \tag{32}
\end{equation}%
where the momentum difference $\Delta P=P-P_{0}$ and the momentum $P_{0}$ is
just the central position of the effective momentum region $[P]$ of a
general momentum wave-packet state. When the carrier frequencies are chosen
according to Eq. (31) and (32), the maximum momentum difference value within
the effective momentum region $[P]$ is minimum. These equations (31) and
(32) can be satisfied when the carrier frequencies $\{\omega _{l}\}$ ($l=0,1$%
) are determined from the two equations: 
\begin{equation}
\frac{\hslash k_{0}^{2}}{2M}-(\omega _{02}-\omega _{0})+c_{0}=-\frac{P_{0}}{M%
}k_{0}  \tag{33a}
\end{equation}%
and%
\begin{equation}
\frac{\hslash k_{1}^{2}}{2M}-(\omega _{12}-\omega _{1})+c_{1}=\frac{P_{0}}{M}%
k_{1}  \tag{33b}
\end{equation}%
where the wave number $k_{l}=\omega _{l}/c$, $c_{0}=\Delta _{p}(P_{0})$, $%
c_{1}=\Delta _{s}(P_{0})$, and $\frac{d}{dt}\phi _{l}(t)=c_{l}+\frac{d}{dt}%
\varphi _{l}(t).$ On the other hand, if the carrier frequencies are given in
advance, then the detunings $c_{0}=\Delta _{p}(P_{0})$ and $c_{1}=\Delta
_{s}(P_{0})$ may also be determined from these two equations (33) ,
respectively. In the decelerating and accelerating processes the momentum
value $P_{0}$ is generally different for each basic STIRAP-based
decelerating or accelerating process. Then one may adjust the carrier
frequencies $\{\omega _{l}\}$ or the detunings $\{c_{l}\}$ so that the
equations (33) can be satisfied. Now using the phase $\alpha _{p}(P,t)$ of
Eq. (31) and $\alpha _{s}(P,t)$ of Eq. (32) one can rewrite the phase
difference of Eq. (20) in the simple form%
\begin{equation}
\alpha _{p}(P,t)-\alpha _{s}(P,t)=\frac{\Delta P}{M}(k_{0}+k_{1})t+\varphi
_{0}(t)-\varphi _{1}(t).  \tag{34}
\end{equation}

If in the STIRAP experiments at the initial time $t_{0}$ the Rabi frequency $%
\Omega _{s}(t_{0})$ of the Stokes pulse is much greater than the one $\Omega
_{p}(t_{0})$ of the pumping pulse, then at the initial time the mixing angle 
$\theta (t)$ in the adiabatic eigenstates of Eqs. (19) satisfies [15, 16,
17, 18] 
\begin{equation}
\cos \theta (t_{0})=\frac{\Omega _{s}(t_{0})}{\Omega (t_{0})}\rightarrow 1%
\text{ or }\theta (t_{0})\rightarrow 0.  \tag{35}
\end{equation}%
In practice both the Rabi frequencies $\Omega _{s}(t_{0})$ and $\Omega
_{p}(t_{0})$ usually could be relatively small at the initial time, but the
Rabi frequency $\Omega _{p}(t_{0})$ of the pumping pulse is much less than $%
\Omega _{s}(t_{0})$ of the Stokes pulse. The initial mixing angle $\theta
(t_{0})\rightarrow 0$ leads the initial adiabatic eigenstates of the
Hamiltonian $H(P,t_{0})$ of Eq. (17) to the asymptotic forms 
\begin{equation}
|g^{0}(P,t_{0})\rangle \rightarrow \exp [-i\gamma (t_{0})]|1\rangle , 
\tag{36a}
\end{equation}%
\begin{equation*}
|g^{\pm }(P,t_{0})\rangle \rightarrow \frac{1}{\sqrt{2}}\exp [-i\delta
(t_{0})]\{\mp \exp [-i\alpha _{p}(P,t_{0})]|2\rangle
\end{equation*}%
\begin{equation}
+\exp [-i(\alpha _{p}(P,t_{0})-\alpha _{s}(P,t_{0}))]|3\rangle \}.  \tag{36b}
\end{equation}%
Notice that the three-state basis vectors $|1\rangle ,$ $|2\rangle ,$ and $%
|3\rangle $ stand for the basis vectors $|P+\hslash k_{0}\rangle
|g_{0}\rangle ,$ $|P\rangle |e\rangle ,$ and $|P-\hslash k_{1}\rangle
|g_{1}\rangle ,$ respectively. The adiabatic eigenstate $|g^{0}(P,t_{0})%
\rangle $ of the Hamiltonian $H(P,t_{0})$ is much simpler than $|g^{\pm
}(P,t_{0})\rangle ,$ the latter is more complex in that their expansion
coefficients are dependent on the momentum. In general, at the initial time
an atomic system may be prepared more easily in the adiabatic eigenstate $%
|g^{0}(P,t_{0})\rangle $ up to a global phase factor. An important example
is that at the initial time the atom is completely in the ground internal
state $|g_{0}\rangle $ and in the Gaussian wave-packet motional state or in
a superposition of the momentum states. Now consider that the initial state
of the atom is prepared in the superposition state $|\Psi (x,r,t_{0})\rangle 
$ given by Eq. (12a). By comparing Eq. (12) with Eq. (12a) with $P^{\prime
}=P+\hslash k_{0}$ one sees that the coefficients of the state $|\Psi
(x,r,t_{0})\rangle $ of Eq. (12) are given by $A_{0}(P,t_{0})=1,$ $%
A_{1}(P,t_{0})=0,$ and $A_{2}(P,t_{0})=0$ at the initial time $t_{0}$. Then
it follows from Eqs. (15) that the coefficients $\{\bar{A}_{l}(P,t_{0})\}$
can be obtained from $\{A_{l}(P,t_{0})\}:$ 
\begin{equation}
\bar{A}_{0}(P,t_{0})=\exp [\frac{i}{\hslash }(\frac{(P+\hslash k_{0})^{2}}{2M%
}+E_{0})t_{0}],\text{ }\bar{A}_{1}(P,t_{0})=\bar{A}_{2}(P,t_{0})=0.  \tag{37}
\end{equation}%
These coefficients associated with the initial mixing angle $\theta
(t_{0})=0 $ (here $\theta (t_{0})$ is so small that it can be taken as zero
without losing generality) are inserted into Eqs. (22) one obtains, by
solving the equations (22), 
\begin{equation}
a_{0}(P,t_{0})=\exp [i\gamma (t_{0})]\exp [\frac{i}{\hslash }(\frac{%
(P+\hslash k_{0})^{2}}{2M}+E_{0})t_{0}],\text{ }%
a_{+}(P,t_{0})=a_{-}(P,t_{0})=0.\newline
\newline
\tag{38}
\end{equation}%
Indeed, at the initial time the atom is completely in the adiabatic
eigenstate $|g^{0}(P,t_{0})\rangle $ of the Hamiltonian $H(P,t_{0})$ of (17)
because both the coefficients $a_{+}(P,t_{0})$ and $a_{-}(P,t_{0})$ are zero
in the initial atomic state $|\Phi (P,t_{0})\rangle $. It follows from Eq.
(25) that at the initial time $t_{0}$ the coefficient $%
a_{l}(P,t_{0})=b_{l}(P,t_{0})$ ($l=0,\pm $). Then at the initial time $t_{0}$
the variables $\{b_{l}(P,t_{0})\}$ of the basic equations (26) can be
obtained from those coefficients of Eq. (38), 
\begin{equation}
b_{0}(P,t_{0})=\exp [i\gamma (t_{0})]\exp [\frac{i}{\hslash }(\frac{%
(P+\hslash k_{0})^{2}}{2M}+E_{0})t_{0}],\text{ }%
b_{+}(P,t_{0})=b_{-}(P,t_{0})=0.  \tag{39}
\end{equation}%
These coefficients $\{b_{l}(P,t_{0})\}$ of Eq. (39) provide the basic
equations (26)\ with the initial values.

Now the atom with the initial wave-packet state $|\Psi (x,r,t_{0})\rangle $
of Eq. (12a) undergoes the STIRAP-based decelerating process (11)\ in the
ideal adiabatic condition (30). Then one can calculate the atomic state at
the end time $t_{f}=t_{0}+T$ of the ideal adiabatic process (11). This
atomic state $|\Psi (x,r,t)\rangle $ is still given by Eq. (12), but the
coefficients in Eq. (12) need to be calculated explicitly. By integrating
the basic equations (26)\ and using the ideal adiabatic condition (30) the
solution to the basic equations (26) is given by 
\begin{equation}
b_{l}(P,t)=b_{l}(P,t_{0}),\text{ }l=0,\pm ;\text{ }t_{0}\leq t\leq t_{0}+T;%
\text{ }P\in \lbrack P].  \tag{40}
\end{equation}%
Here the initial values $\{b_{l}(P,t_{0})\}$ are given by Eq. (39). Once the
coefficients $\{b_{l}(P,t)\}$ are obtained from Eqs. (40) one may use
equation (25) to calculate the coefficients $\{a_{l}(P,t)\}$: 
\begin{equation*}
a_{0}(P,t)=b_{0}(P,t_{0})\exp [i(\gamma (t)-\gamma (t_{0}))]
\end{equation*}%
\begin{equation}
\times \exp \{i\int_{t_{0}}^{t}dt^{\prime }\sin ^{2}\theta (t^{\prime })%
\frac{\partial }{\partial t^{\prime }}[\alpha _{p}(P,t^{\prime })-\alpha
_{s}(P,t^{\prime })]\},  \tag{41a}
\end{equation}%
\begin{equation}
a_{+}(P,t)=a_{-}(P,t)=0.  \tag{41b}
\end{equation}%
Here the frequency parameter $\omega _{0}(P,t)$ of Eq. (24a) has been used.
The fact that the coefficients $a_{+}(P,t)=a_{-}(P,t)=0$ shows that the atom
in the adiabatic eigenstate $|g^{0}(P,t_{0})\rangle $ of the Hamiltonian $%
H(P,t_{0})$ at the initial time $t_{0}$ is transferred adiabatically to the
adiabatic eigenstate $|g^{0}(P,t)\rangle $ of the Hamiltonian $H(P,t)$ and
finally to the desired adiabatic eigenstate $|g^{0}(P,t_{f})\rangle $ of the
Hamiltonian $H(P,t_{f})$ at the end of the adiabatic process. This is just
consistent with the prediction of the adiabatic theorem in quantum mechanics
[25, 30]. By substituting the coefficients $\{a_{l}(P,t),$ $l=0,\pm \}$ of
Eqs. (41) into Eqs. (22) one obtains 
\begin{equation}
\bar{A}_{0}(P,t)=\exp [-i\gamma (t)]a_{0}(P,t)\cos \theta (t),  \tag{42a}
\end{equation}%
\begin{equation}
\bar{A}_{1}(P,t)=0,  \tag{42b}
\end{equation}%
\begin{equation}
\bar{A}_{2}(P,t)=-a_{0}(P,t)\exp [-i\gamma (t)]\exp [-i(\alpha
_{p}(P,t)-\alpha _{s}(P,t))]\sin \theta (t).  \tag{42c}
\end{equation}%
The coefficient $\bar{A}_{1}(P,t)=0$ shows clearly that in the ideal
adiabatic condition the atomic excited internal state $|e\rangle $ indeed
does not appear during the STIRAP\ state-transfer process. Though one has
the coefficients $\{a_{l}(P,t),$ $l=0,\pm \}$ of Eqs. (41)\ at hand, it is
not sufficient from Eqs. (42) to determine uniquely the coefficients $\{\bar{%
A}_{k}(P,t)\}$ if one does not know the mixing angle $\theta (t)$ at the end
time $t_{f}=t_{0}+T$ of the ideal adiabatic process. Suppose that at the end
of the ideal adiabatic process the Rabi frequency $\Omega _{p}(t)$ for the
pumping pulse is much greater than the one $\Omega _{s}(t)$ of the Stokes
pulse. Then the mixing angle $\theta (t)$ at the end of the ideal adiabatic
process takes the asymptotic form [15, 17]%
\begin{equation*}
\sin \theta (t_{f})=\frac{\Omega _{p}(t_{f})}{\Omega (t_{f})}\rightarrow 1%
\text{ or }\theta (t_{f})\rightarrow \pi /2.
\end{equation*}%
Now inserting the mixing angle $\theta (t_{f})=\pi /2$ into Eqs. (42) one
obtains uniquely%
\begin{equation}
\bar{A}_{0}(P,t_{f})=\bar{A}_{1}(P,t_{f})=0,  \tag{43a}
\end{equation}%
\begin{equation}
\bar{A}_{2}(P,t_{f})=-a_{0}(P,t_{f})\exp [-i\gamma (t_{f})]\exp [-i(\alpha
_{p}(P,t_{f})-\alpha _{s}(P,t_{f}))].  \tag{43b}
\end{equation}%
Thus, with the aid of Eqs. (15) and (43) the atomic three-state vector $%
\{A_{0}(P,t),$ $A_{1}(P,t),A_{2}(P,t)\}^{T}$ at the end time $t_{f}$ of the
ideal adiabatic process is determined by 
\begin{equation}
A_{0}(P,t_{f})=A_{1}(P,t_{f})=0,  \tag{44a}
\end{equation}%
\begin{equation*}
A_{2}(P,t_{f})=-\exp [\frac{i}{\hslash }(\frac{(P+\hslash k_{0})^{2}}{2M}%
+E_{0})t_{0}]\exp [-\frac{i}{\hslash }(\frac{(P-\hslash k_{1})^{2}}{2M}%
+E_{1})t_{f}]
\end{equation*}%
\begin{equation}
\times \exp [-i(\alpha _{p}(P,t_{f})-\alpha _{s}(P,t_{f}))]\exp
\{i\int_{t_{0}}^{t_{f}}dt^{\prime }\sin ^{2}\theta (t^{\prime })\frac{%
\partial }{\partial t^{\prime }}[\alpha _{p}(P,t^{\prime })-\alpha
_{s}(P,t^{\prime })]\}.  \tag{44b}
\end{equation}%
The coefficients $A_{0}(P,t_{f})=A_{1}(P,t_{f})=0$ show that at the end of
the ideal adiabatic process the atom is transferred completely to the ground
internal state $|g_{1}\rangle $ from the initial internal state $%
|g_{0}\rangle $ by the basic STIRAP decelerating process (11). This is just
the desired result of the ideal adiabatic process (11). By using the phase
difference in Eq. (34) one can further express the coefficient $%
A_{2}(P,t_{f})$ as 
\begin{equation*}
A_{2}(P,t_{f})=\exp [i\beta (t_{f},t_{0})]\exp [\frac{i}{\hslash }(\frac{%
(P+\hslash k_{0})^{2}}{2M}+E_{0})t_{0}]
\end{equation*}%
\begin{equation*}
\times \exp [-\frac{i}{\hslash }(\frac{(P-\hslash k_{1})^{2}}{2M}%
+E_{1})t_{f}]
\end{equation*}%
\begin{equation}
\times \exp \{-i\frac{\Delta P}{M}(k_{0}+k_{1})[t_{0}+%
\int_{t_{0}}^{t_{f}}dt^{\prime }\cos ^{2}\theta (t^{\prime })]\}  \tag{44c}
\end{equation}%
where the global phase factor is given by 
\begin{equation*}
\exp [i\beta (t_{f},t_{0})]=-\exp [-i(\varphi _{0}(t_{f})-\varphi
_{1}(t_{f}))]
\end{equation*}%
\begin{equation}
\times \exp \{i\int_{t_{0}}^{t_{f}}dt^{\prime }\sin ^{2}\theta (t^{\prime })[%
\dot{\varphi}_{0}(t^{\prime })-\dot{\varphi}_{1}(t^{\prime })]\}.  \tag{45}
\end{equation}%
Once the atomic three-state vector $\{A_{0}(P,t),A_{1}(P,t),A_{2}(P,t)\}^{T}$
is obtained at the end time $t_{f}$ of the basic STIRAP decelerating process
(11), through the equation (12) one can calculate the motional state of the
atom at the end of the ideal adiabatic process. In order to calculate
conveniently the wave-packet motional state of the atom one may use the
momentum $P^{\prime }=P+\hslash k_{0}$ as variable to express the
three-state vector $\{A_{0}(P^{\prime },t),A_{1}(P^{\prime
},t),A_{2}(P^{\prime },t)\}^{T},$ and then the vector at the end time $t_{f}$
can be determined by 
\begin{equation}
A_{0}(P^{\prime },t_{f})=A_{1}(P^{\prime },t_{f})=0,  \tag{46a}
\end{equation}%
\begin{equation*}
A_{2}(P^{\prime },t_{f})=\exp [i\beta ^{\prime }(t_{f},t_{0})]\exp [\frac{i}{%
\hslash }(\frac{P^{\prime }{}^{2}}{2M}+E_{0})t_{0}]
\end{equation*}%
\begin{equation*}
\times \exp [-\frac{i}{\hslash }(\frac{(P^{\prime }-\hslash k_{0}-\hslash
k_{1})^{2}}{2M}+E_{1})t_{f}]
\end{equation*}%
\begin{equation}
\times \exp \{-i\frac{\Delta P^{\prime }}{M}(k_{0}+k_{1})[t_{0}+%
\int_{t_{0}}^{t_{f}}dt^{\prime }\cos ^{2}\theta (t^{\prime })]\}.  \tag{46b}
\end{equation}%
Note that here the momentum difference $\Delta P^{\prime }$ still represents
the deviation of the momentum point $P^{\prime }$ from the central point of
the effective momentum region $[P]$ in the initial wave-packet motional
state.

For the STIRAP-based accelerating process (11a) in the ideal adiabatic
condition the three-state vector $\{A_{0}(P^{\prime },t),A_{1}(P^{\prime
},t),A_{2}(P^{\prime },t)\}^{T}$ at the end of the ideal adiabatic process
(11a)\ should be determined by 
\begin{equation}
A_{0}(P^{\prime },t_{f})=A_{1}(P^{\prime },t_{f})=0,  \tag{47a}
\end{equation}%
\begin{equation*}
A_{2}(P^{\prime },t_{f})=\exp [i\beta ^{\prime }(t_{f},t_{0})]\exp [\frac{i}{%
\hslash }(\frac{P^{\prime }{}^{2}}{2M}+E_{0})t_{0}]
\end{equation*}%
\begin{equation*}
\times \exp [-\frac{i}{\hslash }(\frac{(P^{\prime }+\hslash k_{0}+\hslash
k_{1})^{2}}{2M}+E_{1})t_{f}]
\end{equation*}%
\begin{equation}
\times \exp \{i\frac{\Delta P^{\prime }}{M}(k_{0}+k_{1})[t_{0}+%
\int_{t_{0}}^{t_{f}}dt^{\prime }\cos ^{2}\theta (t^{\prime })]\}.  \tag{47b}
\end{equation}%
This is because the propagating directions of the Raman laser light beams in
the accelerating process are just opposite to those in the decelerating
process, respectively. \newline
\newline
{\large 5. The decelerating and accelerating processes of a Gaussian
wave-packet state in the ideal adiabatic condition}

As a typical example, consider that an atom with a Gaussian wave-packet
state is decelerated by the basic STIRAP-based decelerating sequence (11).
For simplicity, here consider the simple situation that the STIRAP
decelerating process (11) satisfies the ideal adiabatic condition (30). The
theory developed in previous sections can determine the wave-packet motional
state of the atom at the end of the basic decelerating process (11). Suppose
that the initial motional state of the atom is a standard Gaussian
wave-packet state [25, 31] in one-dimensional coordinate space: 
\begin{equation*}
\Psi _{0}(x,t_{0})=\exp (i\varphi _{0})[\frac{(\Delta x)^{2}}{2\pi }]^{1/4}%
\sqrt{\frac{1}{(\Delta x)^{2}+i(\frac{\hslash T_{0}}{2M})}}
\end{equation*}%
\begin{equation}
\times \exp \{-\frac{1}{4}\frac{(x-z_{0})^{2}}{(\Delta x)^{2}+i(\frac{%
\hslash T_{0}}{2M})}\}\exp \{ip_{0}x/\hslash \}.  \tag{48}
\end{equation}%
Here the complex linewidth of the Gaussian wave-packet state $\Psi
_{0}(x,t_{0})$ is defined as%
\begin{equation*}
W(T_{0})=(\Delta x)^{2}+i(\frac{\hslash T_{0}}{2M}).
\end{equation*}%
The probability density of the state $\Psi _{0}(x,t_{0})$ is given by%
\begin{equation*}
|\Psi _{0}(x,t_{0})|^{2}=\frac{1}{\sqrt{\pi }}\sqrt{\frac{1}{2}\frac{1}{%
(\Delta x)^{2}+(\frac{\hslash T_{0}}{2M(\Delta x)})^{2}}}\exp \{-\frac{1}{2}%
\frac{(x-z_{0})^{2}}{(\Delta x)^{2}+(\frac{\hslash T_{0}}{2M(\Delta x)})^{2}}%
\}.
\end{equation*}%
By comparing $|\Psi _{0}(x,t_{0})|^{2}$ with the standard Gaussian function $%
G(x)=[\varepsilon \sqrt{\pi }]^{-1}$ $\times \exp
[-(x-x_{0})^{2}/\varepsilon ^{2}]$ one sees that the center-of-mass position
and the wave-packet spreading of the state $\Psi _{0}(x,t_{0})$ are $z_{0}$
and $\varepsilon _{0}=\sqrt{2[(\Delta x)^{2}+(\frac{\hslash T_{0}}{2M(\Delta
x)})^{2}]},$ respectively. If the atom is a free particle, the Gaussian
wave-packet state $\Psi _{0}(x,t_{0})$ has an explicit physical meaning that
the atom with the Gaussian wave-packet state moves along the direction $+x$
with the velocity $p_{0}/M.$ One may expand the coordinate-space Gaussian
wave-packet state $\Psi _{0}(x,t_{0})$ in terms of the momentum eigenstates $%
\{|p\rangle \}$ of Eq. (7), 
\begin{equation}
\Psi _{0}(x,t_{0})=\sum_{p}\rho _{0}(p,t_{0})|p\rangle .  \tag{49}
\end{equation}%
The expansion coefficient or the amplitude $\rho _{0}(p,t_{0})$ is
determined by 
\begin{equation}
\rho _{0}(p,t_{0})=\int dx\Psi _{0}(x,t_{0})\psi _{p}(x)^{\ast }  \tag{49a}
\end{equation}%
where $\psi _{p}(x)=\frac{1}{\sqrt{2\pi }}\exp (ipx/\hslash )$ is just the
momentum eigenstate $|\psi _{p}(x)\rangle $ or $|p\rangle $ of Eq. (7). By
substituting the wave-packet state $\Psi _{0}(x,t_{0})$ of Eq. (48) and the
momentum eigenstate $|\psi _{p}(x)\rangle $ of Eq. (7) into the amplitude $%
\rho _{0}(p,t_{0})$ of Eq. (49a) one obtains, by a complex calculation, 
\begin{equation*}
\rho _{0}(p,t_{0})=\exp (i\varphi _{0})[\frac{2(\Delta x)^{2}}{\pi }%
]^{1/4}\exp \{-(\Delta x)^{2}(\frac{p-p_{0}}{\hslash })^{2}\}
\end{equation*}%
\begin{equation}
\times \exp \{-i(\frac{\hslash T_{0}}{2M})(\frac{p-p_{0}}{\hslash }%
)^{2}\}\exp \{-i(\frac{p-p_{0}}{\hslash })z_{0}\}.  \tag{50}
\end{equation}%
The amplitude $\rho _{0}(p,t_{0})$ is really the Fourier transform state of
the coordinate-space Gaussian wave-packet state $\Psi _{0}(x,t_{0})$. It
also has a Gaussian shape and this can be seen more clearly from its
absolute square: 
\begin{equation}
|\rho _{0}(p,t_{0})|^{2}=[\frac{2(\Delta x)^{2}}{\pi }]^{1/2}\exp
\{-2(\Delta x)^{2}(\frac{p-p_{0}}{\hslash })^{2}\}.  \tag{50a}
\end{equation}%
This is a standard Gaussian function with the propagation-vector variable $%
k=p/\hslash $. Therefore, it satisfies the normalization, 
\begin{equation}
\int_{-\infty }^{+\infty }dk|\rho _{0}(k,t_{0})|^{2}=\int_{-\infty }^{\infty
}dx|\Psi _{0}(x,t_{0})|^{2}=1.  \tag{51}
\end{equation}%
The center-of-mass position of the Gaussian function is $p_{0}$ for the
momentum variable or $p_{0}/\hslash $ for the propagation-vector variable.
The wave-packet spreading of the Gaussian function for the momentum $p$ is
determined by $(\Delta p)=\hslash /[\sqrt{2}(\Delta x)]$ and for the
propagation-vector variable $k$ is given by $\Delta k=[\sqrt{2}(\Delta
x)]^{-1}.$ The state $\rho _{0}(p,t_{0})$ is called the momentum-space
Gaussian wave-packet state of the atom. There is a Gaussian decay factor $%
\exp [-(\Delta x)^{2}(\frac{p-p_{0}}{\hslash })^{2}]$ in the momentum-space
wave-packet state $\rho _{0}(p,t_{0}),$ which decides the effective momentum
region $[P]$ of the momentum wave-packet state. Obviously, the probability
density $|\rho _{0}(p,t_{0})|^{2}$ approaches zero rapidly (exponentially)
when the momentum $p$ takes a value such that the deviation $|p-p_{0}|$ is
greater than the wave-packet spreading $(\Delta p)$. Equation (49)\ really
shows that the coordinate-space Gaussian wave-packet state $\Psi
_{0}(x,t_{0})$ can be expanded in terms of the momentum-space Gaussian
wave-packet states $\{\rho _{0}(p,t_{0})\}$.

Consider the superposition of the momentum wave-packet states $\{\rho
_{0}(p,t_{0})\}:$%
\begin{equation}
\Phi _{0}(x,t_{0})=\sum_{|p-p_{0}|\leq \Delta P_{M}/2}\rho
_{0}(p,t_{0})|p\rangle ,  \tag{52}
\end{equation}%
where $\Delta P_{M}$ is just the bandwidth of the effective momentum region $%
[P]$ of the momentum wave-packet state $\rho _{0}(p,t_{0})$. When the
bandwidth $\Delta P_{M}\rightarrow \infty ,$ the superposition state $\Phi
_{0}(x,t_{0})$ is just the Gaussian wave-packet state $\Psi _{0}(x,t_{0}),$
as can be seen from Eqs. (49) and (52). The deviation of the state $\Phi
_{0}(x,t_{0})$ from the state $\Psi _{0}(x,t_{0})$ may be measured by the
probability%
\begin{equation*}
P\{\Psi _{0}(x,t_{0})-\Phi _{0}(x,t_{0})\}=|\sum_{|p-p_{0}|>\Delta
P_{M}/2}\rho _{0}(p,t_{0})|p\rangle |^{2}
\end{equation*}%
\begin{equation*}
=2\int_{[\Delta P_{M}/2+p_{0}]/\hslash }^{\infty }dk|\rho _{0}(k,t_{0})|^{2}
\end{equation*}%
where the momentum eigenstates $|p\rangle $ of Eq. (7) and their orthonormal
relations and the momentum wave-packet state $\rho _{0}(p,t_{0})$ of Eq.
(50) have been used. Now using the probability density $|\rho
_{0}(p,t_{0})|^{2}$ of Eq. (50a)\ one has 
\begin{equation}
P\{\Psi _{0}(x,t_{0})-\Phi _{0}(x,t_{0})\}=\frac{2}{\sqrt{\pi }}%
\int_{y_{M}}^{\infty }dy\exp (-y^{2})  \tag{53}
\end{equation}%
where the lower integral limit is 
\begin{equation*}
y_{M}=\frac{(\Delta P_{M})(\Delta x)}{\sqrt{2}\hslash }.
\end{equation*}%
The probability $P\{\Psi _{0}(x,t_{0})-\Phi _{0}(x,t_{0})\}$ is bounded on
by [32] 
\begin{equation*}
\frac{2}{\sqrt{\pi }}\frac{\exp (-y_{M}^{2})}{y_{M}+\sqrt{y_{M}^{2}+2}}%
<P\{\Psi _{0}(x,t_{0})-\Phi _{0}(x,t_{0})\}\leq \frac{2}{\sqrt{\pi }}\frac{%
\exp (-y_{M}^{2})}{y_{M}+\sqrt{y_{M}^{2}+4/\pi }}.
\end{equation*}%
The important thing is that the probability $P\{\Psi _{0}(x,t_{0})-\Phi
_{0}(x,t_{0})\}$ decays exponentially with the number $y_{M}^{2}.$ If the
bandwidth $\Delta P_{M}$ of effective momentum region $[P]$ of the momentum
wave-packet state $\rho _{0}(p,t_{0})$ is chosen such that the number $%
y_{M}>>1,$ then the probability $P\{\Psi _{0}(x,t_{0})-\Phi _{0}(x,t_{0})\}$
is almost zero. As a result, the superposition state $\Phi _{0}(x,t_{0})$ is
almost equal to the Gaussian wave-packet state $\Psi _{0}(x,t_{0}).$

If an atomic system is in a momentum superposition state which spreads from $%
-\infty $ to $+\infty $ in momentum space$,$ then it is generally hard to
achieve a complete STIRAP state transfer in the atomic system by the basic
decelerating process (11), since the adiabatic condition can not be met over
the whole momentum space $(-\infty ,$ $+\infty ).$ On the other hand, an
almost complete STIRAP state transfer could be achieved for an atomic
wave-packet motional state with a finite wave-packet spreading by the basic
decelerating process (11). This can be illustrated through the momentum
wave-packet state $\Phi _{0}(x,t_{0})$ of Eq. (52). The state $\Phi
_{0}(x,t_{0})$ is also a superposition of the momentum eigenstates of Eq.
(7). All the momentum components of the state $\Phi _{0}(x,t_{0})$ are
within the effective momentum region $[P]=[p_{0}-\Delta P_{M}/2,$ $%
p_{0}+\Delta P_{M}/2]$, as can be seen in Eq. (52). When the bandwidth $%
\Delta P_{M}$ of the effective momentum region $[P]$ is finite and satisfies 
$p_{0}-\Delta P_{M}/2>\hslash k_{0}+\hslash k_{1}$, a complete STIRAP state
transfer could be achieved for the state $\Phi _{0}(x,t_{0})$ within the
effective momentum region $[P]$ by the STIRAP decelerating process (11) if
the ideal adiabatic condition (30) is met within the effective momentum
region $[P]$ for the decelerating process (11). Now using the same STIRAP
pulse sequence (11) one can make an almost complete STIRAP\ state transfer
for the Gaussian wave-packet state $\Psi _{0}(x,t_{0})$ as the state $\Psi
_{0}(x,t_{0})$ is almost equal to the state $\Phi _{0}(x,t_{0})$ when the
number $y_{M}>>1.$ Hereafter suppose that the number $y_{M}>>1$ so that the
Gaussian wave-packet state $\Psi _{0}(x,t_{0})$ can be replaced with the
state $\Phi _{0}(x,t_{0})$ and vice versa without generating a significant
error in evaluating the unitary decelerating and accelerating processes.

Now suppose that at the initial time $t_{0}$ the atom is in the Gaussian
wave-packet motional state $\Psi _{0}(x,t_{0})$ of (48) and in the internal
state $|g_{0}\rangle .$ Then the total product state of the atom at the
initial time is given by 
\begin{equation}
\Psi _{0}(x,r,t_{0})=\Psi _{0}(x,t_{0})|g_{0}\rangle =\sum_{p}\rho
_{0}(p,t_{0})|p\rangle |g_{0}\rangle .  \tag{54}
\end{equation}%
By comparing the product state $\Psi _{0}(x,r,t_{0})$\ with that state of
Eq. (12a) one can see that the amplitude $\rho _{0}(p,t_{0})$ in the state $%
\Psi _{0}(x,r,t_{0})$\ just corresponds to the amplitude $\rho (P^{\prime })$
in the state of Eq. (12a). This means that at the initial time $t_{0}$ the
probability to find the atom in the three-state subspace $\{|p\rangle
|g_{0}\rangle ,$ $|p-\hslash k_{0}\rangle |e\rangle ,$ $|p-\hslash
k_{0}-\hslash k_{1}\rangle |g_{1}\rangle \}$ is given by $|\rho
_{0}(p,t_{0})|^{2}$. Note that during the STIRAP decelerating process (11)
this probability is not changed with time. Obviously, the coefficients $%
A_{0}(p,t_{0})=1,$ $A_{1}(p,t_{0})=0,$ and $A_{2}(p,t_{0})=0$ at the initial
time, as can be seen in Eq. (54). Now the initial atomic product state $\Psi
_{0}(x,r,t_{0})$ of (54)\ undergoes the basic STIRAP decelerating process
(11). Then at the end time $t_{f}=t_{0}+T$ of the decelerating process (11)
the total product state of the atom is given by Eq. (12), 
\begin{equation*}
\Psi _{0}(x,r,t_{f})=\sum_{p}\rho _{0}(p,t_{0})\{A_{0}(p,t_{f})|p\rangle
|g_{0}\rangle
\end{equation*}%
\begin{equation}
+A_{1}(p,t_{f})|p-\hslash k_{0}\rangle |e\rangle +A_{2}(p,t_{f})|p-\hslash
k_{0}-\hslash k_{1}\rangle |g_{1}\rangle \}  \tag{55}
\end{equation}%
where in the ideal adiabatic condition (30) the coefficients $%
\{A_{2}(p,t_{f}),$ $l=0,1,2\}$ are given by Eqs. (46) with the parameter
settings $P^{\prime }=p$ and $\Delta P^{\prime }=p-p_{0}$. Though in Eq.
(55) the sum for the momentum $p$ runs over only the effective momentum
region $[P]$, it will not generate a significant error if the sum really
runs over the whole momentum region $(-\infty ,+\infty ),$ as pointed out
before. Since in the ideal adiabatic condition (30) the coefficients $%
A_{0}(p,t_{f})=A_{1}(p,t_{f})=0,$ the total product state (55) is reduced to
the simple form 
\begin{equation}
\Psi _{0}(x,r,t_{f})=\sum_{p}\rho _{0}(p,t_{0})A_{2}(p,t_{f})|p-\hslash
k_{0}-\hslash k_{1}\rangle |g_{1}\rangle .  \tag{56}
\end{equation}%
The product state $\Psi _{0}(x,r,t_{f})$ shows that in the ideal adiabatic
condition (30) at the end of the decelerating process (11)\ the atom is
completely in the ground internal state $|g_{1}\rangle $ and also in the
wave-packet motional state: 
\begin{equation}
\Psi _{0}(x,t_{f})=\sum_{p}\rho _{0}(p,t_{0})A_{2}(p,t_{f})|p-\hslash
k_{0}-\hslash k_{1}\rangle .  \tag{57}
\end{equation}%
The initial product state of Eq. (54) and the final product state of Eq.
(56) show that indeed, the atom is transferred completely from the initial
internal state $|g_{0}\rangle $ and the Gaussian wave-packet motional state $%
\Psi _{0}(x,t_{0})$ of (48) to the final internal state $|g_{1}\rangle $ and
the motional state $\Psi _{0}(x,t_{f})$ of (57)$,$ respectively, by the
STIRAP decelerating sequence (11) in the ideal adiabatic condition (30). It
can turn out that the motional state $\Psi _{0}(x,t_{f})$ of Eq. (57) is
still a Gaussian wave-packet state. By the new variable $q=(p-p_{0})/\hslash 
$ the coefficient $A_{2}(p,t_{f})$ of Eq. (46b)\ with $P^{\prime }=p$ and $%
\Delta P^{\prime }=p-p_{0}$ and the amplitude $\rho _{0}(p,t_{0})$ of Eq.
(50) are respectively rewritten as 
\begin{equation*}
A_{2}(p,t_{f})=\exp [i\beta (t_{f},t_{0})]\exp [\frac{i}{\hslash }(\frac{%
p_{0}^{2}}{2M}+E_{0})t_{0}]
\end{equation*}%
\begin{equation*}
\times \exp [-\frac{i}{\hslash }(\frac{(p_{0}-\hslash k_{0}-\hslash
k_{1})^{2}}{2M}+E_{1})t_{f}]\exp [-iq^{2}\frac{\hslash (t_{f}-t_{0})}{2M}]
\end{equation*}%
\begin{equation*}
\times \exp [-iq\frac{(p_{0}-\hslash k_{0}-\hslash k_{1})}{M}%
(t_{f}-t_{0})]\exp \{-iq\frac{\hslash (k_{0}+k_{1})}{M}%
\int_{t_{0}}^{t_{f}}dt^{\prime }\cos ^{2}\theta (t^{\prime })\}
\end{equation*}%
and 
\begin{equation*}
\rho _{0}(p,t_{0})=\exp (i\varphi _{0})[\frac{2(\Delta x)^{2}}{\pi }%
]^{1/4}\exp \{-q^{2}[(\Delta x)^{2}+i(\frac{\hslash T_{0}}{2M})]\}\exp
(-iqz_{0}).
\end{equation*}%
Inserting the two coefficients and the momentum eigenstate $|p-\hslash
k_{0}-\hslash k_{1}\rangle $ of Eq. (7) into Eq. (57) one obtains%
\begin{equation*}
\Psi _{0}(x,t_{f})=\exp (i\varphi _{0})[\frac{2(\Delta x)^{2}}{\pi }%
]^{1/4}\exp [i\beta (t_{f},t_{0})]\exp [\frac{i}{\hslash }(\frac{p_{0}^{2}}{%
2M}+E_{0})t_{0}]
\end{equation*}%
\begin{equation*}
\times \exp [-\frac{i}{\hslash }(\frac{(p_{0}-\hslash k_{0}-\hslash
k_{1})^{2}}{2M}+E_{1})t_{f}]\exp [i(p_{0}-\hslash k_{0}-\hslash
k_{1})x/\hslash ]
\end{equation*}%
\begin{equation*}
\times \frac{1}{\sqrt{2\pi }}\int_{-\infty }^{\infty }dq\exp (-aq^{2}+bq)
\end{equation*}%
where the sum $\sum_{p}$ has been replaced with the integral as the momentum 
$p$ is continuous for a free atom, and the coefficients $a$ and $b$ are
given by 
\begin{equation*}
a=(\Delta x)^{2}+i\frac{\hslash (T_{0}+t_{f}-t_{0})}{2M},
\end{equation*}%
\begin{equation*}
b=i[x-z_{0}-\frac{(p_{0}-\hslash k_{0}-\hslash k_{1})}{M}(t_{f}-t_{0})-\frac{%
\hslash (k_{0}+k_{1})}{M}\int_{t_{0}}^{t_{f}}dt^{\prime }\cos ^{2}\theta
(t^{\prime })].
\end{equation*}%
The Gaussian integral in the state $\Psi _{0}(x,t_{f})$ can be calculated by 
\begin{equation}
\int_{-\infty }^{\infty }dq\exp (-aq^{2}+bq)=\sqrt{\frac{\pi }{a}}\exp (%
\frac{b^{2}}{4a}).  \tag{58}
\end{equation}%
Now by a simple calculation one obtains the final state: 
\begin{equation*}
\Psi _{0}(x,t_{f})=\exp [i\varphi _{1}(t_{f},t_{0})][\frac{(\Delta x)^{2}}{%
2\pi }]^{1/4}\sqrt{\frac{1}{(\Delta x)^{2}+i\frac{\hslash (T_{0}+t_{f}-t_{0})%
}{2M}}}
\end{equation*}%
\begin{equation}
\times \exp \{-\frac{1}{4}\frac{(x-z_{1})^{2}}{[(\Delta x)^{2}+i\frac{%
\hslash (T_{0}+t_{f}-t_{0})}{2M}]}\}\exp \{ip_{1}x/\hslash \}.  \tag{59}
\end{equation}%
Indeed, the motional state $\Psi _{0}(x,t_{f})$ is also a Gaussian
wave-packet state just like the initial motional state $\Psi _{0}(x,t_{0})$
of Eq. (48). Here the Gaussian wave-packet state $\Psi _{0}(x,t_{f})$ has
the center-of-mass position 
\begin{equation}
z_{1}=z_{0}+\frac{(p_{0}-\hslash k_{0}-\hslash k_{1})}{M}(t_{f}-t_{0})+\frac{%
\hslash (k_{0}+k_{1})}{M}\int_{t_{0}}^{t_{f}}dt^{\prime }\cos ^{2}\theta
(t^{\prime }),  \tag{60}
\end{equation}%
the wave-packet spreading 
\begin{equation}
\varepsilon =\sqrt{2(\Delta x)^{2}+2[\frac{\hslash (T_{0}+t_{f}-t_{0})}{%
2M(\Delta x)}]^{2}},  \tag{61}
\end{equation}%
the mean motional momentum 
\begin{equation*}
p_{1}=p_{0}-\hslash k_{0}-\hslash k_{1},
\end{equation*}%
and the global phase factor 
\begin{equation*}
\exp [i\varphi _{1}(t_{f},t_{0})]=\exp (i\varphi _{0})\exp [i\beta
(t_{f},t_{0})]
\end{equation*}%
\begin{equation*}
\times \exp \{\frac{i}{\hslash }(\frac{p_{0}^{2}}{2M}+E_{0})t_{0}\}\exp \{-%
\frac{i}{\hslash }(\frac{(p_{0}-\hslash k_{0}-\hslash k_{1})^{2}}{2M}%
+E_{1})t_{f}\}.
\end{equation*}%
It is interesting to compare the final motional state $\Psi _{0}(x,t_{f})$
with the initial state $\Psi _{0}(x,t_{0})$ of the decelerating process
(11). It needs only three parameters to characterize completely a Gaussian
wave-packet motional state of a free particle: the center-of-mass position,
the mean motional momentum (or velocity), and the complex linewidth. Here
the wave-packet spreading is determined completely by the complex linewidth.
For the first point, the atom is decelerated by $\hslash k_{0}/M+\hslash
k_{1}/M$ by the STIRAP pulse sequence (11) as expected, because the average
motional momentum $p_{1}=p_{0}-\hslash k_{0}-\hslash k_{1}$ of the final
state $\Psi _{0}(x,t_{f})$ is smaller than $p_{0}$ of the initial state $%
\Psi _{0}(x,t_{0})$ and their difference is $(\hslash k_{0}+\hslash k_{1}).$
Here suppose that the initial velocity for the moving atom $p_{0}/M>>\hslash
(k_{0}+k_{1})/M.$ Note that $\hslash k_{0}/M$ and $\hslash k_{1}/M$ are the
atomic recoil velocities in the two Raman laser light beams with the wave
numbers $k_{0}$ and $k_{1}$, respectively. For the second point, the atom
moves a distance $(z_{1}-z_{0})$ along the direction $+x$ during the
decelerating process. If a free atom moved along the direction $+x$ with the
velocities $p_{0}/M$ and $p_{1}/M,$ respectively, then in the time interval $%
T=t_{f}-t_{0}$ the atom would move distances equal to $T\times p_{0}/M$ and $%
T\times p_{1}/M,$ respectively. Here the velocities $p_{0}/M$ and $p_{1}/M$
are the atomic moving velocities before and after the decelerating process,
respectively. One can expect that the distance $(z_{1}-z_{0})$ should lie in
between the distances $T\times p_{1}/M$ and $T\times p_{0}/M,$ that is, $%
T\times p_{1}/M<z_{1}-z_{0}<T\times p_{0}/M.$ Indeed, the equation (60)
shows this point. For the third point, the wave-packet spreading of the atom
at the end of the decelerating process is larger than the initial one. If
one compares the wave-packet spreading of Eq. (61) with the free-atom
wave-packet spreading (see section 6 below), one can see that though the
atom is irradiated by the Raman laser light beams, the wave-packet spreading
of the atom during the decelerating process is not really affected by the
Raman laser light beams and is just the same as that one of the atom in the
absence of the Raman laser light beams. This point is important as the
wave-packet spreading of the atom after the decelerating (or accelerating)
process can be calculated easily. Note that the wave-packet spreading of a
free atom becomes larger and larger as time increases, as can be seen in
section 6 below.

From the experimental viewpoint one does not expect that after the
decelerating process (11) the wave-packet spreading of the atomic momentum
wave-packet state could become larger, because this may make a trouble for
the design of the STIRAP pulse sequence (11). Fortunately, it can turn out
that in the ideal adiabatic condition (30) the wave-packet spreading of the
momentum wave-packet state is not really affected by the STIRAP pulse
sequence (11). One can expand the wave-packet motional state of Eq. (59) in
terms of the momentum eigenstates $\{|p\rangle \}$ of Eq. (7)$:$ $\Psi
_{0}(x,t_{f})=\sum_{p}\rho _{1}(p,t_{f})|p\rangle $ just like the expansion
(49), and just like the momentum wave-packet state $\rho _{0}(p,t_{0})$ the
momentum wave-packet state $\rho _{1}(p,t_{f})$ can be calculated from Eq.
(49a), \newline
\begin{equation*}
\rho _{1}(p,t_{f})=\exp [i\varphi _{1}(t_{f},t_{0})][\frac{2(\Delta x)^{2}}{%
\pi }]^{1/4}\exp \{-(\Delta x)^{2}(\frac{p-p_{1}}{\hslash })^{2}\}
\end{equation*}%
\begin{equation}
\times \exp \{-i\frac{\hslash (T_{0}+t_{f}-t_{0})}{2M}(\frac{p-p_{1}}{%
\hslash })^{2}\}\exp [-i\frac{(p-p_{1})}{\hslash }z_{1}].  \tag{62}
\end{equation}%
Indeed, the probability density $|\rho _{1}(p,t_{f})|^{2}$ is also a
Gaussian function and it is really equal to $|\rho _{0}(p,t_{0})|^{2}$ of
the initial state $\Psi _{0}(x,t_{0})$ if the initial momentum $p_{0}$ is
replaced with $p_{1}.$ Thus, the wave-packet spreading of the Gaussian
function $|\rho _{1}(p,t_{f})|^{2}$ is equal to that one of $|\rho
_{0}(p,t_{0})|^{2}$ and is also given by $(\Delta p)=\hslash /[\sqrt{2}%
(\Delta x)].$ This shows that the effective momentum region $[P]$ of the
initial momentum wave-packet state $\rho _{0}(p,t_{0})$ is not changed after
the atom is decelerated by the STIRAP pulse sequence (11), although the
center-of-mass position of the momentum wave-packet state is changed to $%
p_{1}=p_{0}-\hslash k_{0}-\hslash k_{1}$ from the initial one $p_{0}$ after
the atom is decelerated.

In the quantum control process the halting-qubit atom usually needs to be
decelerated continuously [22] because each STIRAP pulse sequence (11)
usually can decelerate the atom only by a small velocity value. As shown
above, after the STIRAP decelerating process (11) the decelerated atom is in
the product state $\Psi _{0}(x,r,t_{f})$ of Eq. (56), that is, the atom is
in the internal state $|g_{1}\rangle $ and the Gaussian wave-packet state $%
\Psi _{0}(x,t_{f})$ of Eq. (59). Now the atom needs to be decelerated
further by another STIRAP pulse sequence. This basic STIRAP-based
decelerating process may be expressed as 
\begin{equation}
|P+\hslash l_{0}\rangle |g_{1}\rangle \rightarrow |P\rangle |e\rangle
\rightarrow |P-\hslash l_{1}\rangle |g_{0}\rangle .  \tag{63}
\end{equation}%
In this decelerating process the atom is changed from the internal state $%
|g_{1}\rangle $ to $|g_{0}\rangle .$ This is opposite to the previous
decelerating process (11). Therefore, the experimental parameters for the
STIRAP pulse sequence (63) needs to be set suitably. Now the two Raman laser
light beams in Eq. (9) for the STIRAP pulse sequence (63) should have the
parameter settings: $(E_{L0}(t),k_{L0},\omega
_{L0})=(E_{01}^{l}(t),l_{0},\omega _{0}^{l})$ and $(E_{L1}(t),k_{L1},\omega
_{L1})=(E_{12}^{l}(t),l_{1},\omega _{1}^{l})$, where the first Raman laser
light beam $(E_{01}^{l}(t),l_{0},\omega _{0}^{l})$ (the pumping pulse)
couples the two internal states $|g_{1}\rangle $ and $|e\rangle $ and it
propagates along the opposite motional direction to the atom, while the
second beam $(E_{12}^{l}(t),l_{1},\omega _{1}^{l})$ (the Stokes pulse)
connects the two internal states $|g_{0}\rangle $ and $|e\rangle $ and it
travels along the motional direction of the atom. The atomic three-state
subspace $\{|\psi _{k}(r)\rangle \}$ in the STIRAP decelerating process (63)
should be set by $|\psi _{0}(r)\rangle =|g_{1}\rangle ,$ $|\psi
_{1}(r)\rangle =|g_{0}\rangle ,$ and $|\psi _{2}(r)\rangle =|e\rangle ,$ and
the transition frequencies between the atomic internal energy levels in the
decelerating process (63) should be defined by $\omega _{02}^{l}=\omega
_{12}=(E_{2}-E_{1})/\hslash $ and $\omega _{12}^{l}=\omega
_{02}=(E_{2}-E_{0})/\hslash ,$ respectively. If one makes an exchange $%
E_{0}\leftrightarrow E_{1}$ in all those results obtained in the previous
decelerating process (11), then these results can be adopted in the present
decelerating process (63).

Now suppose that at the initial time $t_{0}$ the atom is in the product
state $\Psi _{0}(x,r,t_{0})$ of Eq. (54). The atom first undergoes the
STIRAP decelerating process (11) and hence the product state $\Psi
_{0}(x,r,t_{0})$ is completely transferred to the product state $\Psi
_{0}(x,r,t_{f})$ of Eq. (56) at the end time $t_{f}=t_{0}+T$ of the
decelerating process (11). Then the atom undergoes the second STIRAP
decelerating process (63). Now one wants to calculate the atomic wave-packet
product state at the end of the second decelerating process (63). At the
initial time $t_{1}=t_{f}$ of the second decelerating process (63) the
atomic product state is given by $\Psi _{1}(x,r,t_{1})=\Psi
_{1}(x,t_{1})|g_{1}\rangle .$ Obviously, this product state is just the
product state of the atom at the end of the first decelerating process (11).
Thus, the initial motional state $\Psi _{1}(x,t_{1})$ is the Gaussian
wave-packet state of Eq. (59): $\Psi _{1}(x,t_{1})=\Psi _{0}(x,t_{f}).$ Then
in the ideal adiabatic condition (30) at the end time $t_{f}^{\prime
}=t_{1}+T$ of the second decelerating process (63) the atom is completely in
the product state: 
\begin{equation}
\Psi _{1}(x,r,t_{f}^{\prime })=\Psi _{1}(x,t_{f}^{\prime })|g_{0}\rangle 
\tag{64}
\end{equation}%
where the wave-packet motional state $\Psi _{1}(x,t_{f}^{\prime })$ can be
calculated by 
\begin{equation}
\Psi _{1}(x,t_{f}^{\prime })=\sum_{p}\rho _{1}(p,t_{1})A_{2}(p,t_{f}^{\prime
})|p-\hslash l_{0}-\hslash l_{1}\rangle .  \tag{65}
\end{equation}%
Here the amplitude $\rho _{1}(p,t_{1})$ is given by Eq. (62) with the time $%
t_{f}=t_{1},$ the momentum eigenstate $|p-\hslash l_{0}-\hslash l_{1}\rangle 
$ is still given by Eq. (7)$,$ and the coefficient $A_{2}(p,t_{f}^{\prime })$
with the center-of-mass momentum $P_{0}=p_{1}$ and $\Delta p=p-p_{1}$ is
written as 
\begin{equation*}
A_{2}(p,t_{f}^{\prime })=\exp [i\beta _{l}(t_{f}^{\prime },t_{1})]\exp [%
\frac{i}{\hslash }(\frac{p^{2}}{2M}+E_{1})t_{1}]
\end{equation*}%
\begin{equation*}
\times \exp [-\frac{i}{\hslash }(\frac{(p-\hslash l_{0}-\hslash l_{1})^{2}}{%
2M}+E_{0})t_{f}^{\prime }]
\end{equation*}%
\begin{equation}
\times \exp \{-i\frac{\Delta p}{M}(l_{0}+l_{1})[t_{1}+\int_{t_{1}}^{t_{f}^{%
\prime }}dt^{\prime }\cos ^{2}\theta _{l}(t^{\prime })]\},  \tag{66}
\end{equation}%
where the global phase $\beta _{l}(t_{f}^{\prime },t_{1})$ is still
calculated by Eq. (45) with the related parameter settings such as the
mixing angle $\theta _{l}(t)$ and the phase modulation functions $\varphi
_{l0}(t)$ and $\varphi _{l1}(t)$ of the present STIRAP pulse sequence (63).
Then by a complex calculation one can obtain from Eq. (65)\ the Gaussian
wave-packet motional state at the end of the decelerating process (63): 
\begin{equation*}
\Psi _{1}(x,t_{f}^{\prime })=\exp [i\varphi _{2}(t_{f}^{\prime },t_{1})][%
\frac{(\Delta x)^{2}}{2\pi }]^{1/4}\sqrt{\frac{1}{(\Delta x)^{2}+i\frac{%
\hslash (T_{0}+(t_{1}-t_{0})+(t_{f}^{\prime }-t_{1}))}{2M}}}
\end{equation*}%
\begin{equation}
\times \exp \{-\frac{1}{4}\frac{(x-z_{2})^{2}}{(\Delta x)^{2}+i\frac{\hslash
(T_{0}+(t_{1}-t_{0})+(t_{f}^{\prime }-t_{1}))}{2M}}\}\exp [ip_{2}x/\hslash ]
\tag{67}
\end{equation}%
where the center-of-mass position $z_{2}$ is given by 
\begin{equation}
z_{2}=z_{1}+\frac{(p_{1}-\hslash l_{0}-\hslash l_{1})}{M}(t_{f}^{\prime
}-t_{1})+\frac{\hslash (l_{0}+l_{1})}{M}\int_{t_{1}}^{t_{f}^{\prime
}}dt^{\prime }\cos ^{2}\theta _{l}(t^{\prime }),  \tag{68}
\end{equation}%
the wave-packet spreading by 
\begin{equation}
\varepsilon =\sqrt{2(\Delta x)^{2}+2[\frac{\hslash
(T_{0}+(t_{1}-t_{0})+(t_{f}^{\prime }-t_{1}))}{2M(\Delta x)}]^{2}},  \tag{69}
\end{equation}%
the mean momentum by%
\begin{equation*}
p_{2}=p_{1}-\hslash l_{0}-\hslash l_{1},
\end{equation*}%
and the global phase factor by 
\begin{equation*}
\exp [i\varphi _{2}(t_{f}^{\prime },t_{1})]=\exp [i\varphi
_{1}(t_{1},t_{0})]\exp [i\beta _{l}(t_{f}^{\prime },t_{1})]
\end{equation*}%
\begin{equation*}
\times \exp [\frac{i}{\hslash }(\frac{p_{1}^{2}}{2M}+E_{1})t_{1}]\exp [-%
\frac{i}{\hslash }(\frac{(p_{1}-\hslash l_{0}-\hslash l_{1})^{2}}{2M}%
+E_{0})t_{f}^{\prime }].
\end{equation*}%
Here both the basic decelerating sequences (11) and (63) are studied in
detail as they are the basic STIRAP-based decelerating processes. Any
unitary decelerating process in the quantum control process [22] may be
constructed with a train of these two basic decelerating sequences.

When the atom is in the Gaussian wave-packet motional state $\Psi
_{0}(x,t_{0})$ of Eq. (48) at the initial time $t_{0}$, the complex
linewidth of the motional state is $W(T_{0})=(\Delta x)^{2}+i\hslash
T_{0}/(2M).$ After the first basic STIRAP\ decelerating process (11)\ the
atom is in the Gaussian wave-packet motional state $\Psi _{0}(x,t_{1})$ ($%
t_{1}=t_{0}+T$) of Eq. (59) and the state has the complex linewidth $%
W(t_{1}-t_{0}+T_{0})=(\Delta x)^{2}+i\hslash (T_{0}+t_{1}-t_{0})/(2M).$ Then
after the second basic STIRAP decelerating process (63) the atom is in the
Gaussian wave-packet motional state $\Psi _{1}(x,t_{f}^{\prime })$ ($%
t_{f}^{\prime }=t_{1}+T$) of Eq. (67) and the complex linewidth of the state
is $W(t_{f}^{\prime }-t_{0}+T_{0})=(\Delta x)^{2}+i\hslash \lbrack
T_{0}+(t_{1}-t_{0})+(t_{f}^{\prime }-t_{1})]/(2M).$ Thus, one can see that
the real part of the complex linewidth of the Gaussian wave-packet motional
state of the atom keeps unchanged during these decelerating processes (11)
and (63), while the imaginary part increases linearly with the time periods
of these decelerating processes. This result is found not only in the
decelerating processes but also in the accelerating processes and the
free-particle motional process.

In a general case a\ unitary decelerating process may consist of a train of
the two basic STIRAP decelerating processes (11) and (63). For convenience,
here each basic STIRAP decelerating process is set to have the same time
period $T=t_{d}$ and suppose that at the initial time of the unitary
decelerating process the atom is in the internal state $|g_{0}\rangle $ and
has a large motional momentum such that the atom still moves along the
initial direction $+x$ even after the unitary decelerating process. The
basic decelerating sequences (11) and (63) may form a basic STIRAP
decelerating cycle in such a way that first the decelerating sequence (11)
and then the sequence (63) is applied to the decelerated atom. The unitary
decelerating process may consist of many these basic STIRAP decelerating
cycles. Denote $U^{d}(11)$ and $U^{d}(63)$ as the unitary propagators of the
basic decelerating processes (11) and (63), respectively. Then a general
unitary decelerating process may be expressed as 
\begin{equation}
U_{D}(2N)=U_{2N}^{d}(63)U_{2N-1}^{d}(11)...U_{2n}^{d}(63)U_{2n-1}^{d}(11)...U_{2}^{d}(63)U_{1}^{d}(11)
\tag{70a}
\end{equation}%
or%
\begin{equation}
U_{D}(2N-1)=U_{2N-1}^{d}(11)...U_{2n}^{d}(63)U_{2n-1}^{d}(11)...U_{2}^{d}(63)U_{1}^{d}(11)
\tag{70b}
\end{equation}%
where $U_{2n-1}^{d}(11)$ is the propagator of the $(2n-1)-$th basic
decelerating unit for $n=1,2,...,N$ in the unitary decelerating process $%
U_{D}(2N)$ or $U_{D}(2N-1)$, while $U_{2n}^{d}(63)$ is the propagator of the 
$2n-$th basic decelerating unit. Each basic decelerating unit with an even
index $2n$ in the unitary decelerating process is taken as the basic
decelerating process (63), while that with an odd index $(2n-1)$ is taken as
the basic decelerating process (11). Thus, $U_{2n-1}^{d}(11)=U^{d}(11)$ and $%
U_{2n}^{d}(63)=U^{d}(63)$ for $n=1,2,...,N$. In particular, $U_{D}(0)=E$
(the unit operator), $U_{D}(1)=U^{d}(11),$ and $U_{D}(2)=U^{d}(63)U^{d}(11)$%
. The unitary decelerating processes $U_{D}(1)$ and $U_{D}(2)$ have been
investigated in detail in the preceding paragraphs. Obviously, the unitary
decelerating process $U_{D}(2N)$ consists of $2N$ basic decelerating
processes (11) and (63) alternately or $N$ basic decelerating cycles, while $%
U_{D}(2N-1)$ consists of $N$ basic decelerating process (11) and $N-1$ basic
decelerating process (63) alternately.

The time evolution process of the atom in the presence of the unitary
decelerating sequence $U_{D}(2N)$ or $U_{D}(2N-1)$ can be calculated exactly
in the ideal adiabatic condition (30). For the simplest cases $N=1$ and $2$
the time evolution processes of the atom have already calculated in the
ideal adiabatic condition (30) in the previous paragraphs. In order to
calculate the time evolution process of the atom in a general unitary
decelerating process one may first set up the recursive relation between the
two atomic product states at the end of the unitary decelerating processes $%
U_{D}(2n-1)$ and $U_{D}(2n)$ for $n=1,2,...,N.$ Since the initial internal
state of the atom is $|g_{0}\rangle $ in both the unitary decelerating
processes $U_{D}(2N)$ and $U_{D}(2N-1),$ after the unitary decelerating
process $U_{D}(2n)$ (or $U_{D}(2n-1))$ ($1\leq n\leq N$) the final internal
state of the atom is clearly $|g_{0}\rangle $ (or $|g_{1}\rangle )$. Then
the initial internal states of the atom in the basic decelerating processes $%
U_{2n+1}^{d}(11)$ and $U_{2n}^{d}(63)$ should be $|g_{0}\rangle $ and $%
|g_{1}\rangle $, respectively. It is known that at the initial time $t_{0}$
the atom is in the Gaussian wave-packet motional state $\Psi _{0}(x,t_{0})$
of Eq. (48) and the product state $\Psi _{0}(x,r,t_{0})$ of Eq. (54). It is
also known that after the unitary decelerating sequences $U_{D}(1)$ and $%
U_{D}(2)$ act on the initial product state $\Psi _{0}(x,r,t_{0})$ of Eq.
(54) the initial motional state $\Psi _{0}(x,t_{0})$ of Eq. (48) is
converted into the Gaussian wave-packet motional states $\Psi _{0}(x,t_{f})$
of Eq. (59) and $\Psi _{1}(x,t_{f}^{\prime })$ of Eq. (67), respectively.
This means that the unitary decelerating sequences $U_{D}(1)$ and $U_{D}(2)$
do not change the Gaussian shape of the atomic motional state. Therefore, it
is reasonable to deduce that after the unitary decelerating process $%
U_{D}(2n)$ for $n=0,1,2,...,N$ the atom is in the Gaussian wave-packet
motional state:\newline
\begin{equation*}
\Psi _{2n}^{d}(x,t_{2n-1}^{e})=\exp (i\varphi _{2n}^{d})[\frac{(\Delta x)^{2}%
}{2\pi }]^{1/4}\sqrt{\frac{1}{(\Delta x)^{2}+i\frac{\hslash (T_{d}+2nt_{d})}{%
2M}}}
\end{equation*}%
\begin{equation}
\times \exp \{-\frac{1}{4}\frac{(x-z_{2n}^{d})^{2}}{(\Delta x)^{2}+i\frac{%
\hslash (T_{d}+2nt_{d})}{2M}}\}\exp \{iP_{2n}^{d}x/\hslash \}  \tag{71}
\end{equation}%
and also in the atomic wave-packet product state: 
\begin{equation}
\Psi _{2n}^{d}(x,r,t_{2n-1}^{e})=\Psi _{2n}^{d}(x,t_{2n-1}^{e})|g_{0}\rangle
=\sum_{p}\rho (p,t_{2n-1}^{e})|p\rangle |g_{0}\rangle .  \tag{72}
\end{equation}%
where $t_{2n-1}^{e}$ is the end time of the unitary decelerating process $%
U_{D}(2n).$ In an analogous way to calculating the amplitude $\rho
_{0}(p,t_{0})$ via the equation (49a) one can calculate the amplitude $\rho
(p,t_{2n-1}^{e})$ of Eq. (72) from the motional state $\Psi
_{2n}^{d}(x,t_{2n-1}^{e})$ of Eq. (71). The result is 
\begin{equation*}
\rho (p,t_{2n-1}^{e})=\exp (i\varphi _{2n}^{d})[\frac{2(\Delta x)^{2}}{\pi }%
]^{1/4}\exp \{-(\Delta x)^{2}(\frac{p-P_{2n}^{d}}{\hslash })^{2}\}
\end{equation*}%
\begin{equation}
\times \exp \{-i(\frac{\hslash (T_{d}+2nt_{d})}{2M})(\frac{p-P_{2n}^{d}}{%
\hslash })^{2}\}\exp \{-i(\frac{p-P_{2n}^{d}}{\hslash })z_{2n}^{d}\}. 
\tag{73}
\end{equation}%
It will prove below that the states $\Psi _{2n}^{d}(x,t_{2n-1}^{e})$ of (71)
and $\Psi _{2n}^{d}(x,r,t_{2n-1}^{e})$ of (72) are indeed the wave-packet
motional state and product state of the atom at the end of the unitary
decelerating process $U_{D}(2n),$ respectively.

First of all, the product state $\Psi _{0}(x,r,t_{0})$ of Eq. (54) and the
motional state $\Psi _{0}(x,t_{0})$ of Eq. (48) are the initial product
state and motional state of the atom in the presence of the unitary
decelerating process $U_{D}(2n)$ (or $U_{D}(2n-1))$, respectively. Of
course, these two states may also be formally thought of as the final
wave-packet product state and motional state of the atom after the unitary
'decelerating' process $U_{D}(0)=E$ (the unity operator), respectively. This
means that the atomic wave-packet product state $\Psi
_{0}^{d}(x,r,t_{-1}^{e})$ of Eq. (72) and the motional state $\Psi
_{0}^{d}(x,t_{-1}^{e})$ of Eq. (71) with $n=0$ should be equal to $\Psi
_{0}(x,r,t_{0})$ of Eq. (54) and $\Psi _{0}(x,t_{0})$ of Eq. (48),
respectively, 
\begin{equation*}
\Psi _{0}^{d}(x,r,t_{-1}^{e})=\Psi _{0}(x,r,t_{0}),\text{ }\Psi
_{0}^{d}(x,t_{-1}^{e})=\Psi _{0}(x,t_{0}),
\end{equation*}%
while the momentum wave-packet state $\rho (p,t_{-1}^{e})$ of Eq. (73) thus
is just $\rho _{0}(p,t_{0})$ of Eq. (50). Indeed, these equations (71),
(72), and (73) show this point, if in Eqs. (71), (72), and (73) one sets the
parameters: $n=0,$ $t_{-1}^{e}=t_{0}^{d}=t_{0},$ $\varphi _{0}^{d}=\varphi
_{0},$ $T_{d}=T_{0},$ $z_{0}^{d}=z_{0},$ $P_{0}^{d}=p_{0},$ and$\ t_{d}=T,$
where $T$ is the time period of the basic decelerating sequence (11). Next,
the product state $\Psi _{1}(x,r,t_{f}^{\prime })$ of Eq. (64) and the
motional state $\Psi _{1}(x,t_{f}^{\prime })$ of Eq. (67) with the time $%
t_{f}^{\prime }=t_{1}^{e}$ are just the product state $\Psi
_{2}^{d}(x,r,t_{1}^{e})$ of Eq. (72) and the motional state $\Psi
_{2}^{d}(x,t_{1}^{e})$ of Eq. (71) with $n=1$ at the end time $t_{1}^{e}$ of
the unitary decelerating process $U_{D}(2)$, respectively, 
\begin{equation*}
\Psi _{2}^{d}(x,r,t_{1}^{e})=\Psi _{1}(x,r,t_{f}^{\prime }),\text{ }\Psi
_{2}^{d}(x,t_{1}^{e})=\Psi _{1}(x,t_{f}^{\prime }).
\end{equation*}%
This can be confirmed by setting the following parameters in Eqs. (71) and
(72): $t_{1}^{d}=t_{f}=t_{1}=t_{0}+t_{d},$ $t_{1}^{e}=t_{f}^{\prime
}=t_{1}^{d}+t_{d}=t_{0}+2t_{d},$ $\varphi _{2}^{d}=\varphi
_{2}(t_{1}^{e},t_{1}^{d}),$ $z_{2}^{d}=z_{2},$ and $P_{2}^{d}=p_{2},$ One
therefore shows that the motional state $\Psi _{2n}^{d}(x,t_{2n-1}^{e})$ of
Eq. (71) and the product state $\Psi _{2n}^{d}(x,r,t_{2n-1}^{e})$ of Eq.
(72) are correct for both the unitary decelerating processes $U_{D}(0)$ ($%
n=0 $) and $U_{D}(2)$ ($n=1$)$.$ It will prove below that both the motional
state $\Psi _{2n}^{d}(x,t_{2n-1}^{e})$ of Eq. (71)\ and the product state $%
\Psi _{2n}^{d}(x,r,t_{2n-1}^{e})$ of Eq. (72) are also correct for the
unitary decelerating process $U_{D}(2n)$ with $n=0,1,...,N$.

Suppose that the states $\Psi _{2n}^{d}(x,t_{2n-1}^{e})$ of Eq. (71) and $%
\Psi _{2n}^{d}(x,r,t_{2n-1}^{e})$ of Eq. (72) are correct for the unitary
decelerating process $U_{D}(2n)$. It is known that the internal state of the
atom is $|g_{0}\rangle $ at the end of the unitary decelerating process $%
U_{D}(2(n+1)).$ Then one needs only to prove that the motional state of Eq.
(71) is also correct for the unitary decelerating process $U_{D}(2(n+1))$.
The propagator of the unitary decelerating process $U_{D}(2(n+1))$ can be
written as 
\begin{equation*}
U_{D}(2(n+1))=U_{2n+2}^{d}(63)U_{D}(2n+1)=U_{2n+2}^{d}(63)U_{2n+1}^{d}(11)U_{D}(2n).
\end{equation*}%
\newline
According to the assumption the motional state $\Psi
_{2n}^{d}(x,t_{2n-1}^{e})$ is just the final motional state of the atom when
the atom is acted on by the unitary propagator $U_{D}(2n)$. Obviously, the
motional state $\Psi _{2n}^{d}(x,t_{2n}^{d})=\Psi _{2n}^{d}(x,t_{2n-1}^{e})$
is also the initial motional state of the $(2n+1)-$th basic decelerating
process (11) with the propagator $U_{2n+1}^{d}(11)$ in the unitary
decelerating process $U_{D}(2(n+1))$, where the initial time is denoted as $%
t_{2n}^{d}=t_{2n-1}^{e}$ and there are the recursive relations: $%
t_{-1}^{e}=t_{0}^{d}=t_{0},$ $t_{0}^{e}=t_{1}^{d}=t_{0}^{d}+t_{d}$, $%
t_{1}^{e}=t_{2}^{d}=t_{0}^{d}+2t_{d},$ and $t_{2n}^{d}=t_{2n-1}^{d}+t_{d}$ ($%
n>0$). Then the initial product state of the basic decelerating process $%
U_{2n+1}^{d}(11)$ is given by $\Psi _{2n}^{d}(x,r,t_{2n}^{d})=\Psi
_{2n}^{d}(x,r,t_{2n-1}^{e})$ of Eq. (72). Now the initial product state $%
\Psi _{2n}^{d}(x,r,t_{2n}^{d})$ is acted on by the unitary propagator $%
U_{2n+1}^{d}(11).$ Then it can turn out that at the end time $%
t_{2n}^{e}=t_{2n}^{d}+t_{d}$ of the basic decelerating process $%
U_{2n+1}^{d}(11)$ the initial motional state $\Psi _{2n}^{d}(x,t_{2n}^{d})$
and product state $\Psi _{2n}^{d}(x,r,t_{2n}^{d})$\ are respectively
transferred into the motional state: 
\begin{equation*}
\Psi _{2n+1}^{d}(x,t_{2n}^{e})=\exp (i\varphi _{2n+1}^{d})[\frac{(\Delta
x)^{2}}{2\pi }]^{1/4}\sqrt{\frac{1}{(\Delta x)^{2}+i\frac{\hslash
(T_{d}+(2n+1)t_{d})}{2M}}}
\end{equation*}%
\begin{equation}
\times \exp \{-\frac{1}{4}\frac{(x-z_{2n+1}^{d})^{2}}{(\Delta x)^{2}+i\frac{%
\hslash (T_{d}+(2n+1)t_{d})}{2M}}\}\exp \{iP_{2n+1}^{d}x/\hslash \}  \tag{74}
\end{equation}%
and the product state: 
\begin{equation}
\Psi _{2n+1}^{d}(x,r,t_{2n}^{e})=\Psi _{2n+1}^{d}(x,t_{2n}^{e})|g_{1}\rangle
\tag{75}
\end{equation}%
where 
\begin{equation}
P_{2n+1}^{d}=P_{2n}^{d}-\hslash k_{0}-\hslash k_{1},  \tag{76}
\end{equation}%
\begin{equation}
z_{2n+1}^{d}=z_{2n}^{d}+\frac{P_{2n+1}^{d}}{M}t_{d}+\frac{\hslash
(k_{0}+k_{1})}{M}\int_{t_{2n}^{d}}^{t_{2n}^{e}}dt^{\prime }\cos ^{2}\theta
(t^{\prime }),  \tag{77}
\end{equation}%
\begin{equation*}
\exp (i\varphi _{2n+1}^{d})=\exp (i\varphi _{2n}^{d})\exp [i\beta
(t_{2n}^{e},t_{2n}^{d})]\exp \{\frac{i}{\hslash }(\frac{(P_{2n}^{d})^{2}}{2M}%
+E_{0})t_{2n}^{d}\}
\end{equation*}%
\begin{equation}
\times \exp \{-\frac{i}{\hslash }(\frac{(P_{2n+1}^{d})^{2}}{2M}%
+E_{1})t_{2n}^{e}\}.  \tag{78}
\end{equation}%
The computational process from the initial state $\Psi
_{2n}^{d}(x,t_{2n-1}^{e})$ to the final state $\Psi
_{2n+1}^{d}(x,t_{2n}^{e}) $ is the same as the previous one from the initial
state $\Psi _{0}(x,t_{0})$ of (48) to the final state $\Psi _{0}(x,t_{f})$
of (59). There are the relations: 
\begin{equation*}
\Psi _{2n+1}^{d}(x,r,t_{2n}^{e})=U_{2n+1}^{d}(11)\Psi
_{2n}^{d}(x,r,t_{2n}^{d})=U_{D}(2n+1)\Psi _{0}(x,r,t_{0}).
\end{equation*}%
These relations show that both the states $\Psi _{2n+1}^{d}(x,r,t_{2n}^{e})$
and $\Psi _{2n+1}^{d}(x,t_{2n}^{e})$ are also the product state and the
motional state of the atom at the end time $t_{2n}^{e}=t_{2n}^{d}+t_{d}$ of
the unitary decelerating process $U_{D}(2n+1),$ respectively.

The atomic product state $\Psi _{2n+1}^{d}(x,r,t_{2n}^{e})$ of Eq. (75) at
the end of the unitary decelerating process $U_{D}(2n+1)$ will be used to
calculate the atomic product state at the end of the unitary decelerating
process $U_{D}(2n+2).$ This computational process is the same as the
previous one from the initial state $\Psi _{1}(x,t_{1})=\Psi _{0}(x,t_{f})$
of (59) to the final state $\Psi _{1}(x,t_{f}^{\prime })$ of (67). There are
the relations:%
\begin{equation*}
\Psi _{2n+2}^{d}(x,r,t_{2n+1}^{e})=U_{2n+2}^{d}(63)\Psi
_{2n+1}^{d}(x,r,t_{2n}^{e})=U_{D}(2n+2)\Psi _{0}(x,r,t_{0})..
\end{equation*}%
These relations show that the atomic product state $\Psi
_{2n+2}^{d}(x,r,t_{2n+1}^{e})$ at the end of the unitary decelerating
process $U_{D}(2n+2)$ can be obtained by applying the propagator $%
U_{2n+2}^{d}(63)$ to the atomic product state $\Psi
_{2n+1}^{d}(x,r,t_{2n}^{e})$ of Eq. (75). For convenient calculation, the
atomic motional state $\Psi _{2n+1}^{d}(x,t_{2n}^{e})$ of Eq. (74) is
rewritten as ($n^{\prime }=n+1$)%
\begin{equation*}
\Psi _{2n^{\prime }-1}^{d}(x,t_{2n^{\prime }-1}^{d})=\exp (i\varphi
_{2n^{\prime }-1}^{d})[\frac{(\Delta x)^{2}}{2\pi }]^{1/4}\sqrt{\frac{1}{%
(\Delta x)^{2}+i\frac{\hslash (T_{d}+(2n^{\prime }-1)t_{d})}{2M}}}
\end{equation*}%
\begin{equation}
\times \exp \{-\frac{1}{4}\frac{(x-z_{2n^{\prime }-1}^{d})^{2}}{(\Delta
x)^{2}+i\frac{\hslash (T_{d}+(2n^{\prime }-1)t_{d})}{2M}}\}\exp
\{iP_{2n^{\prime }-1}^{d}x/\hslash \}.  \tag{79}
\end{equation}%
Then the atomic product state of Eq. (75) can be rewritten as%
\begin{equation}
\Psi _{2n^{\prime }-1}^{d}(x,r,t_{2n^{\prime }-1}^{d})=\Psi _{2n^{\prime
}-1}^{d}(x,t_{2n^{\prime }-1}^{d})|g_{1}\rangle =\sum_{p}\rho
(p,t_{2n^{\prime }-1}^{d})|p\rangle |g_{1}\rangle  \tag{80}
\end{equation}%
where the amplitude $\rho (p,t_{2n^{\prime }-1}^{d})$ can be calculated from
the motional state of Eq. (79) and is given by%
\begin{equation*}
\rho (p,t_{2n^{\prime }-1}^{d})=\exp (i\varphi _{2n^{\prime }-1}^{d})[\frac{%
2(\Delta x)^{2}}{\pi }]^{1/4}\exp \{-(\Delta x)^{2}(\frac{p-P_{2n^{\prime
}-1}^{d}}{\hslash })^{2}\}
\end{equation*}%
\begin{equation}
\times \exp \{-i\frac{\hslash (T_{d}+(2n^{\prime }-1)t_{d})}{2M}(\frac{%
p-P_{2n^{\prime }-1}^{d}}{\hslash })^{2}\}\exp [-i\frac{(p-P_{2n^{\prime
}-1}^{d})}{\hslash }z_{2n^{\prime }-1}^{d}].  \tag{81}
\end{equation}%
\newline
Now the atomic product state $\Psi _{2n^{\prime }-1}^{d}(x,r,t_{2n^{\prime
}-1}^{d})$ of Eq. (80) is applied by the unitary propagator $U_{2n^{\prime
}}^{d}(63)$ ($n^{\prime }=n+1$). Then it can turn out that at the end time $%
t_{2n^{\prime }-1}^{e}=t_{2n^{\prime }-1}^{d}+t_{d}$ of the unitary
decelerating process $U_{D}(2n+2)$ the atomic wave-packet motional state
takes the form 
\begin{equation*}
\Psi _{2n^{\prime }}^{d}(x,t_{2n^{\prime }-1}^{e})=\exp (i\varphi
_{2n^{\prime }}^{d})[\frac{(\Delta x)^{2}}{2\pi }]^{1/4}\sqrt{\frac{1}{%
(\Delta x)^{2}+i\frac{\hslash (T_{d}+2n^{\prime }t_{d})}{2M}}}
\end{equation*}%
\begin{equation}
\times \exp \{-\frac{1}{4}\frac{(x-z_{2n^{\prime }}^{d})^{2}}{(\Delta
x)^{2}+i\frac{\hslash (T_{d}+2n^{\prime }t_{d})}{2M}}\}\exp \{iP_{2n^{\prime
}}^{d}x/\hslash \}.  \tag{82}
\end{equation}%
and the atomic product state is%
\begin{equation}
\Psi _{2n^{\prime }}^{d}(x,r,t_{2n^{\prime }-1}^{e})=\Psi _{2n^{\prime
}}^{d}(x,t_{2n^{\prime }-1}^{e})|g_{0}\rangle ,  \tag{83}
\end{equation}%
where 
\begin{equation}
P_{2n^{\prime }}^{d}=P_{2n^{\prime }-1}^{d}-\hslash l_{0}-\hslash l_{1}, 
\tag{84}
\end{equation}%
\begin{equation}
z_{2n^{\prime }}^{d}=z_{2n^{\prime }-1}^{d}+\frac{P_{2n^{\prime }}^{d}}{M}%
t_{d}+\frac{\hslash (l_{0}+l_{1})}{M}\int_{t_{2n^{\prime
}-1}^{d}}^{t_{2n^{\prime }-1}^{e}}dt^{\prime }\cos ^{2}\theta _{l}(t^{\prime
}),  \tag{85}
\end{equation}%
\begin{equation*}
\exp (i\varphi _{2n^{\prime }}^{d})=\exp (i\varphi _{2n^{\prime
}-1}^{d})\exp [i\beta _{l}(t_{2n^{\prime }-1}^{e},t_{2n^{\prime }-1}^{d})]
\end{equation*}%
\begin{equation}
\times \exp [\frac{i}{\hslash }(\frac{(P_{2n^{\prime }-1}^{d})^{2}}{2M}%
+E_{1})t_{2n^{\prime }-1}^{d}]\exp [-\frac{i}{\hslash }(\frac{(P_{2n^{\prime
}}^{d})^{2}}{2M}+E_{0})t_{2n^{\prime }-1}^{e}].  \tag{86}
\end{equation}%
Now by comparing the motional state $\Psi _{2n^{\prime
}}^{d}(x,t_{2n^{\prime }-1}^{e})$ of Eq. (82) with the motional state $\Psi
_{2n}^{d}(x,t_{2n-1}^{e})$ of Eq. (71) and the product state $\Psi
_{2n^{\prime }}^{d}(x,r,t_{2n^{\prime }-1}^{e})$ of Eq. (83) with the
product state $\Psi _{2n}^{d}(x,r,t_{2n-1}^{e})$ of Eq. (72) one can
conclude by the mathematical principle of induction that the motional state $%
\Psi _{2n}^{d}(x,t_{2n-1}^{e})$ of Eq. (71) and the product state $\Psi
_{2n}^{d}(x,r,t_{2n-1}^{e})$ of Eq. (72) are indeed the states of the atom
at the end of the unitary decelerating process $U_{D}(2n)$ for $%
n=0,1,2,...,N $. In an analogous way, one can prove that the product state $%
\Psi _{2n+1}^{d}(x,r,t_{2n+1}^{d})$ of Eq. (75) and the motional state $\Psi
_{2n+1}^{d}(x,t_{2n+1}^{d})$ of Eq. (74) are the states of the atom at the
end of the unitary decelerating process $U_{D}(2n+1)$ for $n=0,1,2,...,N-1$.

Now one can prove that the atomic motional momentum $P_{2n+1}^{d}$ of Eq.
(76) and $P_{2n^{\prime }}^{d}$ of Eq. (84) are given by, respectively,%
\begin{equation*}
P_{2n+1}^{d}=P_{0}^{d}-(n+1)(\hslash k_{0}+\hslash k_{1})-n(\hslash
l_{0}+\hslash l_{1}),
\end{equation*}%
\begin{equation*}
P_{2n^{\prime }}^{d}=P_{0}^{d}-n^{\prime }(\hslash k_{0}+\hslash
k_{1})-n^{\prime }(\hslash l_{0}+\hslash l_{1}).
\end{equation*}%
It is known that the recursive relations for the atomic motional momentum
are given by $P_{2n+1}^{d}=P_{2n}^{d}-\hslash k_{0}-\hslash k_{1}$ for $%
n=0,1,...,N-1,$ and $P_{2n^{\prime }}^{d}=P_{2n^{\prime }-1}^{d}-\hslash
l_{0}-\hslash l_{1}$ for $n^{\prime }=1,2,...,N,$ which are obtained from
Eqs. (76) and (84), respectively. The two recursive relations together can
lead directly to the two formulae for the atomic motional momentum $%
P_{2n+1}^{d}$ and $P_{2n^{\prime }}^{d}$.

One also can calculate exactly the time evolution process of the atom in the
unitary STIRAP-based accelerating process in the ideal adiabatic condition
(30). There are also two basic STIRAP-based accelerating sequences which
correspond to the basic decelerating sequences (11) and (63), respectively.
One of which is already expressed as (11a). The basic accelerating sequence
(11a)\ corresponds to the basic decelerating sequence (11). Another may be
expressed in an intuitive form 
\begin{equation}
|P-\hslash l_{0}\rangle |g_{1}\rangle \rightarrow |P\rangle |e\rangle
\rightarrow |P+\hslash l_{1}\rangle |g_{0}\rangle .  \tag{63a}
\end{equation}%
This basic accelerating sequence corresponds to the basic decelerating
sequence (63). In an analogous way to constructing the unitary decelerating
processes $U_{D}(2N)$ and $U_{D}(2N-1)$ one may build up the unitary
accelerating processes $U_{A}(2N)$ and $U_{A}(2N-1)$ out of the basic
accelerating sequences (11a) and (63a), 
\begin{equation}
U_{A}(2N)=U_{2N}^{a}(63a)U_{2N-1}^{a}(11a)...U_{2n}^{a}(63a)U_{2n-1}^{a}(11a)...U_{2}^{a}(63a)U_{1}^{a}(11a)
\tag{87a}
\end{equation}%
or%
\begin{equation}
U_{A}(2N-1)=U_{2N-1}^{a}(11a)...U_{2n}^{a}(63a)U_{2n-1}^{a}(11a)...U_{2}^{a}(63a)U_{1}^{a}(11a)
\tag{87b}
\end{equation}%
where $U_{2n-1}^{a}(11a)$ and $U_{2n}^{a}(63a)$ for $n=1,2,...,N$ are the
unitary propagators of the $(2n-1)-$th basic accelerating sequence (11a) and 
$(2n)-$th basic accelerating sequence (63a), respectively. Here also suppose
that the atom is in the internal state $|g_{0}\rangle $ at the initial time
in both the unitary accelerating processes $U_{A}(2N)$ and $U_{A}(2N-1).$

The time evolution process of the atom in the unitary accelerating process $%
U_{A}(2N)$ (and $U_{A}(2N-1))$ can be calculated exactly in the ideal
adiabatic condition (30) in a similar way to that one in the unitary
decelerating process $U_{D}(2N)$ (and $U_{D}(2N-1)).$ Actually, the
recursive relations (71)--(78) and (79)--(86) of the unitary decelerating
process $U_{D}(2N)$ or $U_{D}(2N-1)$ can be used as well for the unitary
accelerating process $U_{A}(2N)$ or $U_{A}(2N-1)$ if one makes
transformations: $k_{0}\rightarrow -k_{0}^{a},$ $k_{1}\rightarrow
-k_{1}^{a}, $ and $\theta (t)\rightarrow \theta _{a}(t)$ in those recursive
equations (71)--(78) and $l_{0}\rightarrow -l_{0}^{a},$ $l_{1}\rightarrow
-l_{1}^{a},$ and $\theta _{l}(t)\rightarrow \theta _{la}(t)$ in those
recursive equations (79)--(86). As an example, suppose that the initial
wave-packet motional state for the atom in the unitary accelerating process $%
U_{A}(2n)$ is given by 
\begin{equation*}
\Psi _{0}^{a}(x,t_{0}^{a})=\exp (i\varphi _{0}^{a})[\frac{(\Delta x)^{2}}{%
2\pi }]^{1/4}\sqrt{\frac{1}{[(\Delta x)^{2}+i\frac{\hslash T_{a}}{2M}]}}
\end{equation*}%
\begin{equation}
\times \exp \{-\frac{1}{4}\frac{[x-z_{0}^{a}]^{2}}{[(\Delta x)^{2}+i\frac{%
\hslash T_{a}}{2M}]}\}\exp [iP_{0}^{a}x/\hslash ]  \tag{88}
\end{equation}%
and the atomic wave-packet product state by 
\begin{equation}
\Psi _{0}^{a}(x,r,t_{0}^{a})=\Psi _{0}^{a}(x,t_{0}^{a})|g_{0}\rangle
=\sum_{p}\rho (p,t_{0}^{a})|p\rangle |g_{0}\rangle .  \tag{89}
\end{equation}%
Then it can turn out that the momentum wave-packet state $\rho (p,t_{0}^{a})$
of the motional state $\Psi _{0}^{a}(x,t_{0}^{a})$ can be written as 
\begin{equation*}
\rho (p,t_{0}^{a})=\exp (i\varphi _{0}^{a})[\frac{2(\Delta x)^{2}}{\pi }%
]^{1/4}\exp \{-(\Delta x)^{2}(\frac{p-P_{0}^{a}}{\hslash })^{2}\}
\end{equation*}%
\begin{equation}
\times \exp \{-i(\frac{\hslash T_{a}}{2M})(\frac{p-P_{0}^{a}}{\hslash }%
)^{2}\}\exp \{-i(\frac{p-P_{0}^{a}}{\hslash })z_{0}^{a}\}.  \tag{90}
\end{equation}%
Now the initial wave-packet product state $\Psi _{0}^{a}(x,r,t_{0}^{a})$ of
the atom undergoes the unitary accelerating process $U_{A}(2n)$. Then it can
be proved that at the end of the unitary accelerating process $U_{A}(2n)$
the atomic wave-packet motional state is given by%
\begin{equation*}
\Psi _{2n}^{a}(x,t_{2n-1}^{e})=\exp (i\varphi _{2n}^{a})[\frac{(\Delta x)^{2}%
}{2\pi }]^{1/4}\sqrt{\frac{1}{(\Delta x)^{2}+i\frac{\hslash (T_{a}+2nt_{a})}{%
2M}}}
\end{equation*}%
\begin{equation}
\times \exp \{-\frac{1}{4}\frac{[x-z_{2n}^{a}]^{2}}{(\Delta x)^{2}+i\frac{%
\hslash (T_{a}+2nt_{a})}{2M}}\}\exp \{iP_{2n}^{a}x/\hslash \}  \tag{91}
\end{equation}%
and the atomic wave-packet product state by 
\begin{equation}
\Psi _{2n}^{a}(x,r,t_{2n}^{a})=\Psi _{2n}^{a}(x,t_{2n-1}^{e})|g_{0}\rangle 
\tag{92}
\end{equation}%
where the end time of the unitary accelerating process $U_{A}(2n)$ is $%
t_{2n-1}^{e}=t_{2n}^{a}=t_{0}^{a}+2nt_{a}$ for $n=0,1,2,...N,$ the atomic
motional momentum $P_{2n}^{a}$ is given by%
\begin{equation}
P_{2n}^{a}=P_{0}^{a}+n(\hslash k_{0}^{a}+\hslash k_{1}^{a})+n(\hslash
l_{0}^{a}+\hslash l_{1}^{a}),  \tag{93}
\end{equation}%
and the center-of-mass position $z_{2n}^{a}$ is determined from these
recursive relations:\newline
\begin{equation}
z_{2k-1}^{a}=z_{2k-2}^{a}+\frac{P_{2k-1}^{a}}{M}t_{a}-\frac{\hslash
(k_{0}^{a}+k_{1}^{a})}{M}%
\int_{t_{0}^{a}+(2k-2)t_{a}}^{t_{0}^{a}+(2k-1)t_{a}}dt^{\prime }\cos
^{2}\theta _{a}(t^{\prime }),  \tag{94a}
\end{equation}%
\begin{equation}
z_{2k}^{a}=z_{2k-1}^{a}+\frac{P_{2k}^{a}}{M}t_{a}-\frac{\hslash
(l_{0}^{a}+l_{1}^{a})}{M}%
\int_{t_{0}^{a}+(2k-1)t_{a}}^{t_{0}^{a}+2kt_{a}}dt^{\prime }\cos ^{2}\theta
_{la}(t^{\prime }),  \tag{94b}
\end{equation}%
\begin{equation*}
P_{2k-1}^{a}=P_{2k-2}^{a}+\hslash (k_{0}^{a}+k_{1}^{a}),\text{ }%
P_{2k}^{a}=P_{2k-1}^{a}+\hslash (l_{0}^{a}+l_{1}^{a}),
\end{equation*}%
where the initial center-of-mass position and momentum are $z_{0}^{a}$ and $%
P_{0}^{a},$ respectively, the index $1\leq k\leq n,$ both the basic
accelerating sequence (11a) and (63a) have the same duration $t_{a}$, and
the global phase factor $\exp (i\varphi _{2n}^{a})$ can also be calculated
through the recursive relations similar to Eq. (78) and (86). In an
analogous way, one also can calculate exactly the time evolution process of
the atom in the unitary accelerating process $U_{A}(2n-1)$ in the ideal
adiabatic condition (30). \newline
\newline
{\large 6. The space- and time-compressing processes based on the unitary
decelerating and accelerating processes}

Suppose that in the quantum control process [22] the first unitary
decelerating process consists of $2n_{d}$ basic STIRAP\ decelerating
sequences (11) and (63) alternately, which is given by $U_{D}(2n_{d})$ of
Eq. (70a), and the ideal adiabatic condition (30) is met for all these basic
decelerating sequences. According to the quantum control process the unitary
decelerating sequence is applied selectively in the given spatial region $%
[D_{L},$ $D_{R}]$ in the right-hand potential well of the double-well
potential field, where $D_{L}$ and $D_{R}$ are the left- and right-boundary
positions of the spatial region in the coordinate axis, respectively. The
halting-qubit atom can be decelerated by the unitary decelerating sequence $%
U_{D}(2n_{d})$ only when the atom enters into the spatial region $[D_{L},$ $%
D_{R}]$. Thus, the spatial region $[D_{L},$ $D_{R}]$ may be called the
decelerating spatial region. The decelerating spatial region must cover
sufficiently the whole wave-packet motional state of the atom during the
whole unitary decelerating process when the atom is decelerated in the
decelerating region. The decelerating region is so wide that for the
wave-packet motional state of the atom the Raman laser light beams of the
unitary decelerating sequence $U_{D}(2n_{d})$ can be thought of as infinite
plane-wave electromagnetic fields. Suppose that the halting-qubit atom is in
the product state $\Psi _{0}(x,r,t_{0})$ of Eq. (54) (or in the motional
state $\Psi _{0}(x,t_{0})$ of Eq. (48) and the internal state $|g_{0}\rangle
)$ and in the decelerating region $[D_{L},$ $D_{R}]$ when the unitary
decelerating sequence $U_{D}(2n_{d})$ is turned on at the initial time $%
t_{0} $. Here the spatial position of an atom is defined as the
center-of-mass position of the atomic wave-packet motional state. Now the
center-of-mass position and wave-packet spread of the initial motional state 
$\Psi _{0}(x,t_{0})$ are $z_{0}$ and $\varepsilon _{0},$ respectively. Since
the halting-qubit atom is in the decelerating region $[D_{L},$ $D_{R}]$ at
the time $t_{0},$ that is, $z_{0}\in \lbrack D_{L},$ $D_{R}]$, both the
distances $(z_{0}-D_{L})$ and $(D_{R}-z_{0})$ must be much greater than the
wave-packet spreading $\varepsilon _{0},$ that is, $%
(D_{R}-z_{0})>(z_{0}-D_{L})>>\varepsilon _{0},$ meaning that the
decelerating region $[D_{L},$ $D_{R}]$ covers sufficiently the whole initial
wave-packet motional state $\Psi _{0}(x,t_{0}).$ The halting-qubit atom
starts to be decelerated by the unitary decelerating process $U_{D}(2n_{d})$
at the initial time $t_{0}$ and in the decelerating region $[D_{L},$ $D_{R}]$%
. With the help of the recursive relations (71)--(78) and (79)--(86) one can
prove that at the end time $t_{2n_{d}}^{d}=t_{0}+2n_{d}t_{d}$ of the unitary
decelerating process $U_{D}(2n_{d})$ the motional state of the halting-qubit
atom is given by 
\begin{equation*}
\Psi _{2n_{d}}^{d}(x,t_{2n_{d}}^{d})=\exp (i\varphi _{2n_{d}}^{d})[\frac{%
(\Delta x)^{2}}{2\pi }]^{1/4}\sqrt{\frac{1}{(\Delta x)^{2}+i\frac{\hslash
(T_{d}+2n_{d}t_{d})}{2M}}}
\end{equation*}%
\begin{equation}
\times \exp \{-\frac{1}{4}\frac{(x-z_{2n_{d}}^{d})^{2}}{(\Delta x)^{2}+i%
\frac{\hslash (T_{d}+2n_{d}t_{d})}{2M}}\}\exp \{iP_{2n_{d}}^{d}x/\hslash \} 
\tag{95}
\end{equation}%
and the atomic product state by 
\begin{equation}
\Psi _{2n_{d}}^{d}(x,r,t_{2n_{d}}^{d})=\Psi
_{2n_{d}}^{d}(x,t_{2n_{d}}^{d})|g_{0}\rangle ,  \tag{96}
\end{equation}%
where the atomic motional momentum $P_{2n_{d}}^{d}$ is given by 
\begin{equation}
P_{2n_{d}}^{d}=p_{0}-n_{d}(\hslash k_{0}+\hslash k_{1})-n_{d}(\hslash
l_{0}+\hslash l_{1}),  \tag{97}
\end{equation}%
and the center-of-mass position $z_{2n_{d}}^{d}$ can be calculated by the
recursive relations (77) and (85),%
\begin{equation}
z_{2n-1}^{d}=z_{2n-2}^{d}+\frac{P_{2n-1}^{d}}{M}t_{d}+\frac{\hslash
(k_{0}+k_{1})}{M}\int_{t_{2n-2}^{d}}^{t_{2n-2}^{e}}dt^{\prime }\cos
^{2}\theta (t^{\prime }),  \tag{98a}
\end{equation}%
\begin{equation}
z_{2n}^{d}=z_{2n-1}^{d}+\frac{P_{2n}^{d}}{M}t_{d}+\frac{\hslash (l_{0}+l_{1})%
}{M}\int_{t_{2n-1}^{d}}^{t_{2n-1}^{e}}dt^{\prime }\cos ^{2}\theta
_{l}(t^{\prime }),  \tag{98b}
\end{equation}%
\begin{equation*}
P_{2n-1}^{d}=P_{2n-2}^{d}-(\hslash k_{0}+\hslash k_{1}),\text{ }%
P_{2n}^{d}=P_{2n-1}^{d}-(\hslash l_{0}+\hslash l_{1}),
\end{equation*}%
where $1\leq n\leq n_{d};$ $z_{0}^{d}=z_{0}$, $P_{0}^{d}=p_{0}$, $%
T_{d}=T_{0};$ $t_{0}^{d}=t_{0},$ $t_{k+1}^{d}=t_{k}^{d}+t_{d}$, and $%
t_{k}^{e}=t_{k+1}^{d}$ for $0\leq k\leq 2n_{d}-1,$ and the global phase
factor $\exp (i\varphi _{2n_{d}}^{d})$ can be calculated by Eq. (78) and
(86) with the initial phase $\varphi _{0}^{d}=\varphi _{0}.$ Both the
initial atomic product state $\Psi _{0}(x,r,t_{0})$ of Eq. (54) and the
final product state $\Psi _{2n_{d}}^{d}(x,r,t_{2n_{d}}^{d})$ of Eq. (96)
show that before and after the unitary decelerating process $U_{D}(2n_{d})$
the halting-qubit atom is in the same internal state $|g_{0}\rangle ,$ while
its initial motional state $\Psi _{0}(x,t_{0})$ of Eq. (48) is changed to
the motional state $\Psi _{2n_{d}}^{d}(x,t_{2n_{d}}^{d})$ of Eq. (95) after
the unitary decelerating process. The final motional state $\Psi
_{2n_{d}}^{d}(x,t_{2n_{d}}^{d})$ of Eq. (95) has the center-of-mass position 
$z_{2n_{d}}^{d}$ and the wave-packet spreading 
\begin{equation*}
\varepsilon =\sqrt{2(\Delta x)^{2}+2[\frac{\hslash (T_{d}+2n_{d}t_{d})}{%
2M(\Delta x)}]^{2}}.
\end{equation*}%
Though the atom moves a distance $(z_{2n_{d}}^{d}-z_{0})$ during the unitary
decelerating process, the center-of-mass position $z_{2n_{d}}^{d}\in \lbrack
D_{L},$ $D_{R}]$ as the atom is still in the decelerating region $[D_{L},$ $%
D_{R}]$ at the end of the unitary decelerating process. Then both the
distances $(z_{2n_{d}}^{d}-D_{L})$ and $(D_{R}-z_{2n_{d}}^{d})$ must be much
greater than the wave-packet spreading $\varepsilon ,$ that is, $%
(z_{2n_{d}}^{d}-D_{L})>(D_{R}-z_{2n_{d}}^{d})>>\varepsilon .$ This means
that the decelerating region $[D_{L},$ $D_{R}]$ also covers sufficiently the
whole final motional state $\Psi _{2n_{d}}^{d}(x,t_{2n_{d}}^{d}).$ There are
two extra constraint conditions on the decelerating region $[D_{L},$ $D_{R}]$%
. If the atom has not yet entered into the decelerating region when the
unitary decelerating sequence is switched on or it leaves the decelerating
region after the unitary decelerating process is switched off, then it will
not be affected by the unitary decelerating sequence or by next unitary
decelerating sequences. The two constraint conditions are stated below.

In the quantum control process [22] the halting-qubit atom may enter into
the right-hand potential well from the left-hand one in any $i-$th cycle of
the quantum program for $i=1,2,...,m_{r}.$ The $i-$th possible wave-packet
motional state of the halting-qubit atom is just defined as the atomic
motional state when the atom enters into the right-hand potential well in
the $i-$th cycle of the quantum program. Thus, there is a different time for
any possible atomic motional state such as the $i-$th wave-packet motional
state to enter into the right-hand potential well. If the time period of
each cycle of the quantum program is $\Delta T,$ then the time difference
between the $i-$th and $j-$th ($i<j$) wave-packet motional states to enter
into the right-hand potential well is given by $\Delta T(i,j)=(j-i)\Delta T$
for $i<j$ and $i,j=1,2,...,m_{r}.$ This time difference results in a
center-of-mass distance in space between the two wave-packet motional
states. If the halting-qubit atom moves along the direction $+x$ with the
velocity $p_{0}/M$, then the distance is given by $\Delta L(i,j)=(j-i)\Delta
T(p_{0}/M).$ Now examine two consecutive possible wave-packet motional
states: the $i-$th and $(i+1)-$th wave-packet motional states. Here for
convenience the $i-$th wave-packet motional state is set to the motional
state $\Psi _{0}(x,t_{0})$ of Eq. (48). Then at the time $t_{0}$ the $i-$th
wave-packet motional state is in the decelerating region $[D_{L},$ $D_{R}]$
and its center-of-mass position is $z_{0},$ while the center-of-mass
position of the $(i+1)-$th wave-packet motional state is clearly $%
[z_{0}-\Delta T(p_{0}/M)]$. It is known that the total duration for the
unitary decelerating sequence $U_{D}(2n_{d})$ is $2n_{d}t_{d}$. Then at the
end time $t_{0}+2n_{d}t_{d}$ of the unitary decelerating process the
center-of-mass position of the $(i+1)-$th wave-packet motional state becomes 
$[z_{0}-(\Delta T-2n_{d}t_{d})(p_{0}/M)].$ Obviously, the distance between
this center-of-mass position and the left-end position of the decelerating
region $[D_{L},$ $D_{R}]$ is $D_{L}-[z_{0}-(\Delta T-2n_{d}t_{d})(p_{0}/M)].$
Denote $\varepsilon _{i+1}(t_{0}+2n_{d}t_{d})$ as the wave-packet spreading
of the $(i+1)-$th wave-packet motional state at the time $t_{0}+2n_{d}t_{d}$%
. The wave-packet spreading $\varepsilon _{i+1}(t_{0}+2n_{d}t_{d})$ may be
calculated with the help of the $i-$th wave-packet state $\Psi _{0}(x,t_{0})$
and the free-particle propagator. Then this distance must be much greater
than $\varepsilon _{i+1}(t_{0}+2n_{d}t_{d}),$ that is, $D_{L}-[z_{0}-(\Delta
T-2n_{d}t_{d})(p_{0}/M)]>>\varepsilon _{i+1}(t_{0}+2n_{d}t_{d}),$ so that
the $(i+1)-$th wave-packet motional state is not affected by the unitary
decelerating sequence $U_{D}(2n_{d})$ during the whole unitary decelerating
process. This is a constraint condition on the decelerating region $[D_{L},$ 
$D_{R}].$

It is known that at the end time $t_{0}+2n_{d}t_{d}$ of the unitary
decelerating process $U_{D}(2n_{d})$ the $i-$th wave-packet motional state
is the state $\Psi _{2n_{d}}^{d}(x,t_{2n_{d}}^{d})$ of Eq. (95), which has
the center-of-mass position $z_{2n_{d}}^{d}$ and the motional momentum $%
P_{2n_{d}}^{d}$. After the unitary decelerating process the $i-$th
wave-packet motional state (i.e., the halting-qubit atom) moves along the
direction $+x$ with the velocity $P_{2n_{d}}^{d}/M$. Since the atom is
usually decelerated greatly by the decelerating sequence $U_{D}(2n_{d})$ the
atomic velocity $P_{2n_{d}}^{d}/M$ is much less than the original velocity $%
p_{0}/M$. Obviously, the $i-$th wave-packet motional state moves to the
position $z_{2n_{d}}^{d}+(\Delta T-2n_{d}t_{d})P_{2n_{d}}^{d}/M$ when next
unitary decelerating sequence starts to work at the time $t_{0}+\Delta T.$
Then the distance between this position and the right-end position of the
decelerating region $[D_{L},$ $D_{R}]$ is given by $z_{2n_{d}}^{d}+(\Delta
T-2n_{d}t_{d})P_{2n_{d}}^{d}/M-D_{R}.$ Denote $\varepsilon _{i}(t_{0}+\Delta
T)$ as the wave-packet spreading of the $i-$th wave-packet motional state at
the time $t_{0}+\Delta T.$ The wave-packet spreading $\varepsilon
_{i}(t_{0}+\Delta T)$ can be calculated with the help of the motional state $%
\Psi _{2n_{d}}^{d}(x,t_{2n_{d}}^{d})$ of Eq. (95) and the free-particle
propagator. Then this distance must be much greater than $\varepsilon
_{i}(t_{0}+\Delta T),$ that is, $z_{2n_{d}}^{d}+(\Delta
T-2n_{d}t_{d})P_{2n_{d}}^{d}/M-D_{R}>>\varepsilon _{i}(t_{0}+\Delta T),$ so
that, from the time $t_{0}+\Delta T$ on, the $i-$th wave-packet motional
state is no longer affected by the unitary decelerating sequences. This is
another constraint condition on the decelerating region $[D_{L},$ $D_{R}].$

The $i-$th wave-packet motional state is decelerated from the initial time $%
t_{0}$ to the end time $t_{0}+2n_{d}t_{d}$ by the unitary decelerating
process $U_{D}(2n_{d})$ in the decelerating region $[D_{L},$ $D_{R}]$. It
moves a distance $z_{2n_{d}}^{d}-z_{0}$ along the direction $+x$ and spends
the time $2n_{d}t_{d}$ and it is decelerated down to $P_{2n_{d}}^{d}/M$ from
the initial velocity $p_{0}/M$ during the unitary decelerating process.
According to the quantum control process [22] the $(i+1)-$th wave-packet
motional state arrives at the position $z_{0}$ in the decelerating region $%
[D_{L},$ $D_{R}]$ at the time $t_{0}+\Delta T.$ Then the $(i+1)-$th
wave-packet motional state at the time $t_{0}+\Delta T$ is really equal to
the $i-$th wave-packet motional state at the time $t_{0}$ up to a global
phase factor, indicating that the $(i+1)-$th wave-packet motional state at
the time $t_{0}+\Delta T$ is also equal to the motional state $\Psi
_{0}(x,t_{0})$ of Eq. (48) up to a global phase factor. Generally, according
to the quantum control process each of these $m_{r}$ possible wave packet
motional states is really equal to the motional state $\Psi _{0}(x,t_{0})$
of Eq. (48) up to a global phase factor when the wave-packet motional state
arrives at the same position $z_{0}$ in the decelerating region $[D_{L},$ $%
D_{R}]$. Just like the $i-$th wave-packet motional state at the time $t_{0}$
the $(i+1)-$th wave-packet motional state at the time $t_{0}+\Delta T$ is
decelerated by the unitary decelerating process $U_{D}(2n_{d})$. It also
moves the distance $z_{2n_{d}}^{d}-z_{0}$ along the direction $+x$ and
spends the time $2n_{d}t_{d}$ and it is also decelerated down to $%
P_{2n_{d}}^{d}/M$ from the initial velocity $p_{0}/M$ during the unitary
decelerating process. Generally, each of these $m_{r}$ possible wave-packet
motional states moves the same distance $z_{2n_{d}}^{d}-z_{0}$ along the
direction $+x$ and also spends the same time $2n_{d}t_{d}$ and it is also
decelerated down to the same velocity $P_{2n_{d}}^{d}/M$ from the same
initial velocity $p_{0}/M$ during the unitary decelerating process. The
difference among these $m_{r}$ possible wave-packet motional states is that
the starting time is different to decelerate each one of these wave-packet
motional states by the unitary decelerating process $U_{D}(2n_{d})$.

In the quantum control process [22] the unitary decelerating sequence is
used to decelerate the halting-qubit atom so that the center-of-mass
distances between these $m_{r}$ possible wave-packet motional states of the
atom can be narrowed greatly. Thus, the unitary decelerating process $%
U_{D}(2n_{d})$ is really a space-compressing process for these possible
wave-packet motional states. Since each one of these $m_{r}$ possible
wave-packet motional states spends the same time $2n_{d}t_{d}$ in the
unitary decelerating process $U_{D}(2n_{d}),$ the time difference $\Delta
T(i,j)=(j-i)\Delta T$ between the $i-$th and $j-$th wave-packet motional
states ($i<j;$ $i,j=1,2,...,m_{r}$) does not change before and after the
unitary decelerating process. It is known that each possible wave-packet
motional state has the initial moving velocity $p_{0}/M$ before the unitary
decelerating process and the final moving velocity $P_{2n_{d}}^{d}/M>0$
after the unitary decelerating process. Here the atomic moving velocity $%
P_{2n_{d}}^{d}/M$ can be obtained from Eq. (97),%
\begin{equation}
P_{2n_{d}}^{d}/M=[p_{0}-n_{d}(\hslash k_{0}+\hslash k_{1})-n_{d}(\hslash
l_{0}+\hslash l_{1})]/M.  \tag{99}
\end{equation}%
If the number $n_{d}$ of the unitary decelerating process $U_{D}(2n_{d})$ is
chosen suitably, then the velocity $P_{2n_{d}}^{d}/M$ can be much less than
the initial one $p_{0}/M.$ Before the unitary decelerating process the
distance between the $i-$th and $j-$th wave-packet motional states is $%
\Delta L_{0}(i,j)=(j-i)\Delta T(p_{0}/M)$, since the velocity is $p_{0}/M$
and the time difference is $\Delta T(i,j)=(j-i)\Delta T$ before the unitary
decelerating process. After the unitary decelerating process the atomic
moving velocity is $P_{2n_{d}}^{d}/M$ and the time difference is still $%
\Delta T(i,j)=(j-i)\Delta T$. Then after the unitary decelerating process
the distance between the $i-$th and $j-$th ($i<j$) wave-packet motional
states is equal to 
\begin{equation}
\Delta L(i,j)=(j-i)\Delta T(P_{2n_{d}}^{d}/M),  \tag{100}
\end{equation}%
where\ $i<j;$ $i,j=1,2,...,m_{r}.$ Since the velocity $%
(P_{2n_{d}}^{d}/M)<<p_{0}/M,$ the distance $\Delta L(i,j)<<\Delta
L_{0}(i,j), $ indicating that the spatial region to cover all these $m_{r}$
possible wave-packet motional states is greatly compressed after the unitary
decelerating process. The distance $\Delta L(i,j)$ of Eq. (100)\ has been
obtained in the previous paper [22], where the atomic velocity $%
(P_{2n_{d}}^{d}/M)$ is denoted as $v_{0}$ after the unitary decelerating
process. Then the ratio of the two distances $\Delta L(i,j)$ and $\Delta
L_{0}(i,j)$ is the space-compressing factor for these wave-packet motional
states after and before the unitary decelerating process, which can be
calculated by 
\begin{equation}
R_{s}=\frac{\Delta L(i,j)}{\Delta L_{0}(i,j)}=\frac{[p_{0}-n_{d}(\hslash
k_{0}+\hslash k_{1})-n_{d}(\hslash l_{0}+\hslash l_{1})]}{p_{0}}.  \tag{101}
\end{equation}%
The space-compressing factor is not dependent upon the indices $i$ and $j$,
since the time difference $\Delta T(i,j)$ does not change before and after
the unitary decelerating process and since all these possible wave-packet
motional states have the same initial motional momentum $p_{0}$ and also the
same final motional momentum $P_{2n_{d}}^{d}$ after each of these possible
wave-packet motional states undergoes the same unitary decelerating process $%
U_{D}(2n_{d})$ in the same decelerating region $[D_{L},$ $D_{R}]$.

Before the unitary accelerating process comes to making a real action on the
halting-qubit atom, the atom needs to stay in the right-hand potential well
for a time period to wait for the quantum program running to the end
according to the quantum control process [22]. The time period during which
the halting-qubit atom stays in the right-hand potential well is different
and dependent upon how early the halting-qubit atom enters into the
right-hand potential well from the left-hand one. When the atom enters into
the right-hand potential well at an earlier time, it will stay in the
right-hand potential well for a longer time. Denote that $%
T_{s}(i)=T_{s}-(i-1)\Delta T$ with the index $i=1,2,...,m_{r}$ is the time
period during which the atom moves freely along the direction $+x$ in the
right-hand potential well after the atom is decelerated by the unitary
decelerating sequence $U_{D}(2n_{d})$ and before the atom starts to be
accelerated at the end time of the quantum program. The index $i$ indicates
that the halting-qubit atom enters into the right-hand potential well from
the left-hand one in the $i-$th cycle of the quantum program. Here suppose
that the last unitary decelerating process $U_{D}(2n_{d})$ is turned off
before the quantum program comes to the end. The calculation for the time
evolution process of the halting-qubit atom moving freely during the time
period $T_{s}(i)$ needs to use the free-particle unitary propagator. Now the
unitary propagator of a free particle is written as [25]%
\begin{equation}
G(x^{\prime },t^{\prime };x,t)=\sqrt{\frac{M}{2\pi i\hslash (t^{\prime }-t)}}%
\exp [\frac{iM(x^{\prime }-x)^{2}}{2\hslash (t^{\prime }-t)}].  \tag{102}
\end{equation}%
Then the time evolution process of an atom in a free-particle motion with
the time period $T=t^{\prime }-t$ can be calculated by 
\begin{equation}
\Psi (x^{\prime },t^{\prime })=\int dxG(x^{\prime },t^{\prime };x,t)\Psi
(x,t).  \tag{103}
\end{equation}%
It is known that the $i-$th wave-packet motional state of the atom is given
by the motional state $\Psi _{2n_{d}}^{d}(x,t_{2n_{d}}^{d})$ of Eq. (95) at
the end time $t_{2n_{d}}^{d}=t_{0}+2n_{d}t_{d}$ of the unitary decelerating
process. When the wave-packet motional state $\Psi
_{2n_{d}}^{d}(x,t_{2n_{d}}^{d})$ moves freely along the direction $+x$ for
the time period $T_{s}(i)$ from the time $t_{2n_{d}}^{d}$ to the time $%
t_{2n_{d}}^{d}+T_{s}(i),$ it will change to another Gaussian wave-packet
motional state. This Gaussian wave-packet motional state can be calculated
from the equation (103) by taking the initial state $\Psi (x,t)$ as $\Psi
_{2n_{d}}^{d}(x,t_{2n_{d}}^{d})$ of Eq. (95), using the free-particle
propagator $G(x^{\prime },t^{\prime };x,t)$ of Eq. (102), and\ denoting $%
t^{\prime }=t_{2n_{d}}^{d}+T_{s}(i)$ and $t=t_{2n_{d}}^{d}.$ By a complex
calculation, in which the Gaussian integral (58) has been used, the final
Gaussian wave-packet state $\Psi (x^{\prime },t^{\prime })$ can be obtained
explicitly, which now is renamed $\Psi _{i}^{F}(x,t_{2n_{d}}^{d}+T_{s}(i))$, 
\begin{equation*}
\Psi _{i}^{F}(x,t_{2n_{d}}^{d}+T_{s}(i))=\exp (i\varphi _{2n_{d}}^{d})\exp
\{-i\frac{(P_{2n_{d}}^{d})^{2}T_{s}(i)}{2\hslash M}\}
\end{equation*}%
\begin{equation*}
\times \lbrack \frac{(\Delta x)^{2}}{2\pi }]^{1/4}\sqrt{\frac{1}{[(\Delta
x)^{2}+i\frac{\hslash (T_{d}+2n_{d}t_{d}+T_{s}(i))}{2M}]}}
\end{equation*}%
\begin{equation}
\times \exp \{-\frac{1}{4}\frac{%
[x-z_{2n_{d}}^{d}-(P_{2n_{d}}^{d}/M)T_{s}(i)]^{2}}{[(\Delta x)^{2}+i\frac{%
\hslash (T_{d}+2n_{d}t_{d}+T_{s}(i))}{2M}]}\}\exp \{iP_{2n_{d}}^{d}x/\hslash
\}.  \tag{104}
\end{equation}%
On the other hand, the atomic internal state $|g_{0}\rangle $ and the
motional momentum $P_{2n_{d}}^{d}$ keep unchanged during the free-particle
motion of the atom. Therefore, before the unitary accelerating process
starts at the end of the quantum program, these $m_{r}$ possible wave-packet
motional states are given by $\Psi _{i}^{F}(x,t_{2n_{d}}^{d}+T_{s}(i))$ of
Eq. (104) for $i=1,2,...,m_{r}$ and each of them has a different
center-of-mass position: $z_{2n_{d}}^{d}+(P_{2n_{d}}^{d}/M)T_{s}(i),$ a
different global phase factor:%
\begin{equation*}
\exp (i\varphi _{2n_{d}}^{d})\exp \{-i(P_{2n_{d}}^{d})^{2}T_{s}(i)/(2\hslash
M)\},
\end{equation*}%
and a different complex linewidth: 
\begin{equation}
W_{i}(T_{d}+2n_{d}t_{d}+T_{s}(i))=(\Delta x)^{2}+i[\frac{\hslash
(T_{d}+2n_{d}t_{d}+T_{s}(i))}{2M}].  \tag{105}
\end{equation}%
An important fact is that the imaginary part of the complex linewidth of the
motional state $\Psi _{i}^{F}(x,t_{2n_{d}}^{d}+T_{s}(i))$ increases linearly
with the time period $T_{s}(i),$ while the real part keeps unchanged. Though
each one of these $m_{r}$ possible wave-packet motional states has the same
wave-packet spreading and the same complex linewidth $W(T_{d}+2n_{d}t_{d})=(%
\Delta x)^{2}+i\frac{\hslash (T_{d}+2n_{d}t_{d})}{2M}$ before the
free-particle motion, as can seen from the state $\Psi
_{2n_{d}}^{d}(x,t_{2n_{d}}^{d})$ of Eq. (95), each possible wave-packet
motional state has a larger wave-packet spreading and a different complex
linewidth $W_{i}(T_{d}+2n_{d}t_{d}+T_{s}(i))$ when the unitary accelerating
sequence starts to act on the halting-qubit atom at the end of the quantum
program. Obviously, the first wave-packet motional state $\Psi
_{1}^{F}(x,t_{2n_{d}}^{d}+T_{s})$ has the largest wave-packet spreading,
while the last motional state $\Psi
_{m_{r}}^{F}(x,t_{2n_{d}}^{d}+T_{s}(m_{r}))$ has the least one. These show
that the free-particle motion of the halting-qubit atom leads to the
difference among the wave-packet spreads of these $m_{r}$ possible
wave-packet motional states and makes these wave-packet motional states
broader. This difference may have a significant impact on the quantum
control process [22]. On the other hand, the free-particle motion of the
halting-qubit atom does not change the time differences and the distances in
space between these $m_{r}$ possible wave-packet motional states. This is
because the motional momentum $P_{2n_{d}}^{d}$ is the same for all these $%
m_{r}$ possible wave-packet motional states and keeps unchanged during the
free-particle motion. Thus, the distance between the $i-$th and $j-$th ($i<j$%
) wave-packet motional states is still given by $\Delta L(i,j)$ of Eq. (100)
and their time difference by $\Delta T(i,j)=(j-i)\Delta T$. Particularly,
the distance between two nearest wave-packet motional states is given by $%
\Delta T(P_{2n_{d}}^{d}/M).$ Obviously, the halting-qubit atom moves a
distance equal to $(P_{2n_{d}}^{d}/M)T_{s}(i)$ along the direction $+x$ in
the time period $T_{s}(i)$ of the free-particle motion. This distance is
dependent upon the index $i$. The first wave-packet motional state $\Psi
_{1}^{F}(x,t_{2n_{d}}^{d}+T_{s})$ moves the largest distance $%
(P_{2n_{d}}^{d}/M)T_{s}$ which decides mainly the dimensional size of the
right-hand potential well, while the last wave-packet motional state $\Psi
_{m_{r}}^{F}(x,t_{2n_{d}}^{d}+T_{s}(m_{r}))$ moves the shortest distance $%
(P_{2n_{d}}^{d}/M)[T_{s}-(m_{r}-1)\Delta T]$.

The atomic wave-packet states $\{\Psi _{i}^{F}(x,t_{2n_{d}}^{d}+T_{s}(i))\}$
of Eq. (104) show that just before the unitary accelerating sequence is
switched on, all these $m_{r}$ possible wave-packet states of Eq. (104) are
in the spatial region $[x_{0}(m_{r})-\varepsilon _{d}(m_{r}),$ $%
x_{0}(1)+\varepsilon _{d}(1)],$ where $x_{0}(j)$ and $\varepsilon _{d}(j)$ ($%
j=1,2,...,m_{r}$) are the center-of-mass position and the wave-packet
spreading of the $j-$th wave-packet state $\Psi
_{j}^{F}(x,t_{2n_{d}}^{d}+T_{s}(j)),$ respectively. Suppose that all these $%
m_{r}$ possible wave-packet states are accelerated uniformly by the unitary
accelerating sequence and each possible wave-packet state moves the same
distance $L_{A}$ during the unitary accelerating process. The distance $%
L_{A} $ will be obtained later. Obviously, after the unitary accelerating
process all these $m_{r}$ possible wave-packet motional states are in the
spatial region $[x_{0}(m_{r})+L_{A}-\varepsilon _{a}(m_{r}),$ $%
x_{0}(1)+L_{A}+\varepsilon _{a}(1)],$ where $\varepsilon _{a}(j)$ ($%
j=1,2,...,m_{r}$) is the wave-packet spreading of the $j-$th wave-packet
motional state of the atom after the unitary accelerating process.
Therefore, during the unitary accelerating process any possible wave-packet
motional state of the halting-qubit atom is within the effective spatial
region:%
\begin{equation*}
\lbrack A_{L},\text{ }A_{R}]=[x_{0}(m_{r})-\varepsilon _{d},\text{ }%
x_{0}(1)+L_{A}+\varepsilon _{a}]\newline
\end{equation*}%
where $\varepsilon _{a}>>\varepsilon _{a}(1)$ and $\varepsilon
_{d}>>\varepsilon _{d}(m_{r})$. The effective spatial region $[A_{L},$ $%
A_{R}]$ covers all these $m_{r}$ possible wave-packet motional states of the
atom during the whole unitary accelerating process. Now the spatial region
of the unitary accelerating sequence must encompass sufficiently the whole
effective spatial region $[A_{L},$ $A_{R}],$ so that for all these $m_{r}$
possible wave-packet motional states the Raman laser light beams of the
unitary accelerating sequence can be thought of as infinite plane-wave
electromagnetic fields, and the most important is that the unitary
accelerating sequence can act on all these $m_{r}$ possible wave-packet
motional states simultaneously and uniformly during the whole unitary
accelerating process. The spatial region $[A_{L},$ $A_{R}]$ may be called
the accelerating spatial region.

According to the quantum control process [22] the halting-qubit atom is
accelerated by a unitary accelerating sequence at the end time of the
quantum program. Here the unitary accelerating sequence may be given by $%
U_{A}(2n_{a})$ of Eq. (87a), which consists of $n_{a}$ pairs of the basic
STIRAP\ accelerating sequences (11a) and (63a) in an alternate form and each
basic accelerating sequence has the same time period $t_{a}$. The unitary
accelerating process $U_{A}(2n_{a})$ has a total time period $2n_{a}t_{a}.$
The ideal adiabatic condition (30) is also met in the unitary accelerating
process. Now one may use the recursive relations (88)---(94) to obtain the
final wave-packet motional state of the halting-qubit atom after the atom is
accelerated by the unitary accelerating sequence $U_{A}(2n_{a})$. Here the
starting time of the unitary accelerating process is the end time $t_{m_{r}}$
of the quantum program. At the initial time $t_{m_{r}}$ each possible
wave-packet motional state of the halting-qubit atom is given by $\Psi
_{j}^{F}(x,t_{2n_{d}}^{d}+T_{s}(j))$ of Eq. (104) for $j=1,2,...,m_{r}$. All
these $m_{r}$ possible wave-packet motional states start to undergo the same
unitary accelerating process $U_{A}(2n_{a})$ at the initial time $t_{m_{r}}$
simultaneously. In order to use the recursive relations (88)--(94) the
initial motional state $\Psi _{0}^{a}(x,t_{0}^{a})$ of Eq. (88) needs first
to be obtained from the state $\Psi _{j}^{F}(x,t_{2n_{d}}^{d}+T_{s}(j))$ of
Eq. (104). By comparing the initial state $\Psi _{0}^{a}(x,t_{0}^{a})$ of
Eq. (88) with $\Psi _{j}^{F}(x,t_{2n_{d}}^{d}+T_{s}(j))$ of Eq. (104) one
can see that at the initial time $t_{0}^{a}=t_{m_{r}}$ the center-of-mass
position, momentum, and global phase factor of the initial state $\Psi
_{0}^{a}(x,t_{0}^{a})$ are given by $z_{0}^{a}\equiv
z_{0}^{a}(j)=z_{2n_{d}}^{d}+(P_{2n_{d}}^{d}/M)T_{s}(j),$ $%
P_{0}^{a}=P_{2n_{d}}^{d},$ and $\exp [i\varphi _{0}^{a}]\equiv \exp
[i\varphi _{0}^{a}(j)]=\exp (i\varphi _{2n_{d}}^{d})\exp
[-i(P_{2n_{d}}^{d})^{2}T_{s}(j)/(2\hslash M)],$ respectively, and in the
complex linewidth $W(T_{a})$ of the initial state $\Psi
_{0}^{a}(x,t_{0}^{a}) $ the time interval $T_{a}\equiv
T_{a}(j)=T_{d}+2n_{d}t_{d}+T_{s}(j).$ It is known that the initial internal
state is $|g_{0}\rangle .$ The initial atomic wave-packet product state then
is given by $\Psi _{0}^{a}(x,r,t_{0}^{a})=\Psi
_{j}^{F}(x,t_{2n_{d}}^{d}+T_{s}(j))|g_{0}\rangle $. After the unitary
accelerating process $U_{A}(2n_{a})$ the wave-packet motional state of the
halting-qubit atom will take the form, according to the recursive relations
(88)--(94), 
\begin{equation*}
\Psi _{2n_{a},j}^{a}(x,t_{2n_{a}}^{a})=\exp [i\varphi _{2n_{a}}^{a}(j)][%
\frac{(\Delta x)^{2}}{2\pi }]^{1/4}\sqrt{\frac{1}{(\Delta x)^{2}+i\frac{%
\hslash (T_{a}(j)+2n_{a}t_{a})}{2M}}}
\end{equation*}%
\begin{equation}
\times \exp \{-\frac{1}{4}\frac{[x-z_{2n_{a}}^{a}(j)]^{2}}{(\Delta x)^{2}+i%
\frac{\hslash (T_{a}(j)+2n_{a}t_{a})}{2M}}\}\exp \{iP_{2n_{a}}^{a}x/\hslash
\}  \tag{106}
\end{equation}%
and the atomic wave-packet product state is given by 
\begin{equation}
\Psi _{2n_{a},j}^{a}(x,r,t_{2n_{a}}^{a})=\Psi
_{2n_{a},j}^{a}(x,t_{2n_{a}}^{a})|g_{0}\rangle ,  \tag{107}
\end{equation}%
where the end time of the unitary accelerating process is $%
t_{2n_{a}}^{a}=t_{m_{r}}+2n_{a}t_{a},$ and the atomic motional momentum is
given by 
\begin{equation}
P_{2n_{a}}^{a}=P_{2n_{d}}^{d}+n_{a}(\hslash k_{0}^{a}+\hslash
k_{1}^{a})+n_{a}(\hslash l_{0}^{a}+\hslash l_{1}^{a}),  \tag{108}
\end{equation}%
and the center-of-mass position $z_{2n_{a}}^{a}(j)$ can be determined from
the recursive relations:%
\begin{equation}
z_{2k-1}^{a}(j)=z_{2k-2}^{a}(j)+\frac{P_{2k-1}^{a}}{M}t_{a}-\frac{\hslash
(k_{0}^{a}+k_{1}^{a})}{M}%
\int_{t_{0}^{a}+(2k-2)t_{a}}^{t_{0}^{a}+(2k-1)t_{a}}dt^{\prime }\cos
^{2}\theta _{a}(t^{\prime }),  \tag{109a}
\end{equation}%
\begin{equation}
z_{2k}^{a}(j)=z_{2k-1}^{a}(j)+\frac{P_{2k}^{a}}{M}t_{a}-\frac{\hslash
(l_{0}^{a}+l_{1}^{a})}{M}%
\int_{t_{0}^{a}+(2k-1)t_{a}}^{t_{0}^{a}+2kt_{a}}dt^{\prime }\cos ^{2}\theta
_{la}(t^{\prime }),  \tag{109b}
\end{equation}%
\begin{equation*}
P_{2k-1}^{a}=P_{2k-2}^{a}+\hslash (k_{0}^{a}+k_{1}^{a}),\text{ }%
P_{2k}^{a}=P_{2k-1}^{a}+\hslash (l_{0}^{a}+l_{1}^{a}),
\end{equation*}%
where $1\leq k\leq n_{a}.$ The global phase factor $\exp [i\varphi
_{2n_{a}}^{a}(j)]$ in Eq. (106) can also be obtained from the recursive
relations similar to Eq. (78) and (86). Now one can find from the final
motional state $\Psi _{2n_{a},j}^{a}(x,t_{2n_{a}}^{a})$ of Eq. (106) that
the moving distance $L_{A}$ of the halting-qubit atom is $%
L_{A}=z_{2n_{a}}^{a}(j)-z_{0}^{a}(j)$ during the unitary accelerating
process, which appears in the accelerating region $[A_{L},$ $A_{R}]$ above.
Note that the distance $L_{A}$ is the same for each one of these $m_{r}$
possible wave-packet motional states.

The unitary accelerating process tells ones some facts. For the first point,
the halting-qubit atom indeed is accelerated by $n_{a}(\hslash
k_{0}^{a}+\hslash k_{1}^{a})+n_{a}(\hslash l_{0}^{a}+\hslash l_{1}^{a})$ and
this accelerating process is uniform, that is, the accelerating process is
the same for each one of these $m_{r}$ possible wave-packet motional states $%
\{\Psi _{j}^{F}(x,t_{2n_{d}}^{d}+T_{s}(j))\}$. Thus, after the unitary
accelerating process the atom is accelerated to the velocity $%
(P_{2n_{a}}^{a}/M).$ For the second point, it can be seen from the motional
states $\{\Psi _{2n_{a},j}^{a}(x,t_{2n_{a}}^{a})\}$ of Eq. (106) that in the
complex linewidth the imaginary part increases linearly with the time period
of the unitary accelerating process and is increased by $\hslash
(2n_{a}t_{a})/(2M),$ which is also independent of any index value $j$, while
the real part keeps unchanged in the unitary accelerating process. For the
third point, the distances between these $m_{r}$ possible wave-packet
motional states keep unchanged during the unitary accelerating process. This
fact can be deduced from the recursive relations (109a) and (109b) because
the motional momentum $P_{l}^{a}$ ($l=0,1,2,...,2n_{a}$)$,$ the mixing
angles $\theta _{a}(t)$ and $\theta _{la}(t),$ and the wave numbers $%
(k_{0}^{a}+k_{1}^{a})$ and $(l_{0}^{a}+l_{1}^{a})$ all are independent of
the index value $j$. This means that each one of these $m_{r}$ possible
wave-packet motional states moves the same spatial distance during the
unitary accelerating process. Since the distance $\Delta L(i,j)$ between the 
$i-$th and $j-$th ($i<j$) wave-packet motional states is still given by Eq.
(100) and the atomic moving velocity is $(P_{2n_{a}}^{a}/M)$ after the
unitary accelerating process, the time difference between the two
wave-packet states $\Psi _{2n_{a},i}^{a}(x,t_{2n_{a}}^{a})$ and $\Psi
_{2n_{a},j}^{a}(x,t_{2n_{a}}^{a})$ then is given by%
\begin{equation}
\Delta T(i,j)=\Delta L(i,j)/(P_{2n_{a}}^{a}/M)=(j-i)\Delta
T(P_{2n_{d}}^{d}/P_{2n_{a}}^{a}),  \tag{110}
\end{equation}%
where $i<j;$ $i,j=1,2,...,m_{r}.$ It is known that the time difference $%
\Delta T_{0}(i,j)=(j-i)\Delta T$ before the unitary accelerating process.
Since the atomic velocity $(P_{2n_{a}}^{a}/M)$ after the accelerating
process is much greater than the velocity $(P_{2n_{d}}^{d}/M)$ before the
accelerating process, the time difference $\Delta T(i,j)<<\Delta T_{0}(i,j),$
indicating that the time differences are compressed greatly for these $m_{r}$
possible wave-packet motional states after the unitary accelerating process.
Then the time-compressing factor for these possible wave-packet motional
states after and before the unitary accelerating process $U_{A}(2n_{a})$ can
be calculated by%
\begin{equation}
R_{t}=\frac{\Delta T(i,j)}{\Delta T_{0}(i,j)}=\frac{P_{2n_{d}}^{d}}{%
P_{2n_{d}}^{d}+n_{a}(\hslash k_{0}^{a}+\hslash k_{1}^{a})+n_{a}(\hslash
l_{0}^{a}+\hslash l_{1}^{a})}.  \tag{111}
\end{equation}%
The time-compressing factor $R_{t}$ is independent of the indices $i$ and $j$%
. Thus, the time-compressing process is uniform. The time-compressing factor 
$R_{t}$ has been obtained in the previous paper [22], where $R_{t}=(v_{0}/v)$
and $v_{0}$ and $v$ are denoted as the atomic moving velocities $%
P_{2n_{d}}^{d}/M$ and $P_{2n_{a}}^{a}/M$ before and after the unitary
accelerating process, respectively.\newline
\newline
{\large 7. General adiabatic conditions and the error estimation for the
decelerating and accelerating processes}

The starting point to set up a general adiabatic condition for a basic
STIRAP decelerating or accelerating process is to solve the basic equations
(23) to find the coefficients $\{a_{k}(P,t)\}$ or to solve the basic
equations (26) to obtain the coefficients $\{b_{k}(P,t)\}$. Then it is to
seek under what experimental conditions a real adiabatic condition for the
basic STIRAP decelerating or accelerating process can be sufficiently close
to the ideal adiabatic condition (30). This is a routine procedure in
quantum mechanics [25]. There are three basic parameters to affect the real
adiabatic condition of a STIRAP experiment: the time period of the STIRAP
experiment, the Rabi frequencies and the phase-modulation functions of the
Raman laser light beams. From the point of view of quantum computation one
usually does not expect the quantum control process to consume a long time.
However, a long time period of the STIRAP experiment usually can lead to
that the adiabatic condition for the STIRAP experiment is met better [30].
If the time period of each basic STIRAP pulse sequence in the STIRAP-based
unitary decelerating and accelerating processes is not long enough, then the
adiabatic condition could not be met well. Then in this situation one may
use jointly the time period, the Rabi frequencies, and even the
phase-modulation functions to achieve a better adiabatic condition for these
decelerating and accelerating processes. Actually, the Rabi frequencies of
the Raman laser light beams are very important to achieve a better adiabatic
condition for the STIRAP\ experiment [15, 18b]. Without losing generality
here take the basic STIRAP decelerating sequence (11) as an example to
discuss a general adiabatic condition. The obtained results can be used as
well for other basic STIRAP decelerating and accelerating processes. The
STIRAP adiabatic conditions have been discussed in detail in many references
[15, 16, 17, 18] in the conventional STIRAP experiments without considering
explicitly the atomic or molecular momentum distribution. The conventional
adiabatic conditions [4, 15, 17, 18] usually are based on the first-order
approximation solution to the basic equations similar to the present basic
differential equations (26). These adiabatic conditions are usually a
qualitative and approximate description to the adiabatic theorem. In the
following two strict and different general adiabatic conditions are derived
analytically. They are a quantitative description to the adiabatic theorem.
The first general adiabatic condition is based on the Dyson series solution
(29) of the basic differential equations (26). The second is based on a new
method to solve the basic differential equations (26). This new method uses
the equivalent transformations to solve the basic differential equations
(26). That is, by making repeatedly the equivalent transformations the three
basic differential equations (26) are transformed to the three equivalent
linear algebra equations. Though the final solution to the basic
differential equations (26) obtained with the new method is approximate, the
truncation error of the approximation solution can be controlled as desired.
The two general adiabatic conditions may be used to set up the conventional
STIRAP experiments. Thus, they may be used to design the STIRAP pulse
sequence to realize the perfect state (or population) transfer for a quantum
ensemble of the atoms or molecules. But their more important application is
that they may be used to set up the basic STIRAP unitary decelerating and
accelerating processes for a free atom and an atomic or molecular ensemble.

The basic differential equations (26) or their matrix form (28) can be
integrated formally. The formal solution to the basic equations (28) may be
expressed as the Dyson series (29). Here one needs to use the initial
condition of the basic STIRAP decelerating sequence (11).\ At the initial
time $t_{0}$ of the basic STIRAP\ decelerating sequence (11)\ the
three-state vector $%
B(P,t_{0})=(b_{0}(P,t_{0}),b_{+}(P,t_{0}),b_{-}(P,t_{0}))^{T}$ is given by
Eq. (39). The initial condition (39) has been used to set up the ideal
adiabatic condition (30). For a real adiabatic condition the initial
condition may be generally given in (130) below. At first the formal
solution (29) may be rewritten as 
\begin{equation}
B(P,t)=B(P,t_{0})+E_{r}(P,t)  \tag{112}
\end{equation}%
where $t_{0}\leq t\leq t_{0}+T$ and $T$ is the time period of the basic
STIRAP decelerating process, and the error term $E_{r}(P,t)$ measures the
deviation of a real adiabatic condition from the ideal adiabatic condition
and it may be expressed as%
\begin{equation*}
E_{r}(P,t)=\{(\frac{1}{i})\int_{t_{0}}^{t}dt_{1}M(P,t_{1})+(\frac{1}{i}%
)^{2}\int_{t_{0}}^{t}\int_{t_{0}}^{t_{1}}dt_{1}dt_{2}M(P,t_{1})M(P,t_{2})
\end{equation*}%
\begin{equation}
+(\frac{1}{i})^{3}\int_{t_{0}}^{t}\int_{t_{0}}^{t_{1}}%
\int_{t_{0}}^{t_{2}}dt_{1}dt_{2}dt_{3}M(P,t_{1})M(P,t_{2})M(P,t_{3})+...%
\}B(P,t_{0}).  \tag{113}
\end{equation}%
The upper bound of the error term is evaluated accurately below. Denote the
maximum norm of the hermitian matrix $M(P,t_{n})$ which is given in (28) in
the time region $[t_{0},t_{n-1}]$ as 
\begin{equation*}
||M(P,t_{n})||_{\max }=\max_{t_{0}\leq t_{n}\leq t_{n-1}}\{||M(P,t_{n})||\},
\end{equation*}%
where $t_{0}<...<t_{n}<t_{n-1}<...<t_{1}<t_{0}+T.$ Obviously, there are the
following relations for the maximum norms $\{||M(P,t_{n})||_{\max }\}:$%
\begin{equation*}
||M(P,t_{0})||_{\max }\leq ...\leq ||M(P,t_{n})||_{\max }\leq
||M(P,t_{n-1})||_{\max }
\end{equation*}%
\begin{equation}
\leq ||M(P,t_{2})||_{\max }\leq ||M(P,t_{1})||_{\max }=||M(P,t)||_{\max }. 
\tag{114}
\end{equation}%
Here the maximum norm $||M(P,t)||_{\max }$ is defined as%
\begin{equation}
||M(P,t)||_{\max }=\max_{t_{0}\leq t\leq t_{0}+T}\{||M(P,t)||\}.  \tag{115}
\end{equation}%
Then with the help of (113) and (114) it can turn out that the upper bound
of the deviation $E_{r}(P,t)$ may be determined from 
\begin{equation}
||E_{r}(P,t)||\leq \exp [(||M(P,t)||_{\max })T]\times ||B^{(1)}(P,t)||_{\max
},  \tag{116}
\end{equation}%
where the first-order approximation solution $B^{(1)}(P,t)$ to the basic
differential equations (26) is given by%
\begin{equation}
B^{(1)}(P,t)=(\frac{1}{i})\int_{t_{0}}^{t}dt_{1}M(P,t_{1})B(P,t_{0}),\text{ }%
(t_{0}\leq t\leq t_{0}+T),  \tag{117}
\end{equation}%
while $||B^{(1)}(P,t)||_{\max }$ is the maximum norm of the solution $%
B^{(1)}(P,t)$ in the time region $t_{0}\leq t\leq t_{0}+T$. This norm $%
||B^{(1)}(P,t)||_{\max }$ is written as%
\begin{equation}
||B^{(1)}(P,t)||_{\max }=\sqrt{%
(|b_{0}^{(1)}(P,t)|^{2}+|b_{+}^{(1)}(P,t)|^{2}+|b_{-}^{(1)}(P,t)|^{2})_{\max
}}.  \tag{118}
\end{equation}%
On the other hand, it follows from the matrix $M(P,t)$ in (28) that the
maximum norm $||M(P,t)||_{\max }$ is bounded by%
\begin{equation}
||M(P,t)||_{\max }\leq \sqrt{\sum_{i,j}|M_{ij}(P,t)|^{2}}\leq \frac{1}{\sqrt{%
2}}\{2|\Theta (P,t)|+|\Gamma (P,t)|\}_{\max }.  \tag{119}
\end{equation}%
The adiabatic condition (116) is strict because it is required that at any
instant of time in the whole STIRAP decelerating or accelerating process the
deviation from the ideal adiabatic condition (30) be limited within a given
small value, that is, the upper bound of the error term $E_{r}(P,t)$ is less
than a given small value at any instant of time. Notice that in theory at
the initial time $t_{0}$ the atom is prepared to be in the trapped state $%
|g^{0}(P,t_{0})\rangle $ of (19a) completely. If the error term $E_{r}(P,t)$
is large, then this will mean that during the STIRAP decelerating or
accelerating process there is a large probability for the atom to be excited
to the two eigenstates $|g^{\pm }(P,t)\rangle $ of the instantaneous
Hamiltonian $H(P,t)$ of (17). It is known from (19b) that any one of the two
eigenstates $|g^{\pm }(P,t)\rangle $ contains the excited internal state of
the atom. Then the atom could be easily affected due to the atomic
spontaneous emission if it is in any one of the eigenstates $|g^{\pm
}(P,t)\rangle .$ On the other hand, a high probability for the atom to stay
in the trapped state $|g^{0}(P,t)\rangle $ may lead to that the atom is not
easily affected by environment and may avoid the spontaneous emission. The
adiabatic condition (116) indicates that the probability for the atom to
leave the trapped state $|g^{0}(P,t)\rangle $ may be limited to a small
value as desired during the STIRAP decelerating or accelerating process.
Therefore, it ensures that the atom is almost completely in the trapped
state $|g^{0}(P,t)\rangle $ during the STIRAP decelerating or accelerating
process. The adiabatic condition (116) is more severe than those in the
conventional STIRAP experiments [15, 17, 18]. The latter usually require
that the probability for the atoms or molecules under investigation in the
two eigenstates $|g^{\pm }(P,t)\rangle $ be much smaller than one. This is a
qualitative description for the adiabatic theorem. The present adiabatic
condition (116) is closely related to the requirement that Gaussian shape of
the Gaussian wave-packet motional state of the decelerated or accelerated
atom keep unchanged before and after the basic STIRAP decelerating and
accelerating processes. It measures the deviation of a real adiabatic
condition from the ideal adiabatic condition (30), while the deviation may
occur not only in the two eigenstates $|g^{\pm }(P,t)\rangle $ but also in
the trapped state $|g^{0}(P,t)\rangle .$ The present adiabatic condition
(116) limits the upper bound of the deviation to a given small value. This
is a quantitative description for the adiabatic theorem. This results in
that the present adiabatic condition (116) is more severe.

When the adiabatic condition (116) is met, the error term $%
||E_{r}(P,t_{0}+T)||$ of the final state at the time $t=t_{0}+T$ is clearly
not more than the upper bound (116) and the real error term $%
||E_{r}(P,t_{0}+T)||$ could be much less than the upper bound (116). It may
be required in theory that the real error term $||E_{r}(P,t_{0}+T)||$ of the
final state be less than some given value which is much less than the upper
bound (116). This requirement is not severe in theory with respect to the
adiabatic condition (116). It may be met by setting the suitable
experimental parameters at the final time $t=t_{0}+T$ for the STIRAP
decelerating or accelerating process. However, in practice the lower bound
of the error term $E_{r}(P,t_{0}+T)$ of the final state is generally
affected by the adiabatic condition (116). If the upper bound (116) is
large, then the lower bound of the error term $E_{r}(P,t_{0}+T)$ usually is
large too.

According to the superposition principle in quantum mechanics in a real
adiabatic condition the atomic product state at any instant of time $t$ ($%
t_{0}\leq t\leq t_{0}+T$) in the basic STIRAP decelerating process (11) may
be calculated from Eq. (12) $(P=P^{\prime }-\hslash k_{0}),$%
\begin{equation*}
|\Psi _{r}(x,r,t)\rangle =\sum_{P}\rho (P)\{[A_{0}^{i}(P,t)+\delta
_{0}^{A}(P,t)]|P+\hslash k_{0}\rangle |g_{0}\rangle
\end{equation*}%
\begin{equation}
+[A_{1}^{i}(P,t)+\delta _{1}^{A}(P,t)]|P\rangle |e\rangle
+[A_{2}^{i}(P,t)+\delta _{2}^{A}(P,t)]|P-\hslash k_{1}\rangle |g_{1}\rangle
\}  \tag{120}
\end{equation}%
where the coefficients $\{A_{k}^{i}(P,t)\}$ for $k=0,1,2$ are obtained in
the ideal adiabatic condition (30), while the coefficients $\{\delta
_{k}^{A}(P,t)\}$ measure the deviation of the real adiabatic condition from
the ideal one. The product state (120) may be rewritten as%
\begin{equation*}
|\Psi _{r}(x,r,t)\rangle =|\Psi _{i}(x,r,t)\rangle +E_{r}(x,r,t)
\end{equation*}%
where the wave-packet state $|\Psi _{i}(x,r,t)\rangle $ is the atomic state
at the time $t$ in the basic STIRAP decelerating process (11) in the ideal
adiabatic condition and it may be written as%
\begin{equation*}
|\Psi _{i}(x,r,t)\rangle =\sum_{P}\rho (P)\{A_{0}^{i}(P,t)|P+\hslash
k_{0}\rangle |g_{0}\rangle
\end{equation*}%
\begin{equation}
+A_{1}^{i}(P,t)|P\rangle |e\rangle +A_{2}^{i}(P,t)|P-\hslash k_{1}\rangle
|g_{1}\rangle \},  \tag{121}
\end{equation}%
and the error term $E_{r}(x,r,t)$ is given by%
\begin{equation*}
E_{r}(x,r,t)=\sum_{P}\rho (P)\{\delta _{0}^{A}(P,t)|P+\hslash k_{0}\rangle
|g_{0}\rangle
\end{equation*}%
\begin{equation}
+\delta _{1}^{A}(P,t)|P\rangle |e\rangle +\delta _{2}^{A}(P,t)|P-\hslash
k_{1}\rangle |g_{1}\rangle \}.  \tag{122}
\end{equation}%
It turns out in the preceding section 5 that the final state $|\Psi
_{i}(x,r,t)\rangle $ with $t=t_{0}+T$ in the ideal adiabatic condition is a
perfect Gaussian wave-packet state if the initial state of the basic STIRAP
decelerating process (11) is a Gaussian wave-packet state. Obviously, it
follows from (122) that the probability for the error term $E_{r}(x,r,t)$ at
any time $t$ may be calculated by%
\begin{equation}
||E_{r}(x,r,t)||^{2}=\sum_{P}|\rho (P)|^{2}\{|\delta
_{0}^{A}(P,t)|^{2}+|\delta _{1}^{A}(P,t)|^{2}+|\delta _{2}^{A}(P,t)|^{2}\}. 
\tag{123}
\end{equation}%
(Notice that the error probability (121) in the previous versions of this
paper which is denoted as $E_{r}(P,t)$ is just equal to $%
2||E_{r}(x,r,t)||^{2}$ of (123)). In order to use directly the solution to
the basic differential equations (26) or their matrix form (28) to calculate
the upper bound of the error term $E_{r}(x,r,t)$ one may use the
coefficients $b_{0}(P,t)$ and $b_{\pm }(P,t)$ to express the error
probability $||E_{r}(x,r,t)||^{2}$ of (123). Notice that there is the
unitary transformation between the two three-state vectors $%
(b_{0}(P,t),b_{+}(P,t),b_{-}(P,t))^{T}$ and $(A_{0}(P,t),A_{1}(P,t),$ $%
A_{2}(P,t))^{T}.$ The three-state vector $%
(A_{0}(P,t),A_{1}(P,t),A_{2}(P,t))^{T}$ is first converted into the
three-state vector $(\bar{A}_{0}(P,t),\bar{A}_{1}(P,t),\bar{A}_{2}(P,t))^{T}$
by the unitary transformation of (15a)-(15c), then into the three-state
vector $(a_{0}(P,t),a_{+}(P,t),$ $a_{-}(P,t))^{T}$ by the unitary
transformation of (22a)-(22c), and finally into the three-state vector $%
(b_{0}(P,t),$ $b_{+}(P,t),b_{-}(P,t))^{T}$ by the unitary transformation
(25). Thus, under these unitary transformations there is the relation: 
\begin{equation}
(A_{0}(P,t),A_{1}(P,t),A_{2}(P,t))^{T}=U_{Ab}(b_{0}(P,t),b_{+}(P,t),b_{-}(P,t))^{T}
\tag{124}
\end{equation}%
where $U_{Ab}$ is the unitary transformation between the two three-state
vectors $(b_{0}(P,t),b_{+}(P,t),b_{-}(P,t))^{T}$ and $%
(A_{0}(P,t),A_{1}(P,t),A_{2}(P,t))^{T}.$ If now the solution to the basic
equations (26) in the ideal adiabatic condition is given by $%
(b_{0}^{i}(P,t),b_{+}^{i}(P,t),b_{-}^{i}(P,t))^{T},$ then after the unitary
transformation $U_{Ab}$ one obtains the three-state vector $%
(A_{0}^{i}(P,t),A_{1}^{i}(P,t),A_{2}^{i}(P,t))^{T}$ of the ideal adiabatic
condition and then the state $|\Psi _{i}(x,r,t)\rangle $ can be calculated
from Eq. (121) by using the three-state vector. If the solution to the basic
equations (26) in a real adiabatic condition is given by $%
(b_{0}(P,t),b_{+}(P,t),b_{-}(P,t))^{T}$ with $b_{k}(P,t)=b_{k}^{i}(P,t)+%
\delta _{k}^{b}(P,t)$ for $k=0,+,-,$ then after the unitary transformation $%
U_{Ab}$ one obtains the three-state vector $%
(A_{0}(P,t),A_{1}(P,t),A_{2}(P,t))^{T}$ of the real adiabatic condition,
where $A_{k}(P,t)=A_{k}^{i}(P,t)+\delta _{k}^{A}(P,t).$ Thus, there is the
unitary transformation between the two three-state deviation vectors: 
\begin{equation}
(\delta _{0}^{A}(P,t),\delta _{1}^{A}(P,t),\delta
_{2}^{A}(P,t))^{T}=U_{Ab}(\delta _{0}^{b}(P,t),\delta _{+}^{b}(P,t),\delta
_{-}^{b}(P,t))^{T}.  \tag{125}
\end{equation}%
It is well known that the unitary transformation $U_{Ab}$\ does not change
the norm of the three-state deviation vector $(\delta _{0}^{b}(P,t),\delta
_{+}^{b}(P,t),\delta _{-}^{b}(P,t))^{T}.$ This indicates that there is the
relation: 
\begin{equation*}
|\delta _{0}^{A}(P,t)|^{2}+|\delta _{1}^{A}(P,t)|^{2}+|\delta
_{2}^{A}(P,t)|^{2}=|\delta _{0}^{b}(P,t)|^{2}+|\delta
_{+}^{b}(P,t)|^{2}+|\delta _{-}^{b}(P,t)|^{2}.
\end{equation*}%
This relation leads to that the error probability $||E_{r}(x,r,t)||^{2}$ of
(123) may be expressed as%
\begin{equation}
||E_{r}(x,r,t)||^{2}=\sum_{P}|\rho (P)|^{2}\{|\delta
_{0}^{b}(P,t)|^{2}+|\delta _{+}^{b}(P,t)|^{2}+|\delta _{-}^{b}(P,t)|^{2}\}. 
\tag{126}
\end{equation}%
It is convenient to calculate the error upper bound $||E_{r}(x,r,t)||$ by
using the equation (126), since the deviation vector $(\delta
_{0}^{b}(P,t),\delta _{+}^{b}(P,t),\delta _{-}^{b}(P,t))^{T}$ can be
obtained conveniently by solving the basic differential equations (26).
Thus, an accurate error upper bound $||E_{r}(x,r,t)||$ could be obtained
directly by computing the equation (126) by using the deviation vector $%
(\delta _{0}^{b}(P,t),\delta _{+}^{b}(P,t),\delta _{-}^{b}(P,t))^{T}$ for
the basic STIRAP decelerating or accelerating process. Obviously, the
three-state deviation vector $(\delta _{0}^{b}(P,t),\delta
_{+}^{b}(P,t),\delta _{-}^{b}(P,t))^{T}$ has the maximum norm or length over
the effective momentum distribution region $[P]$ and in the time period $%
t_{0}\leq t\leq t_{0}+T,$%
\begin{equation*}
\sqrt{|\delta _{0}^{b}(P,t)|^{2}+|\delta _{+}^{b}(P,t)|^{2}+|\delta
_{-}^{b}(P,t)|^{2}}
\end{equation*}%
\begin{equation*}
\leq \sqrt{(|\delta _{0}^{b}(P,t)|^{2}+|\delta _{+}^{b}(P,t)|^{2}+|\delta
_{-}^{b}(P,t)|^{2})_{\max }},\text{ for }P\in \lbrack P]\text{ and }%
t_{0}\leq t\leq t_{0}+T.
\end{equation*}%
Then the upper bound of the error term $E_{r}(x,r,t)$ may be determined from%
\begin{equation}
||E_{r}(x,r,t)||\leq \sqrt{(|\delta _{0}^{b}(P,t)|^{2}+|\delta
_{+}^{b}(P,t)|^{2}+|\delta _{-}^{b}(P,t)|^{2})_{\max }},  \tag{127}
\end{equation}%
where the normalization relation $\sum_{P}|\rho (P)|^{2}=1$ is used and the
truncation error is neglected for any momentum components outside the
effective momentum region $[P]$. The inequality (127) is a general adiabatic
condition for the basic STIRAP decelerating or accelerating process of a
free atom in a wave-packet motional state. There is also a simpler method to
obtain the error upper bound $||E_{r}(x,r,t)||,$ as stated below. It uses
the general adiabatic condition (116). It is known that the formal solution
to the basic differential equations (26) or their matrix form (28) may be
expressed as (112), where the solution in the ideal adiabatic condition (30)
is given by $B(P,t)=B(P,t_{0}),$ as shown in Eq. (40) in the previous
section 4. Then the equation (112) shows that the three-state deviation
vector is just $E_{r}(P,t)$ and hence one has $E_{r}(P,t)=(\delta
_{0}^{b}(P,t),\delta _{+}^{b}(P,t),\delta _{-}^{b}(P,t))^{T}.$ Furthermore,
the adiabatic condition (116) and the equation (126) show that there are the
relations: 
\begin{equation*}
||E_{r}(x,r,t)||=\{\sum_{P}|\rho (P)|^{2}||E_{r}(P,t)||^{2}\}^{1/2}
\end{equation*}%
\begin{equation*}
\leq \{\sum_{P}|\rho (P)|^{2}\exp [2(||M(P,t)||_{\max })T]\times
||B^{(1)}(P,t)||_{\max }^{2}\}^{1/2}
\end{equation*}%
\begin{equation}
\leq \exp [(||\hat{M}(P,t)||_{\max })T]\times ||\hat{B}^{(1)}(P,t)||_{\max },
\tag{128}
\end{equation}%
where the relation $\sum_{P}|\rho (P)|^{2}=1$ is used and the truncation
error has been neglected for any momentum components outside the effective
momentum region $[P]$, and the maximum norms $||\hat{M}(P,t)||_{\max }$ and $%
||\hat{B}^{(1)}(P,t)||_{\max }^{2}$ are respectively defined as%
\begin{equation*}
(||\hat{M}(P,t)||_{\max })=\max_{P\in \lbrack P]}(||M(P,t)||_{\max }),
\end{equation*}%
\begin{equation*}
||\hat{B}^{(1)}(P,t)||_{\max }=\max_{P\in \lbrack P]}||B^{(1)}(P,t)||_{\max
}.
\end{equation*}%
The last inequality in (128) is a real adiabatic condition of the basic
STIRAP decelerating or accelerating process for a free atom in a wave-packet
motional state. It could be useful to design the basic STIRAP decelerating
or accelerating process.

At first the adiabatic condition (128) requires one to calculate the norm $%
(||M(P,t)||_{\max })$ and the first-order approximation solution $%
B^{(1)}(P,t).$ It is easy to calculate the first-order approximation
solution to the basic differential equations (26). Actually, the first-order
approximation solution may be obtained from the equation (117). Here for
convenience setting the global phases $\gamma (t_{0})=\delta (t_{0})=0$ in
the basic equations (26) and the phase-modulation functions of the Raman
laser light beams\ to be $\varphi _{0}(t)=0$ in Eq. (31) and $\varphi
_{1}(t)=0$ in Eq. (32). It should be pointed out that the following methods
are available as well for the phase-modulation Raman laser light beams. It
follows from Eqs. (31) and (32) that 
\begin{equation*}
\frac{d}{dt}\alpha _{p}(P,t)=\frac{\Delta P}{M}k_{0},\text{ \ }\frac{d}{dt}%
\alpha _{s}(P,t)=-\frac{\Delta P}{M}k_{1}.
\end{equation*}%
By inserting these two equations into Eqs. (27a)--(27c) one obtains%
\begin{equation}
\Omega _{\pm }(P,t)=\Omega (t)\pm K_{0}(t)\Delta P,  \tag{129a}
\end{equation}%
\begin{equation}
\Theta (P,t)=-\dot{\theta}(t)+iK_{1}(t)\Delta P,\text{ }\Gamma (P,t)=\frac{%
k_{0}}{M}\Delta P-K_{2}(t)\Delta P,  \tag{129b}
\end{equation}%
where 
\begin{equation*}
K_{0}(t)=\frac{1}{4M}[(k_{0}-k_{1})+3(k_{0}+k_{1})\cos 2\theta (t)],
\end{equation*}%
\begin{equation*}
K_{1}(t)=\frac{k_{0}+k_{1}}{2M}\sin 2\theta (t),\text{ }K_{2}(t)=\frac{%
(k_{0}+k_{1})}{M}\cos ^{2}\theta (t).
\end{equation*}%
The initial condition for the basic equations (26) is given by (39). If the
initial mixing angle $\theta (t_{0})$ is very small $(\theta (t_{0})<<1)$
but not equal to zero, then the initial condition (39) may be changed to the
general form%
\begin{equation}
b_{0}(P,t_{0})=\exp [\frac{i}{\hslash }(\frac{(P+\hslash k_{0})^{2}}{2M}%
+E_{0})t_{0}]\cos \theta (t_{0}),  \tag{130a}
\end{equation}%
\begin{equation}
b_{+}(P,t_{0})=b_{-}(P,t_{0})=\frac{1}{\sqrt{2}}\exp [\frac{i}{\hslash }(%
\frac{(P+\hslash k_{0})^{2}}{2M}+E_{0})t_{0}]\sin \theta (t_{0}).  \tag{130b}
\end{equation}%
Then in the initial condition (130) the first-order approximation solution
to the basic equations (26) for the coefficient $b_{0}(P,t)$ may be written
as, by integrating by parts the integral (117), 
\begin{equation}
b_{0}^{(1)}(P,t)=b_{0}(P,t)-b_{0}(P,t_{0})=C_{0}^{(1)}(P,t)+C_{0T}^{(2)}(P,t),
\tag{131a}
\end{equation}%
where the main term $C_{0}^{(1)}(P,t)$ that is proportional to $\Theta
(P,t)^{\ast }/\Omega (t)$ is written as%
\begin{equation}
C_{0}^{(1)}(P,t)=\sqrt{2}b_{+}(P,t_{0})\frac{\Theta (P,t)^{\ast }}{\Omega (t)%
}\exp [i\Delta P\int_{t_{0}}^{t}dt^{\prime }K_{0}(t^{\prime })]\sin
[i\int_{t_{0}}^{t}dt^{\prime }\Omega (t^{\prime })],  \tag{131b}
\end{equation}%
and the secondary term $C_{0T}^{(2)}(P,t)$ is given by%
\begin{equation*}
C_{0T}^{(2)}(P,t)=i\sqrt{2}b_{+}(P,t_{0})\int_{t_{0}}^{t}dt_{1}\{[i\frac{%
\partial }{\partial t_{1}}(\frac{\Theta (P,t_{1})^{\ast }}{\Omega (t_{1})})-%
\frac{K_{0}(t_{1})\Theta (P,t_{1})^{\ast }}{\Omega (t_{1})}\Delta P]
\end{equation*}%
\begin{equation}
\times \exp [i\Delta P\int_{t_{0}}^{t_{1}}dt^{\prime }K_{0}(t^{\prime
})]\sin [i\int_{t_{0}}^{t_{1}}dt^{\prime }\Omega (t^{\prime })]\}. 
\tag{131c}
\end{equation}%
The first-order solution for the coefficients $b_{\pm }(P,t)$ is given by%
\begin{equation}
b_{\pm }^{(1)}(P,t)=b_{\pm }(P,t)-b_{\pm }(P,t_{0})=C_{\pm
}^{(1)}(P,t)+C_{\pm T}^{(2)}(P,t)+F_{\pm }^{(1)}(P,t)+F_{\pm T}^{(2)}(P,t) 
\tag{132a}
\end{equation}%
where the main terms $C_{\pm }^{(1)}(P,t)$ are given by 
\begin{equation*}
C_{\pm }^{(1)}(P,t)=\mp i\frac{1}{\sqrt{2}}b_{0}(P,t_{0})\frac{\Theta (P,t)}{%
\Omega (t)}\exp [-i\Delta P\int_{t_{0}}^{t}dt^{\prime }K_{0}(t^{\prime
})]\exp [\mp i\int_{t_{0}}^{t}dt^{\prime }\Omega (t^{\prime })]
\end{equation*}%
\begin{equation}
\pm i\frac{1}{\sqrt{2}}b_{0}(P,t_{0})\frac{\Theta (P,t_{0})}{\Omega (t_{0})},
\tag{132b}
\end{equation}%
and the secondary terms are%
\begin{equation*}
C_{\pm T}^{(2)}(P,t)=\pm \frac{1}{\sqrt{2}}b_{0}(P,t_{0})%
\int_{t_{0}}^{t}dt_{1}\{[i\frac{\partial }{\partial t_{1}}(\frac{\Theta
(P,t_{1})}{\Omega (t_{1})})+\frac{K_{0}(t_{1})\Theta (P,t_{1})}{\Omega
(t_{1})}\Delta P]
\end{equation*}%
\begin{equation}
\times \exp [-i\Delta P\int_{t_{0}}^{t_{1}}dt^{\prime }K_{0}(t^{\prime
})]\exp [\mp i\int_{t_{0}}^{t_{1}}dt^{\prime }\Omega (t^{\prime })]\}, 
\tag{132c}
\end{equation}%
\begin{equation}
F_{\pm }^{(1)}(P,t)=\pm \frac{1}{4}b_{\mp }(P,t_{0})\frac{\Gamma (P,t)}{%
\Omega (t)}\exp [\mp i\int_{t_{0}}^{t}dt^{\prime }2\Omega (t^{\prime })]\mp 
\frac{1}{4}b_{\mp }(P,t_{0})\frac{\Gamma (P,t_{0})}{\Omega (t_{0})}, 
\tag{132d}
\end{equation}%
\begin{equation}
F_{\pm T}^{(2)}(P,t)=\mp \frac{1}{4}b_{\mp
}(P,t_{0})\int_{t_{0}}^{t}dt_{1}\{[\frac{\partial }{\partial t_{1}}(\frac{%
\Gamma (P,t_{1})}{\Omega (t_{1})})]\exp [\mp i\int_{t_{0}}^{t_{1}}dt^{\prime
}2\Omega (t^{\prime })]\}.  \tag{132e}
\end{equation}%
The dominating terms in the first-order approximation solution of (131a) and
(132a) are $C_{\pm }^{(1)}(P,t),$ which are proportional to $\Theta
(P,t)/\Omega (t)$ and $b_{0}(P,t_{0}).$ It can turn out by integrating by
parts the integral (132c) that the terms $C_{\pm T}^{(2)}(P,t)$ are
proportional to $\Omega (t)^{-2}.$ Thus, the terms $C_{\pm T}^{(2)}(P,t)$
are secondary in the first-order solution for a large Rabi frequency $\Omega
(t).$ Similarly, it can turn out that $C_{0T}^{(2)}(P,t)\varpropto
b_{+}(P,t_{0})\Omega (t)^{-2}$ and $F_{\pm T}^{(2)}(P,t)\varpropto
b_{+}(P,t_{0})\Omega (t)^{-2},$ by integrating by parts the integrals (131c)
and (132e), respectively. Thus, these terms $C_{0T}^{(2)}(P,t)$ and $F_{\pm
T}^{(2)}(P,t)$ are secondary with respect to the terms $C_{0}^{(1)}(P,t)$
and $F_{\pm }^{(1)}(P,t),$ respectively. On the other hand, the imperfection
for the initial conditions $b_{+}(P,t_{0})=b_{-}(P,t_{0})\neq 0$ could
mainly affect $C_{0}^{(1)}(P,t)$ and $F_{\pm }^{(1)}(P,t).$ Its magnitude is
approximately proportional to the factors $|b_{\pm }(P,t_{0})||\Theta
(P,t)|/\Omega (t)$ or $|b_{\pm }(P,t_{0})||\Gamma (P,t)|/\Omega (t).$ Since $%
|b_{\pm }(P,t_{0})|<<|b_{0}(P,t_{0})|,$ these terms $C_{0}^{(1)}(P,t)$ and $%
F_{\pm }^{(1)}(P,t)$ are secondary with respect to the main terms $C_{\pm
}^{(1)}(P,t).$ Thus, the error upper bound $||\hat{B}^{(1)}(P,t)||_{\max }$
in (128) may be determined from the main terms $C_{\pm }^{(1)}(P,t).$ Now by
inserting $b_{0}^{(1)}(P,t)$ of (131a) and $b_{\pm }^{(1)}(P,t)$ of (132a)
into (118) it can be found that the norm $||B^{(1)}(P,t)||$ is bounded by 
\begin{equation*}
||B^{(1)}(P,t)||\leq \frac{|\Theta (P,t)|}{\Omega (t)}+\frac{|\Theta
(P,t_{0})|}{\Omega (t_{0})}
\end{equation*}%
\begin{equation}
=\frac{\sqrt{\dot{\theta}(t)^{2}+K_{1}(t)^{2}|\Delta P|^{2}}}{\Omega (t)}+%
\frac{\sqrt{\dot{\theta}(t_{0})^{2}+K_{1}(t_{0})^{2}|\Delta P|^{2}}}{\Omega
(t_{0})},  \tag{133}
\end{equation}%
where those secondary terms of the first-order approximation solution are
neglected and only the main terms $C_{\pm }^{(1)}(P,t)$ of (132b) are used
and $|b_{0}(P,t_{0})|\leq 1$ is also used. It is known that the initial
mixing angle $\theta (t_{0})\rightarrow 0,$ as can be seen in (35). Note
that $K_{1}(t)=(k_{0}+k_{1})\sin 2\theta (t)/(2M)$ and the time derivative $%
\dot{\theta}(t)$ of the mixing angle is given by%
\begin{equation*}
\dot{\theta}(t)=\frac{\dot{\Omega}_{p}(t)\cos \theta (t)-\dot{\Omega}%
_{s}(t)\sin \theta (t)}{\Omega (t)}.
\end{equation*}%
If the Rabi frequencies $\Omega _{p}(t)$ and $\Omega _{s}(t)$ are chosen
suitably in experiment such that at the initial and final times the mixing
angle satisfies the relations:%
\begin{equation*}
\tan \theta (t_{0})=\Omega _{p}(t_{0})/\Omega _{s}(t_{0})\thickapprox \dot{%
\Omega}_{p}(t_{0})/\dot{\Omega}_{s}(t_{0})\rightarrow 0,
\end{equation*}%
\begin{equation*}
\dot{\Omega}_{p}(t_{0}+T)\cos \theta (t_{0}+T)-\dot{\Omega}_{s}(t_{0}+T)\sin
\theta (t_{0}+T)\rightarrow 0,
\end{equation*}%
then the time derivative $\dot{\theta}(t_{0})\thickapprox 0$ and $\dot{\theta%
}(t_{0}+T)\thickapprox 0.$ On the other hand, the momentum distribution
satisfies $|\Delta P|\leq \Delta P_{M}/2$ for a momentum wave-packet state
with an effective momentum bandwidth $\Delta P_{M}$. Therefore, at the
initial time $|K_{1}(t_{0})\Delta P|\leq |K_{1}(t_{0})|\Delta P_{M}/2=\Delta
P_{M}(k_{0}+k_{1})|\sin 2\theta (t_{0})|/(4M)\thickapprox 0$ if the momentum
wave-packet state has a finite wave-packet spread. Then in these conditions
the error upper bound $||\hat{B}^{(1)}(P,t)||_{\max }$ may be determined
from, by neglecting the second term on the rightest side of (133), 
\begin{equation}
||B^{(1)}(P,t)||\leq \{\frac{\sqrt{\dot{\theta}(t)^{2}+(\Delta
P_{M})^{2}(k_{0}+k_{1})^{2}\sin ^{2}2\theta (t)/(16M^{2})}}{\Omega (t)}%
\}_{\max }.  \tag{134}
\end{equation}%
Here the subscript $^{\prime }\max^{\prime }$ means that the function on the
right-hand side of (134) is taken as the maximum value in the time period $%
t_{0}\leq t\leq t_{0}+T$ of the STIRAP process. The first-order
approximation adiabatic condition is that the maximum value of the function
on the right-hand side of (134) is controlled to be smaller than some
desired small value. (Notice that this first-order adiabatic condition (134)
is slightly different from that one (128a) in the previous versions of this
paper). As shown below, in the initial and final time periods of the STIRAP
process the adiabatic condition (134) still may be met even if the Rabi
frequency $\Omega (t)$ is small in these time periods. This is a global
adiabatic condition, since it is involved in the whole time period of the
STIRAP process. Here the global adiabatic condition has a different
definition from the conventional one in Ref. [4]. In the conventional STIRAP
experiments [4, 15, 17, 18] the (local) adiabatic condition is defined as
that at any instant of time of the STIRAP process the population or
probability in the two eigenstates $|g^{\pm }(P,t)\rangle $ that contain the
excited internal state is much smaller unity. This is approximately
equivalent to the first inequality of (133) for the first-order
approximation, where the inequality symbol $^{\prime }\leq ^{\prime }$ is
replaced with $^{\prime }<<^{\prime }.$ On the other hand, according to
(119) the maximum matrix norm $(||\hat{M}(P,t)||_{\max })$ is determined
from 
\begin{equation*}
||M(P,t)||\leq \frac{1}{\sqrt{2}}\{2|\Theta (P,t)|+|\Gamma (P,t)|\}
\end{equation*}%
\begin{equation}
\leq \frac{1}{\sqrt{2}}\{2\sqrt{|\dot{\theta}(t)|_{\max
}^{2}+(k_{0}+k_{1})^{2}(\Delta P_{M})^{2}/(16M^{2})}+\frac{(\Delta P_{M})}{2M%
}\max (k_{0},k_{1})\}  \tag{135}
\end{equation}%
where $|\dot{\theta}(t)|_{\max }$ is the maximum value of $|\dot{\theta}(t)|$
in the whole STIRAP decelerating or accelerating process. After the upper
bound $||\hat{B}^{(1)}(P,t)||_{\max }$ and maximum norm $(||\hat{M}%
(P,t)||_{\max })$ are determined from (134) and (135), respectively, one may
determine the upper bound of the error term $E_{r}(x,r,t)$ from (128). It
can be seen from (134), (135), and (128) that the adiabatic condition (128)
may be better satisfied for a small time derivative $\dot{\theta}(t),$ a
short time period $T$, a large Rabi frequency $\Omega (t)$, and a narrow
momentum wave-packet state ($\Delta P_{M}$ is small). If the time derivative 
$\dot{\theta}(t)$ is smaller, then the time period $T$ usually is larger.
Thus, there is a compromise between the settings of the time derivative $%
\dot{\theta}(t)$ and the time period $T$ in experiment. Obviously, one has
the relation for any basic STIRAP\ decelerating or accelerating process: 
\begin{equation*}
\theta (t_{0}+T)-\theta (t_{0})=\int_{t_{0}}^{t_{0}+T}dt^{\prime }\dot{\theta%
}(t^{\prime })=\pi /2.
\end{equation*}%
This relation may be used to determine the time period $T$ if one knows the
time derivative $\dot{\theta}(t)$ of the mixing angle.

It seems that the adiabatic conditions (116) and (128) could not be better
satisfied in the initial and final time periods, since the Rabi frequency $%
\Omega (t)$ takes a smaller value in these time periods. It is well known
that any STIRAP experiment requires that at the initial and final time the
mixing angle $\theta (t)$ satisfy the constraint conditions: 
\begin{equation*}
\theta (t_{0})\rightarrow 0,\text{ }\theta (t_{0}+T)\rightarrow \pi /2.
\end{equation*}%
The two constraint conditions are compatible with the adiabatic conditions
(116) and (128). This can be seen from (134). At the initial time period the
Rabi frequency $\Omega (t)$ takes generally a smaller value, but at the same
time the mixing angle $\theta (t_{0})\rightarrow 0$ and its time derivative $%
\dot{\theta}(t_{0})$ may be set to a value close to zero, leading to that
the value $||B^{(1)}(P,t)||$ may be kept at a smaller value at the initial
time period. At the final time the mixing angle $\theta (t_{0}+T)\rightarrow
\pi /2$ or $\sin ^{2}2\theta (t_{0}+T)\rightarrow 0$ and the time derivative 
$\dot{\theta}(t_{0}+T)$ also may be set to a value close to zero. Then the
value $||B^{(1)}(P,t)||$ still may be kept at a smaller value, although the
Rabi frequency $\Omega (t)$ takes a smaller value at the final time period.
These results show that in theory the adiabatic conditions (116) and (128)
still may be met in the initial and final time periods as long as the Raman
laser light beams of the STIRAP experiment are suitably designed. The
adiabatic conditions (116) and (128) could be better used for a conventional
three-state STIRAP\ experiment that uses a pair of copropagating Raman laser
light beams and those STIRAP-based decelerating and accelerating processes
of the atomic or molecular systems with a narrow momentum distribution.

The deviation $E_{r}(P,t)$ of (113) and its upper bound (116) are obtained
for a single basic STIRAP decelerating sequence (11). If a unitary
decelerating process consists of $n_{d}$ pairs of the basic STIRAP
decelerating sequences (11) and (63), then the total deviation generated in
the unitary decelerating process is bounded by 
\begin{equation*}
||E_{r}(P,t)||\leq n_{d}\exp [(||\hat{M}_{k}(P,t)||_{\max })T]\times ||\hat{B%
}_{k}^{(1)}(P,t)||_{\max }
\end{equation*}%
\begin{equation}
+n_{d}\exp [(||\hat{M}_{l}(P,t)||_{\max })T]\times ||\hat{B}%
_{l}^{(1)}(P,t)||_{\max },  \tag{136}
\end{equation}%
where the subscript $k$ marks the basic STIRAP decelerating sequence (11)
that uses a pair of the Raman laser light beams with the Rabi frequency $%
\Omega (t),$ the mixing angle $\theta (t)$, the wave numbers $k_{1}$ and $%
k_{2}$, and the carrier frequencies $\omega _{01}$ and $\omega _{02},$ while
the subscript $l$ denotes the basic STIRAP\ decelerating sequence (63) that
may use another pair of the Raman laser light beams with the Rabi frequency $%
\Omega _{l}(t),$ the mixing angle $\theta _{l}(t),$ the wave numbers $%
k_{1}^{l}$ and $k_{2}^{l}$, and the carrier frequencies $\omega _{01}^{l}$
and $\omega _{02}^{l}.$ Both the basic STIRAP decelerating processes have
the same time period $T$. The upper bound of the total deviation $E_{r}(P,t)$
on the right-hand side of the inequality (136) for the unitary decelerating
or accelerating process may be controlled by setting suitably the
experimental parameters of the Raman laser light beams, which include the
Rabi frequencies $\Omega (t)$ and $\Omega _{l}(t),$ the mixing angles $%
\theta (t)$ and $\theta _{l}(t),$ and the time derivatives $\dot{\theta}(t)$
and $\dot{\theta}_{l}(t)$, and so on.

The basic STIRAP decelerating or accelerating process is a time-dependent
unitary quantum dynamical problem from the viewpoint of quantum mechanics.
The three-state basic STIRAP decelerating or accelerating process is a quite
simple unitary dynamical process, but it can describe completely the complex
STIRAP-based unitary decelerating and accelerating processes of a free atom.
It is relatively simple to solve approximately such a unitary dynamical
problem as the three-state basic STIRAP decelerating or accelerating process
in quantum mechanics, although this problem is time-dependent and it is
difficult to solve exactly the basic differential equations (26) except for
some special cases [39]. For example, one may use the conventional methods
of successive approximations [33] to solve these basic differential
equations approximately when the Rabi frequencies $\Omega (t)$ is large. As
mentioned before, the Dyson series solution (29) to the basic differential
equations (26) or (28) may be used to calculate the deviation of a real
adiabatic condition from the ideal adiabatic condition. The problem to be
answered is that one needs to calculate how many leading terms in the Dyson
series (29) so that the result obtained is enough accurate. The leading term
number is dependent upon the desired error value and the maximum norm of the
matrix $M(P,t)$ in (28). If one uses the first $n$ terms on the right-hand
side of (113) to calculate the error term $E_{r}(P,t),$ then the residual
term may be given by%
\begin{equation*}
R(P,t)=\{(\frac{1}{i})^{n+1}\int_{t_{0}}^{t}\int_{t_{0}}^{t_{1}}...%
\int_{t_{0}}^{t_{n}}dt_{1}dt_{2}...dt_{n+1}M(P,t_{1})M(P,t_{2})...M(P,t_{n+1})
\end{equation*}%
\begin{equation*}
+(\frac{1}{i})^{n+2}\int_{t_{0}}^{t}\int_{t_{0}}^{t_{1}}...%
\int_{t_{0}}^{t_{n+1}}dt_{1}dt_{2}...dt_{n+2}M(P,t_{1})M(P,t_{2})...M(P,t_{n+2})
\end{equation*}%
\begin{equation}
+......\}B(P,t_{0}).  \tag{137}
\end{equation}%
Then the residual term $R(P,t)$ is bounded by%
\begin{equation*}
||R(P,t)||\leq \sum_{k=n+1}^{\infty }\frac{(t-t_{0})^{k}}{k!}(||\hat{M}%
(P,t)||_{\max })^{k}
\end{equation*}%
\begin{equation}
\leq \frac{(||\hat{M}(P,t)||_{\max }(t-t_{0}))^{n+1}}{(n+1)!}\exp [(||\hat{M}%
(P,t)||_{\max })(t-t_{0})]  \tag{138}
\end{equation}%
where $||B(P,t_{0})||=1$ is used. Now $(n+1)!\thickapprox \sqrt{2\pi (n+1)}%
[(n+1)/e]^{n+1}.$ Then the upper bound of the residual term is determined
from%
\begin{equation}
||R(P,t)||\leq \frac{\{\frac{||\hat{M}(P,t)||_{\max }(t-t_{0})}{(n+1)/e}\exp
[\frac{(||\hat{M}(P,t)||_{\max })(t-t_{0})}{n+1}]\}^{n+1}}{\sqrt{2\pi (n+1)}}%
.  \tag{139}
\end{equation}%
If $(||\hat{M}(P,t)||_{\max })(t-t_{0})<(n+1),$ then $\exp [\frac{(||\hat{M}%
(P,t)||_{\max })(t-t_{0})}{n+1}]<e.$ Thus, if $(n+1)/e>||\hat{M}%
(P,t)||_{\max }(t-t_{0})\exp [\frac{(||\hat{M}(P,t)||_{\max })(t-t_{0})}{n+1}%
],$ then the residual term $R(P,t)$ is exponentially small. Suppose that the
upper bound of the error term $E_{r}(P,t)$ is set to a given value $%
\varepsilon _{r}:||E_{r}(P,t)||\leq \varepsilon _{r}.$ Then this requires
the residual term to satisfy $||R(P,t)||<<\varepsilon _{r}.$ The condition $%
||R(P,t)||<<\varepsilon _{r}$ can be easily met by setting a minimum integer 
$n$ such that 
\begin{equation}
\frac{1}{\sqrt{2\pi (n+1)}}\{\frac{||\hat{M}(P,t)||_{\max }T}{(n+1)/e}\exp [%
\frac{(||\hat{M}(P,t)||_{\max })T}{n+1}]\}^{n+1}<<\varepsilon _{r}. 
\tag{140}
\end{equation}%
Once the minimum integer $n$ is determined, one may use the first $n$ terms
on the right-hand side of (113) to calculate the error term $E_{r}(P,t)$ and
its upper bound, while the residual term $R(P,t)$ does not affect
significantly the final result.

It can be seen from (133) that the first-order approximation solution shows
that the error upper bound $||B^{(1)}(P,t)||$ is mainly dependent upon the
parameter $|\Theta (P,t)|/\Omega (t)$ and almost independent of the
parameters $\Gamma (P,t)$ and $K_{0}(t).$ Actually, in the first-order
approximation solution the parameter $\Gamma (P,t)$ appears in the secondary
terms $F_{\pm }^{(1)}(P,t)$ and $F_{\pm T}^{(2)}(P,t)$ and $K_{0}(t_{1})$ in
the secondary terms $C_{\pm T}^{(2)}(P,t)$ (these parameters may appear in
the phase factors, and if so, they do not make a contribution to the error
upper bound). It is known that the parameter$\ \Theta (P,t)=-\dot{\theta}%
(t)+i\frac{1}{2}(\Delta P/M)(k_{0}+k_{1})\sin 2\theta (t).$ If the two Raman
laser light beams are copropagating, then the wave-number sum $(k_{0}+k_{1})$
will be changed to the wave-number difference $(k_{0}-k_{1})$ in the
parameter $\Theta (P,t).$ Then the effect of the momentum distribution on
the error upper bound $||B^{(1)}(P,t)||$ will be greatly weakened and the
momentum distribution will become a higher-order effect on the STIRAP state
transfer. Therefore, in this sense the copropagating Raman laser light beams
used to construct the STIRAP pulse sequence may be better than the
counterpropagating ones to realize the perfect STIRAP state transfer. Unlike
the parameter $\Theta (P,t)$ these two parameters $\Gamma (P,t)$ and $%
K_{0}(t)$ are dependent on both the wave-number sum and difference. That the
momentum distribution affects the STIRAP state transfer is mainly through
the parameters $\Gamma (P,t)$ and $K_{0}(t),$ which appear in the
higher-order terms in the Dyson series (29), if the copropagating Raman
laser light beams are used in the STIRAP state transfer. In fact, the
maximum norm $(||\hat{M}(P,t)||_{\max })$ determined from (135) shows that
it is proportional to $|\Gamma (P,t)|.$ These parameters make an important
effect for the momentum distribution on the STIRAP state transfer. Thus,
that the conventional three-state STIRAP experiments [4, 15] use the
copropagating Raman laser light beams is favorable for the perfect state
transfer and may minimize the effect of the momentum distribution of the
atomic and molecular systems under investigation on the perfect state
transfer. However, in the basic STIRAP decelerating or accelerating process
the counterpropagating Raman laser light beams are generally used so that
the fast moving atom can be decelerated or accelerated more efficiently. On
the other hand, the first-order solution (117) could not exactly account for
the momentum distribution in any case that either copropagating or
counterpropagating laser light beams is used in the STIRAP experiments. The
adiabatic conditions (116) and (128) are accurate, but due to that there is
an exponential correction factor in (116) and (128) they could not be met
for a broad momentum distribution. A broad momentum distribution is often
met in an atomic or molecular quantum ensemble. It is necessary to consider
the effect of the momentum distribution when these physical ensembles are
decelerated (or accelerated) by the STIRAP decelerating (or accelerating)
pulse sequence. Thus, it is necessary to find a more useful adiabatic
condition that can account for the effect of a broad momentum distribution
on the STIRAP state transfer.

It is still complex to use the Dyson series solution (29) to calculate the
error term $E_{r}(P,t)$ of (113) and its upper bound. In the following an
equivalent transformation method based on the integration by parts to solve
the basic differential equations (26) is proposed so that the error term $%
E_{r}(P,t)$ of (113) and its upper bound can be obtained conveniently in a
high accuracy. On the other hand, by solving the basic differential
equations (26) to obtain an enough accurate solution one may further use the
solution to calculate conveniently the time evolution process for the basic
STIRAP decelerating and accelerating processes for a free atom. Generally,
it is quite inconvenient to calculate the time evolution process of an
atomic decelerating or accelerating process by directly solving the Schr\"{o}%
dinger equation. The present scheme is convenient to calculate the time
evolution process of the basic STIRAP\ decelerating or accelerating process
because it does not solve directly the original Schr\"{o}dinger equation but
solves the three first-order differential equations (26) that are equivalent
to and much simpler than the original Schr\"{o}dinger equation. The
equivalent transformation method to solve the basic equations (26) is based
on the fact that the Rabi frequency $\Omega (t)$ may be set to a large value
in experiment. Though the solution to the basic equations (26) obtained by
this method is approximate, the truncation error of the solution can be
controlled as expected. The procedure to solve the basic equations (26) with
the equivalent transformation method may be described below. By integrating
the basic differential equations (26) one obtains the equivalent integral
equations: 
\begin{equation*}
b_{0}(P,t)-b_{0}(P,t_{0})=\frac{1}{\sqrt{2}}\int_{t_{0}}^{t}dt_{1}%
\{b_{+}(P,t_{1})\Theta (P,t_{1})^{\ast }
\end{equation*}%
\begin{equation*}
\times \exp [i\Delta P\int_{t_{0}}^{t_{1}}dt^{\prime }K_{0}(t^{\prime
})]\exp [i\int_{t_{0}}^{t_{1}}dt^{\prime }\Omega (t^{\prime })]\}
\end{equation*}%
\begin{equation}
+\frac{1}{\sqrt{2}}\int_{t_{0}}^{t}dt_{1}\{b_{-}(P,t_{1})\Theta
(P,t_{1})^{\ast }\exp [i\Delta P\int_{t_{0}}^{t_{1}}dt^{\prime
}K_{0}(t^{\prime })]\exp [-i\int_{t_{0}}^{t_{1}}dt^{\prime }\Omega
(t^{\prime })]\},  \tag{141a}
\end{equation}%
\begin{equation*}
b_{\pm }(P,t)-b_{\pm }(P,t_{0})=-i\frac{1}{2}\int_{t_{0}}^{t}dt_{1}\{b_{\mp
}(P,t_{1})\Gamma (P,t_{1})\exp [\mp i\int_{t_{0}}^{t_{1}}dt^{\prime }2\Omega
(t^{\prime })]\}
\end{equation*}%
\begin{equation}
-\frac{1}{\sqrt{2}}\int_{t_{0}}^{t}dt_{1}\{b_{0}(P,t_{1})\Theta
(P,t_{1})\exp [-i\Delta P\int_{t_{0}}^{t_{1}}dt^{\prime }K_{0}(t^{\prime
})]\exp [\mp i\int_{t_{0}}^{t_{1}}dt^{\prime }\Omega (t^{\prime })]\}. 
\tag{141b}
\end{equation}%
Hereafter the initial condition (39) is used for convenience. The equivalent
transformation method is that by the integration by parts and another
transformation (see below) the integral equations (141) may be approximately
reduced to the three linear algebra equations. These three linear algebra
equations are equivalent to the original integral equations (141) if the
initial condition (39) is taken into account and when the truncation error
can be neglected. At the first step of the equivalent transformation method
the integrals on the right-hand sides of (141) are calculated by the
integration by parts. Then the initial condition (39) is used to simplify
the calculated results \ If there are the time derivatives of the variables $%
b_{0}(P,t)$ and $b_{\pm }(P,t)$ in the integrands after the integration by
parts, then one may use the basic differential equations (26) to replace
these time derivatives. As an example, by integrating by parts the equation
(141b) for the variable $b_{+}(P,t)$ and then using the basic differential
equations (26) and the initial condition (39), one can obtain the following
equation: 
\begin{equation*}
b_{+}^{1}(P,t)\equiv b_{+}(P,t)=b_{-}(P,t)\Gamma _{+1}^{-}(P,t)\exp
[-i\int_{t_{0}}^{t}dt^{\prime }2\Omega (t^{\prime })]
\end{equation*}%
\begin{equation*}
+ib_{0}(P,t)\Gamma _{+1}^{0}(P,t)\exp [-i\Delta P\int_{t_{0}}^{t}dt^{\prime
}K_{0}(t^{\prime })]\exp [-i\int_{t_{0}}^{t}dt^{\prime }\Omega (t^{\prime })]
\end{equation*}%
\begin{equation*}
-ib_{0}(P,t_{0})\Gamma
_{+1}^{0}(P,t_{0})+i\int_{t_{0}}^{t}dt_{1}\{b_{+}(P,t_{1})\Gamma
_{1}^{+}(P,t_{1})\}
\end{equation*}%
\begin{equation*}
+i\int_{t_{0}}^{t}dt_{1}\{b_{-}(P,t_{1})\Theta _{+1}^{-}(P,t_{1})\exp
[-i\int_{t_{0}}^{t_{1}}dt^{\prime }2\Omega (t^{\prime })]\}
\end{equation*}%
\begin{equation}
+\int_{t_{0}}^{t}dt_{1}\{b_{0}(P,t_{1})\Theta _{+1}^{0}(P,t_{1})\exp
[-i\Delta P\int_{t_{0}}^{t_{1}}dt^{\prime }K_{0}(t^{\prime })]\exp
[-i\int_{t_{0}}^{t_{1}}dt^{\prime }\Omega (t^{\prime })]\}  \tag{142}
\end{equation}%
where the five amplitudes are given by 
\begin{equation*}
\Gamma _{+1}^{-}(P,t)=\frac{1}{4}\frac{\Gamma (P,t)}{\Omega (t)},\text{ }%
\Gamma _{+1}^{0}(P,t)=-\frac{1}{\sqrt{2}}\frac{\Theta (P,t)}{\Omega (t)},%
\text{ }
\end{equation*}%
\begin{equation*}
\Gamma _{1}^{+}(P,t)=\frac{1}{8}\frac{\Gamma (P,t)^{2}+4|\Theta (P,t)|^{2}}{%
\Omega (t)},
\end{equation*}%
\begin{equation*}
\Theta _{+1}^{-}(P,t)=\frac{1}{2}\{i\frac{1}{2}\frac{\partial }{\partial t}(%
\frac{\Gamma (P,t)}{\Omega (t)})+\frac{|\Theta (P,t)|^{2}}{\Omega (t)}\},
\end{equation*}%
\begin{equation*}
\Theta _{+1}^{0}(P,t)=\frac{1}{\sqrt{2}}\{i\frac{\partial }{\partial t}(%
\frac{\Theta (P,t)}{\Omega (t)})+\frac{1}{4}\frac{\Gamma (P,t)\Theta (P,t)}{%
\Omega (t)}+\frac{K_{0}(t)\Theta (P,t)}{\Omega (t)}\Delta P\}.
\end{equation*}%
The equation (142) is almost completely equivalent to the original equation
(141b) for the variable $b_{+}(P,t).$ The unique difference between the two
equations is that the equation (142) uses the initial condition (39), while
the original equation does not. The first four terms on the right-hand side
of (142) may be considered as the main terms, since these terms have a
greater contribution to the solution $b_{+}^{1}(P,t)$ of (142). On the other
hand, each one of the last two integrals on the right-hand side of (142)
contains the integrands with the largely oscillatory phase factor $\exp
[-i\int_{t_{0}}^{t_{1}}dt^{\prime }\Omega (t^{\prime })]$ or $\exp
[-i\int_{t_{0}}^{t_{1}}dt^{\prime }2\Omega (t^{\prime })].$ These largely
oscillatory phase factors make the two integrals secondary in the solution
(142). This can be seen by integrating by parts the two integrals once
again. If these two integrals are neglected, then one obtains the
first-order approximation solution to the original equation (141b) for the
variable $b_{+}(P,t),$ since all the five amplitudes including $\Gamma
_{+1}^{-}(P,t),$ $\Gamma _{+1}^{0}(P,t),$ etc., appearing in the equation
(142) are inversely proportional to the Rabi frequency $\Omega (t)$.

One may further obtain a better approximation solution than the first-order
one. This can be done by integrating by parts the last two integrals on the
right-hand side of (142) again. However, the fourth term (or the first
integral) contains the solution $b_{+}(P,t)$ itself on the right-hand side
of (142). While one may substitute the first-order approximation solution of 
$b_{+}^{1}(P,t)$ into the integral to obtain a better approximation, the
calculation process becomes so complex that one can only obtain a
lower-order approximation solution. In order to avoid this complex one may
make a transformation on the solution $b_{+}(P,t)$ to cancel the integral
before integrating by parts the last two integrals. This transformation is
given by%
\begin{equation}
\hat{b}_{+}^{1}(P,t)=b_{+}^{1}(P,t)\exp [-i\int_{t_{0}}^{t}dt_{1}\Gamma
_{1}^{+}(P,t_{1})].  \tag{143}
\end{equation}%
This transformation is the key point to the present equivalent
transformation method to solve the basic differential equations (26). By
this transformation and the initial condition (39) the transformed solution $%
\hat{b}_{+}^{1}(P,t)$ may be written as%
\begin{equation*}
\hat{b}_{+}^{1}(P,t)=b_{-}(P,t)\hat{\Gamma}_{+1}^{-}(P,t)\exp
[-i\int_{t_{0}}^{t}dt^{\prime }2\Omega (t^{\prime })]
\end{equation*}%
\begin{equation*}
+ib_{0}(P,t)\hat{\Gamma}_{+1}^{0}(P,t)\exp [-i\Delta
P\int_{t_{0}}^{t}dt^{\prime }K_{0}(t^{\prime })]\exp
[-i\int_{t_{0}}^{t}dt^{\prime }\Omega (t^{\prime })]
\end{equation*}%
\begin{equation*}
-ib_{0}(P,t_{0})\hat{\Gamma}_{+1}^{0}(P,t_{0})+i\int_{t_{0}}^{t}dt_{1}%
\{b_{-}(P,t_{1})\hat{\Theta}_{+1}^{-}(P,t_{1})\exp
[-i\int_{t_{0}}^{t_{1}}dt^{\prime }2\Omega (t^{\prime })]\}
\end{equation*}%
\begin{equation}
+\int_{t_{0}}^{t}dt_{1}\{b_{0}(P,t_{1})\hat{\Theta}_{+1}^{0}(P,t_{1})\exp
[-i\Delta P\int_{t_{0}}^{t_{1}}dt^{\prime }K_{0}(t^{\prime })]\exp
[-i\int_{t_{0}}^{t_{1}}dt^{\prime }\Omega (t^{\prime })]\}.  \tag{144}
\end{equation}%
Now there is not any integral containing the solution $b_{+}^{1}(P,t)$
itself on the right-hand side of (144). The amplitudes of the solution $\hat{%
b}_{+}^{1}(P,t)$ are related to those amplitudes of the solution $%
b_{+}^{1}(P,t)$ of (142) by the recursive relations: 
\begin{equation}
\hat{\Gamma}_{+1}^{-}(P,t)=\Gamma _{+1}^{-}(P,t)\exp
[-i\int_{t_{0}}^{t}dt_{1}\Gamma _{1}^{+}(P,t_{1})],  \tag{145a}
\end{equation}%
\begin{equation}
\hat{\Gamma}_{+1}^{0}(P,t)=\Gamma _{+1}^{0}(P,t)\exp
[-i\int_{t_{0}}^{t}dt_{1}\Gamma _{1}^{+}(P,t_{1})],  \tag{145b}
\end{equation}%
\begin{equation}
\hat{\Theta}_{+1}^{-}(P,t)=[\Theta _{+1}^{-}(P,t)+\Gamma
_{+1}^{-}(P,t)\Gamma _{1}^{+}(P,t)]\exp [-i\int_{t_{0}}^{t}dt_{1}\Gamma
_{1}^{+}(P,t_{1})],  \tag{145c}
\end{equation}%
\begin{equation}
\hat{\Theta}_{+1}^{0}(P,t)=[\Theta _{+1}^{0}(P,t)-\Gamma
_{+1}^{0}(P,t)\Gamma _{1}^{+}(P,t)]\exp [-i\int_{t_{0}}^{t}dt_{1}\Gamma
_{1}^{+}(P,t_{1})].  \tag{145d}
\end{equation}%
The transformation (143) and the integration by parts may be called the
equivalent transformations as they does not generate any error term in these
transformation processes. The transformation (143) does not improve
essentially the first-order approximation solution, but it does simplify
greatly the calculation process to further obtain a higher-order
approximation solution. Now by integrating by parts the last two integrals
on the right-hand side of (144) and using the basic equations (26) and the
initial condition (39) the solution $\hat{b}_{+}^{1}(P,t)$ may be written as 
\begin{equation*}
b_{+}^{2}(P,t)\equiv \hat{b}_{+}^{1}(P,t)=b_{-}(P,t)\Gamma
_{+2}^{-}(P,t)\exp [-i\int_{t_{0}}^{t}dt^{\prime }2\Omega (t^{\prime })]
\end{equation*}%
\begin{equation*}
+ib_{0}(P,t)\Gamma _{+2}^{0}(P,t)\exp [-i\Delta P\int_{t_{0}}^{t}dt^{\prime
}K_{0}(t^{\prime })]\exp [-i\int_{t_{0}}^{t}dt^{\prime }\Omega (t^{\prime })]
\end{equation*}%
\begin{equation*}
-ib_{0}(P,t_{0})\Gamma
_{+2}^{0}(P,t_{0})+i\int_{t_{0}}^{t}dt_{1}\{b_{+}^{2}(P,t_{1})\Gamma
_{2}^{+}(P,t_{1})\}
\end{equation*}%
\begin{equation*}
+i\int_{t_{0}}^{t}dt_{1}\{b_{-}(P,t_{1})\Theta _{+2}^{-}(P,t_{1})\exp
[-i\int_{t_{0}}^{t_{1}}dt^{\prime }2\Omega (t^{\prime })]\}
\end{equation*}%
\begin{equation}
+\int_{t_{0}}^{t}dt_{1}\{b_{0}(P,t_{1})\Theta _{+2}^{0}(P,t_{1})\exp
[-i\Delta P\int_{t_{0}}^{t_{1}}dt^{\prime }K_{0}(t^{\prime })]\exp
[-i\int_{t_{0}}^{t_{1}}dt^{\prime }\Omega (t^{\prime })]\}  \tag{146}
\end{equation}%
where the amplitudes satisfy the recursive relations:%
\begin{equation}
\Gamma _{+2}^{-}(P,t)=\hat{\Gamma}_{+1}^{-}(P,t)-\frac{1}{2}\frac{\hat{\Theta%
}_{+1}^{-}(P,t)}{\Omega (t)},\text{ }\Gamma _{+2}^{0}(P,t)=\hat{\Gamma}%
_{+1}^{0}(P,t)+\frac{\hat{\Theta}_{+1}^{0}(P,t)}{\Omega (t)},  \tag{147a}
\end{equation}%
\begin{equation}
\Gamma _{2}^{+}(P,t)=-[\frac{1}{4}\frac{\Gamma (P,t)\hat{\Theta}%
_{+1}^{-}(P,t)}{\Omega (t)}+\frac{1}{\sqrt{2}}\frac{\Theta (P,t)^{\ast }\hat{%
\Theta}_{+1}^{0}(P,t)}{\Omega (t)}]\exp [i\int_{t_{0}}^{t}dt_{1}\Gamma
_{1}^{+}(P,t_{1})],  \tag{147b}
\end{equation}%
\begin{equation}
\Theta _{+2}^{-}(P,t)=-i\frac{1}{2}\frac{\partial }{\partial t}(\frac{\hat{%
\Theta}_{+1}^{-}(P,t)}{\Omega (t)})-\frac{1}{\sqrt{2}}\frac{\Theta
(P,t)^{\ast }\hat{\Theta}_{+1}^{0}(P,t)}{\Omega (t)},  \tag{147c}
\end{equation}%
\begin{equation}
\Theta _{+2}^{0}(P,t)=-\frac{1}{2\sqrt{2}}\frac{\Theta (P,t)\hat{\Theta}%
_{+1}^{-}(P,t)}{\Omega (t)}-i\frac{\partial }{\partial t}(\frac{\hat{\Theta}%
_{+1}^{0}(P,t)}{\Omega (t)})-\frac{K_{0}(t)\hat{\Theta}_{+1}^{0}(P,t)}{%
\Omega (t)}\Delta P.  \tag{147d}
\end{equation}%
The unique difference between the equation (146) and the original equation
(141b) for the variable $b_{+}(P,t)$ is that the initial condition (39) has
been used in (146), while it is not used in (141b). Now the amplitudes $%
\Gamma _{2}^{+}(P,t),$ $\Theta _{+2}^{-}(P,t),$ and $\Theta _{+2}^{0}(P,t)$
of the last three integrals on the right-hand side of (146) are inversely
proportional to $\Omega (t)^{2}.$ Thus, these three integrals are secondary
with respect to the first three terms on the right-hand side of (146).
Moreover, the last two integrals contain the integrands with the largely
oscillatory phase factor $\exp [-i\int_{t_{0}}^{t_{1}}dt^{\prime }\Omega
(t^{\prime })]$ or $\exp [-i\int_{t_{0}}^{t_{1}}dt^{\prime }2\Omega
(t^{\prime })].$ Then the integration by parts shows that these two
integrals are secondary with respect to the other four terms on the
right-hand side of (146). Thus, by the integration by parts the last two
integrals becomes less important in the solution than before.

The above equivalent transformations can be repeated many times that the
solution $b_{+}^{k}(P,t)$ ($k=1,2,...,$) is transformed to the solution $%
\hat{b}_{+}^{k}(P,t)$ by the equivalent transformation similar to (143) and
then the solution $\hat{b}_{+}^{k}(P,t)$ is changed to the solution $%
b_{+}^{k+1}(P,t)$ by integrating by parts the last two integrals of the
solution $\hat{b}_{+}^{k}(P,t).$ The equivalent transformation from the
solution $b_{+}^{k}(P,t)$ to the transformed solution $\hat{b}_{+}^{k}(P,t)$
may be generally given by%
\begin{equation}
\hat{b}_{+}^{k}(P,t)=b_{+}^{k}(P,t)\exp [-i\int_{t_{0}}^{t}dt_{1}\Gamma
_{k}^{+}(P,t_{1})],\text{ }k=1,2,....  \tag{148}
\end{equation}%
The solutions $\hat{b}_{+}^{k}(P,t)$ and $b_{+}^{k}(P,t)$ are called the $k-$%
order exact solutions to the equation (141b) for the variable $b_{+}(P,t).$
From the solution $b_{+}^{k}(P,t)$ to the transformed solution $\hat{b}%
_{+}^{k}(P,t)$ the amplitudes $\{\Gamma _{+k}^{\alpha }(P,t),$ $\Theta
_{+k}^{\alpha }(P,t)\}$ $(\alpha =0,$ $-)$ of the solution $b_{+}^{k}(P,t)$
are transformed to the amplitudes $\{\hat{\Gamma}_{+k}^{\alpha }(P,t),$ $%
\hat{\Theta}_{+k}^{\alpha }(P,t)\}$ of the solution $\hat{b}_{+}^{k}(P,t)$
according to the recursive equations (145) if in the recursive relations
(145) one makes the following replacements: $\hat{\Gamma}_{+1}^{\alpha
}(P,t)\leftrightarrow \hat{\Gamma}_{+k}^{\alpha }(P,t)$, $\hat{\Theta}%
_{+1}^{\alpha }(P,t)\leftrightarrow \hat{\Theta}_{+k}^{\alpha }(P,t),$ $%
\Gamma _{+1}^{\alpha }(P,t)\leftrightarrow \Gamma _{+k}^{\alpha }(P,t),$ $%
\Theta _{+1}^{\alpha }(P,t)\leftrightarrow \Theta _{+k}^{\alpha }(P,t),$ for 
$\alpha =0,-,$ and $\Gamma _{1}^{+}(P,t)\leftrightarrow \Gamma
_{k}^{+}(P,t). $ On the other hand, from the solution $\hat{b}_{+}^{k}(P,t)$
to the solution $b_{+}^{k+1}(P,t)$ the recursive relations for their
amplitudes are still given by (147) except the relation (147b), in which one
needs to make the following replacements: $\hat{\Gamma}_{+1}^{\alpha
}(P,t)\leftrightarrow \hat{\Gamma}_{+k}^{\alpha }(P,t)$, $\hat{\Theta}%
_{+1}^{\alpha }(P,t)\leftrightarrow \hat{\Theta}_{+k}^{\alpha }(P,t),$ $%
\Gamma _{+2}^{\alpha }(P,t)\leftrightarrow \Gamma _{+k+1}^{\alpha }(P,t),$ $%
\Theta _{+2}^{\alpha }(P,t)\leftrightarrow \Theta _{+k+1}^{\alpha }(P,t),$
for $\alpha =0,-,$ and $\Gamma _{1}^{+}(P,t)\leftrightarrow \Gamma
_{k}^{+}(P,t)$ and $\Gamma _{2}^{+}(P,t)\leftrightarrow \Gamma
_{k+1}^{+}(P,t).$ The relation (147b) is modified to the form 
\begin{equation*}
\Gamma _{k+1}^{+}(P,t)=-[\frac{1}{4}\frac{\Gamma (P,t)\hat{\Theta}%
_{+k}^{-}(P,t)}{\Omega (t)}+\frac{1}{\sqrt{2}}\frac{\Theta (P,t)^{\ast }\hat{%
\Theta}_{+k}^{0}(P,t)}{\Omega (t)}]
\end{equation*}%
\begin{equation*}
\times \exp [i\int_{t_{0}}^{t}dt_{1}\Gamma _{k}^{+}(P,t_{1})]\exp
[i\int_{t_{0}}^{t}dt_{1}\Gamma _{k-1}^{+}(P,t_{1})]...\exp
[i\int_{t_{0}}^{t}dt_{1}\Gamma _{1}^{+}(P,t_{1})].
\end{equation*}%
It can be found that after integrating by parts the last two integrals of
the solution $\hat{b}_{+}^{k}(P,t)$ many times, the two integrals become
less and less important in the solution. Actually, it can turn out that the
amplitudes $\hat{\Theta}_{+k}^{-}(P,t)$ and $\hat{\Theta}_{+k}^{0}(P,t)$ of
the solution $\hat{b}_{+}^{k}(P,t)$ is inversely proportional to the $k-$th
power of the Rabi frequency $\Omega (t),$ that is, $\hat{\Theta}%
_{+k}^{-}(P,t)\varpropto \Omega (t)^{-k}$ and $\hat{\Theta}%
_{+k}^{0}(P,t)\varpropto \Omega (t)^{-k}.$ Thus, by making only a few
equivalent transformations of the integration by parts one can obtain a
highly accurate solution to the original equation (141b) for the variable $%
b_{+}(P,t)$ even if the last two integrals are neglected.

Now it is easy to obtain the first-order approximation solution to the
original equation (141b) for the variable $b_{+}(P,t)$ from the equation
(144) by neglecting the last two integrals on the right-hand side of (144).
The first-order approximation solution is a linear algebra equation with the
three variables $\{b_{0}(P,t)-b_{0}(P,t_{0}),$ $b_{\pm }(P,t)\}$ and is
given by%
\begin{equation*}
b_{+}(P,t)\equiv b_{+}^{1}(P,t)=b_{-}(P,t)\Gamma _{+1}^{-}(P,t)\exp
[-i\int_{t_{0}}^{t}dt^{\prime }2\Omega (t^{\prime })]
\end{equation*}%
\begin{equation*}
+i(b_{0}(P,t)-b_{0}(P,t_{0}))\Gamma _{+1}^{0}(P,t)\exp [-i\Delta
P\int_{t_{0}}^{t}dt^{\prime }K_{0}(t^{\prime })]\exp
[-i\int_{t_{0}}^{t}dt^{\prime }\Omega (t^{\prime })]
\end{equation*}%
\begin{equation*}
+ib_{0}(P,t_{0})\Gamma _{+1}^{0}(P,t)\exp [-i\Delta
P\int_{t_{0}}^{t}dt^{\prime }K_{0}(t^{\prime })]\exp
[-i\int_{t_{0}}^{t}dt^{\prime }\Omega (t^{\prime })]
\end{equation*}%
\begin{equation}
-ib_{0}(P,t_{0})\Gamma _{+1}^{0}(P,t_{0})\exp [i\int_{t_{0}}^{t}dt_{1}\Gamma
_{1}^{+}(P,t_{1})],  \tag{149}
\end{equation}%
while\ the truncation error is just the last two integrals:%
\begin{equation*}
E_{r1}^{+}(P,t)=\exp [i\int_{t_{0}}^{t}dt_{1}\Gamma _{1}^{+}(P,t_{1})]
\end{equation*}%
\begin{equation*}
\times \{i\int_{t_{0}}^{t}dt_{1}\{b_{-}(P,t_{1})\hat{\Theta}%
_{+1}^{-}(P,t_{1})\exp [-i\int_{t_{0}}^{t_{1}}dt^{\prime }2\Omega (t^{\prime
})]\}
\end{equation*}%
\begin{equation*}
+\int_{t_{0}}^{t}dt_{1}\{b_{0}(P,t_{1})\hat{\Theta}_{+1}^{0}(P,t_{1})\exp
[-i\Delta P\int_{t_{0}}^{t_{1}}dt^{\prime }K_{0}(t^{\prime })]\exp
[-i\int_{t_{0}}^{t_{1}}dt^{\prime }\Omega (t^{\prime })]\}\}.
\end{equation*}%
By the integration by parts it can turn out that the error term $%
E_{r1}^{+}(P,t)$ is bounded by 
\begin{equation*}
|E_{r1}^{+}(P,t)|\leq \frac{1}{2}\frac{|\hat{\Theta}_{+1}^{-}(P,t)|}{\Omega
(t)}+\frac{|\hat{\Theta}_{+1}^{0}(P,t)|}{\Omega (t)}+\frac{|\hat{\Theta}%
_{+1}^{0}(P,t_{0})|}{\Omega (t_{0})}
\end{equation*}%
\begin{equation}
+\int_{t_{0}}^{t}dt_{1}\{|\Gamma _{2}^{+}(P,t_{1})|+|\Theta
_{+2}^{-}(P,t_{1})|+|\Theta _{+2}^{0}(P,t_{1})|\}.  \tag{150}
\end{equation}%
It is clear that this error upper bound is inversely proportional to $\Omega
(t)^{2}.$ The second-order approximation solution is given by%
\begin{equation*}
b_{+}^{2}(P,t)=b_{-}(P,t)\Gamma _{+2}^{-}(P,t)\exp
[-i\int_{t_{0}}^{t}dt^{\prime }2\Omega (t^{\prime })]
\end{equation*}%
\begin{equation*}
+i(b_{0}(P,t)-b_{0}(P,t_{0}))\Gamma _{+2}^{0}(P,t)\exp [-i\Delta
P\int_{t_{0}}^{t}dt^{\prime }K_{0}(t^{\prime })]\exp
[-i\int_{t_{0}}^{t}dt^{\prime }\Omega (t^{\prime })]
\end{equation*}%
\begin{equation*}
+ib_{0}(P,t_{0})\Gamma _{+2}^{0}(P,t)\exp [-i\Delta
P\int_{t_{0}}^{t}dt^{\prime }K_{0}(t^{\prime })]\exp
[-i\int_{t_{0}}^{t}dt^{\prime }\Omega (t^{\prime })]
\end{equation*}%
\begin{equation}
-ib_{0}(P,t_{0})\Gamma _{+2}^{0}(P,t_{0})\exp [i\int_{t_{0}}^{t}dt_{1}\Gamma
_{2}^{+}(P,t_{1})]  \tag{151}
\end{equation}%
and the truncation error is bounded by%
\begin{equation*}
|E_{r2}^{+}(P,t)|\leq \frac{1}{2}\frac{|\hat{\Theta}_{+2}^{-}(P,t)|}{\Omega
(t)}+\frac{|\hat{\Theta}_{+2}^{0}(P,t)|}{\Omega (t)}+\frac{|\hat{\Theta}%
_{+2}^{0}(P,t_{0})|}{\Omega (t_{0})}
\end{equation*}%
\begin{equation}
+\int_{t_{0}}^{t}dt_{1}\{|\Gamma _{3}^{+}(P,t_{1})|+|\Theta
_{+3}^{-}(P,t_{1})|+|\Theta _{+3}^{0}(P,t_{1})|\}.  \tag{152}
\end{equation}%
Obviously, this error upper bound is inversely proportional to $\Omega
(t)^{3}.$ The first- and second-order approximation solutions may be used to
set up the adiabatic condition.

By using the similar equivalent transformations mentioned above one may
obtain the $k-$order exact solution $b_{-}^{k}(P,t)$ ($k=1,2,...$) from the
equation (141b) for the variable $b_{-}(P,t)$, 
\begin{equation*}
b_{-}^{k}(P,t)=b_{+}(P,t)\Gamma _{-k}^{+}(P,t)\exp
[i\int_{t_{0}}^{t}dt^{\prime }2\Omega (t^{\prime })]
\end{equation*}%
\begin{equation*}
+ib_{0}(P,t)\Gamma _{-k}^{0}(P,t)\exp [-i\Delta P\int_{t_{0}}^{t}dt^{\prime
}K_{0}(t^{\prime })]\exp [i\int_{t_{0}}^{t}dt^{\prime }\Omega (t^{\prime })]
\end{equation*}%
\begin{equation*}
-ib_{0}(P,t_{0})\Gamma
_{-k}^{0}(P,t_{0})+i\int_{t_{0}}^{t}dt_{1}\{b_{-}^{k}(P,t_{1})\Gamma
_{k}^{-}(P,t_{1})\}
\end{equation*}%
\begin{equation*}
+i\int_{t_{0}}^{t}dt_{1}\{b_{+}(P,t_{1})\Theta _{-k}^{+}(P,t_{1})\exp
[i\int_{t_{0}}^{t_{1}}dt^{\prime }2\Omega (t^{\prime })]\}
\end{equation*}%
\begin{equation}
+\int_{t_{0}}^{t}dt_{1}\{b_{0}(P,t_{1})\Theta _{-k}^{0}(P,t_{1})\exp
[-i\Delta P\int_{t_{0}}^{t_{1}}dt^{\prime }K_{0}(t^{\prime })]\exp
[i\int_{t_{0}}^{t_{1}}dt^{\prime }\Omega (t^{\prime })]\}.  \tag{153}
\end{equation}%
The equation (153) is equivalent to the original equation (141b) for the
variable $b_{-}(P,t)$ if the initial condition (39) is taken into account.
In particular, the first-order exact solution $b_{-}^{1}(P,t)\equiv
b_{-}(P,t)$ and its five amplitudes are given by%
\begin{equation}
\Gamma _{-1}^{+}(P,t)=-\frac{1}{4}\frac{\Gamma (P,t)}{\Omega (t)},\text{ }%
\Gamma _{-1}^{0}(P,t)=\frac{1}{\sqrt{2}}\frac{\Theta (P,t)}{\Omega (t)},%
\text{ }  \tag{154a}
\end{equation}%
\begin{equation}
\Gamma _{1}^{-}(P,t)=-\frac{1}{8}\frac{\Gamma (P,t)^{2}+4|\Theta (P,t)|^{2}}{%
\Omega (t)},  \tag{154b}
\end{equation}%
\begin{equation}
\Theta _{-1}^{+}(P,t)=-\frac{1}{2}\{i\frac{1}{2}\frac{\partial }{\partial t}(%
\frac{\Gamma (P,t)}{\Omega (t)})+\frac{|\Theta (P,t)|^{2}}{\Omega (t)}\}, 
\tag{154c}
\end{equation}%
\begin{equation}
\Theta _{-1}^{0}(P,t)=\frac{1}{\sqrt{2}}\{i\frac{\partial }{\partial t}(%
\frac{\Theta (P,t)}{\Omega (t)})+\frac{1}{4}\frac{\Gamma (P,t)\Theta (P,t)}{%
\Omega (t)}+\frac{K_{0}(t)\Theta (P,t)}{\Omega (t)}\Delta P\}.  \tag{154d}
\end{equation}%
The equivalent transformation from the solution $b_{-}^{k}(P,t)$ of (153) to
the solution $\hat{b}_{-}^{k}(P,t)$ is given by%
\begin{equation}
\hat{b}_{-}^{k}(P,t)=b_{-}^{k}(P,t)\exp [-i\int_{t_{0}}^{t}dt_{1}\Gamma
_{k}^{-}(P,t_{1})].  \tag{155}
\end{equation}%
By the transformation (155) and the initial condition (39) the fourth term
on the right-hand side of (153) is cancelled and the solution $%
b_{-}^{k}(P,t) $ is changed to the transformed solution $\hat{b}%
_{-}^{k}(P,t) $:%
\begin{equation*}
\hat{b}_{-}^{k}(P,t)=b_{+}(P,t)\hat{\Gamma}_{-k}^{+}(P,t)\exp
[i\int_{t_{0}}^{t}dt^{\prime }2\Omega (t^{\prime })]
\end{equation*}%
\begin{equation*}
+ib_{0}(P,t)\hat{\Gamma}_{-k}^{0}(P,t)\exp [-i\Delta
P\int_{t_{0}}^{t}dt^{\prime }K_{0}(t^{\prime })]\exp
[i\int_{t_{0}}^{t}dt^{\prime }\Omega (t^{\prime })]
\end{equation*}%
\begin{equation*}
-ib_{0}(P,t_{0})\hat{\Gamma}_{-k}^{0}(P,t_{0})+i\int_{t_{0}}^{t}dt_{1}%
\{b_{+}(P,t_{1})\hat{\Theta}_{-k}^{+}(P,t_{1})\exp
[i\int_{t_{0}}^{t_{1}}dt^{\prime }2\Omega (t^{\prime })]\}
\end{equation*}%
\begin{equation}
+\int_{t_{0}}^{t}dt_{1}\{b_{0}(P,t_{1})\hat{\Theta}_{-k}^{0}(P,t_{1})\exp
[-i\Delta P\int_{t_{0}}^{t_{1}}dt^{\prime }K_{0}(t^{\prime })]\exp
[i\int_{t_{0}}^{t_{1}}dt^{\prime }\Omega (t^{\prime })]\}.  \tag{156}
\end{equation}%
The recursive relations for the amplitudes of both the solutions $\hat{b}%
_{-}^{k}(P,t)$ and $b_{-}^{k}(P,t)$ are given by%
\begin{equation}
\hat{\Gamma}_{-k}^{+}(P,t)=\Gamma _{-k}^{+}(P,t)\exp
[-i\int_{t_{0}}^{t}dt_{1}\Gamma _{k}^{-}(P,t_{1})],  \tag{157a}
\end{equation}%
\begin{equation}
\hat{\Gamma}_{-k}^{0}(P,t)=\Gamma _{-k}^{0}(P,t)\exp
[-i\int_{t_{0}}^{t}dt_{1}\Gamma _{k}^{-}(P,t_{1})],  \tag{157b}
\end{equation}%
\begin{equation}
\hat{\Theta}_{-k}^{+}(P,t)=[\Theta _{-k}^{+}(P,t)+\Gamma
_{-k}^{+}(P,t)\Gamma _{k}^{-}(P,t)]\exp [-i\int_{t_{0}}^{t}dt_{1}\Gamma
_{k}^{-}(P,t_{1})],  \tag{157c}
\end{equation}%
\begin{equation}
\hat{\Theta}_{-k}^{0}(P,t)=[\Theta _{-k}^{0}(P,t)-\Gamma
_{-k}^{0}(P,t)\Gamma _{k}^{-}(P,t)]\exp [-i\int_{t_{0}}^{t}dt_{1}\Gamma
_{k}^{-}(P,t_{1})].  \tag{157d}
\end{equation}%
After making the integration by parts on the last two integrals on the
right-hand side of (156) the solution $\hat{b}_{-}^{k}(P,t)$ is changed to
the $(k+1)-$order exact solution $b_{-}^{k+1}(P,t)$ which is also given by
(153). The recursive relations between the amplitudes of both the solutions $%
\hat{b}_{-}^{k}(P,t)$ and $b_{-}^{k+1}(P,t)$ are generally given by%
\begin{equation}
\Gamma _{-(k+1)}^{+}(P,t)=\hat{\Gamma}_{-k}^{+}(P,t)+\frac{1}{2}\frac{\hat{%
\Theta}_{-k}^{+}(P,t)}{\Omega (t)},  \tag{158a}
\end{equation}%
\begin{equation}
\Gamma _{-(k+1)}^{0}(P,t)=\hat{\Gamma}_{-k}^{0}(P,t)-\frac{\hat{\Theta}%
_{-k}^{0}(P,t)}{\Omega (t)},  \tag{158b}
\end{equation}%
\begin{equation*}
\Gamma _{k+1}^{-}(P,t)=[\frac{1}{4}\frac{\Gamma (P,t)\hat{\Theta}%
_{-k}^{+}(P,t)}{\Omega (t)}+\frac{1}{\sqrt{2}}\frac{\Theta (P,t)^{\ast }\hat{%
\Theta}_{-k}^{0}(P,t)}{\Omega (t)}]
\end{equation*}%
\begin{equation}
\times \exp [i\int_{t_{0}}^{t}dt_{1}\Gamma _{k}^{+}(P,t_{1})]\exp
[i\int_{t_{0}}^{t}dt_{1}\Gamma _{k-1}^{+}(P,t_{1})]...\exp
[i\int_{t_{0}}^{t}dt_{1}\Gamma _{1}^{+}(P,t_{1})],  \tag{158c}
\end{equation}%
\begin{equation}
\Theta _{-(k+1)}^{+}(P,t)=i\frac{1}{2}\frac{\partial }{\partial t}(\frac{%
\hat{\Theta}_{-k}^{+}(P,t)}{\Omega (t)})+\frac{1}{\sqrt{2}}\frac{\Theta
(P,t)^{\ast }\hat{\Theta}_{-k}^{0}(P,t)}{\Omega (t)},  \tag{158d}
\end{equation}%
\begin{equation}
\Theta _{-(k+1)}^{0}(P,t)=\frac{1}{2\sqrt{2}}\frac{\Theta (P,t)\hat{\Theta}%
_{-k}^{+}(P,t)}{\Omega (t)}+i\frac{\partial }{\partial t}(\frac{\hat{\Theta}%
_{-k}^{0}(P,t)}{\Omega (t)})+\frac{K_{0}(t)\hat{\Theta}_{-k}^{0}(P,t)}{%
\Omega (t)}\Delta P.  \tag{158e}
\end{equation}%
The recursive relations (157c), (157d), (158d), and (158e) show that in the
last two integrals of the $k-$order exact solution $\hat{b}_{-}^{k}(P,t)$ of
(156) the amplitudes $\hat{\Theta}_{-k}^{+}(P,t)$ and $\hat{\Theta}%
_{-k}^{0}(P,t)$ are inversely proportional to the $k-$th power of the Rabi
frequency $\Omega (t),$ that is, $\hat{\Theta}_{-k}^{+}(P,t)\varpropto
\Omega (t)^{-k}$ and $\hat{\Theta}_{-k}^{0}(P,t)\varpropto \Omega (t)^{-k}.$
Thus, by making a few equivalent transformations one may obtain a high-order
approximation solution from the exact solution $b_{-}^{k}(P,t)$ or $\hat{b}%
_{-}^{k}(P,t).$ The first-order approximation solution may be obtained from
the exact solution $\hat{b}_{-}^{1}(P,t)$ by neglecting the last two
integrals on the right-hand side of (156) ($k=1$), 
\begin{equation*}
b_{-}(P,t)\equiv b_{-}^{1}(P,t)=b_{+}(P,t)\Gamma _{-1}^{+}(P,t)\exp
[i\int_{t_{0}}^{t}dt^{\prime }2\Omega (t^{\prime })]
\end{equation*}%
\begin{equation*}
+i(b_{0}(P,t)-b_{0}(P,t_{0}))\Gamma _{-1}^{0}(P,t)\exp [-i\Delta
P\int_{t_{0}}^{t}dt^{\prime }K_{0}(t^{\prime })]\exp
[i\int_{t_{0}}^{t}dt^{\prime }\Omega (t^{\prime })]
\end{equation*}%
\begin{equation*}
+ib_{0}(P,t_{0})\Gamma _{-1}^{0}(P,t)\exp [-i\Delta
P\int_{t_{0}}^{t}dt^{\prime }K_{0}(t^{\prime })]\exp
[i\int_{t_{0}}^{t}dt^{\prime }\Omega (t^{\prime })]
\end{equation*}%
\begin{equation}
-ib_{0}(P,t_{0})\Gamma _{-1}^{0}(P,t_{0})\exp [i\int_{t_{0}}^{t}dt_{1}\Gamma
_{1}^{-}(P,t_{1})],  \tag{159}
\end{equation}%
and the truncation error is just given by these last two integrals
neglected, and by the integration by parts it can turn out that the
truncation error is bounded by%
\begin{equation*}
|E_{r1}^{-}(P,t)|\leq \frac{1}{2}\frac{|\hat{\Theta}_{-1}^{+}(P,t)|}{\Omega
(t)}+\frac{|\hat{\Theta}_{-1}^{0}(P,t)|}{\Omega (t)}+\frac{|\hat{\Theta}%
_{-1}^{0}(P,t_{0})|}{\Omega (t_{0})}
\end{equation*}%
\begin{equation}
+\int_{t_{0}}^{t}dt_{1}\{|\Gamma _{2}^{-}(P,t_{1})|+|\Theta
_{-2}^{+}(P,t_{1})|+|\Theta _{-2}^{0}(P,t_{1})|\}.  \tag{160}
\end{equation}%
The error upper bound is proportional to $\Omega (t)^{-2}.$ The second-order
approximation solution is obtained from the exact solution $\hat{b}%
_{-}^{2}(P,t)$ by neglecting the last two integrals on the right-hand side
of (156) $(k=2)$, 
\begin{equation*}
b_{-}^{2}(P,t)=b_{+}(P,t)\Gamma _{-2}^{+}(P,t)\exp
[i\int_{t_{0}}^{t}dt^{\prime }2\Omega (t^{\prime })]
\end{equation*}%
\begin{equation*}
+ib_{0}(P,t)\Gamma _{-2}^{0}(P,t)\exp [-i\Delta P\int_{t_{0}}^{t}dt^{\prime
}K_{0}(t^{\prime })]\exp [i\int_{t_{0}}^{t}dt^{\prime }\Omega (t^{\prime })]
\end{equation*}%
\begin{equation}
-ib_{0}(P,t_{0})\Gamma _{-2}^{0}(P,t_{0})\exp [i\int_{t_{0}}^{t}dt_{1}\Gamma
_{2}^{-}(P,t_{1})]  \tag{161}
\end{equation}%
and the truncation error is bounded by 
\begin{equation*}
|E_{r2}^{-}(P,t)|\leq \frac{1}{2}\frac{|\hat{\Theta}_{-2}^{+}(P,t)|}{\Omega
(t)}+\frac{|\hat{\Theta}_{-2}^{0}(P,t)|}{\Omega (t)}+\frac{|\hat{\Theta}%
_{-2}^{0}(P,t_{0})|}{\Omega (t_{0})}
\end{equation*}%
\begin{equation}
+\int_{t_{0}}^{t}dt_{1}\{|\Gamma _{3}^{-}(P,t_{1})|+|\Theta
_{-3}^{+}(P,t_{1})|+|\Theta _{-3}^{0}(P,t_{1})|\}.  \tag{162}
\end{equation}%
The error upper bound is proportional to $\Omega (t)^{-3}.$ The first- and
second-order approximation solutions may be used to set up the adiabatic
condition below.

With the help of the equivalent transformations similar to those used above
one may obtain the $k-$order exact solution for the variable $b_{0}(P,t)$
from the equation (141a). At first the first-order exact solution may be
given by 
\begin{equation*}
b_{0}(P,t)-b_{0}(P,t_{0})=ib_{+}(P,t)\Gamma _{01}^{+}(P,t)\exp [i\Delta
P\int_{t_{0}}^{t}dt^{\prime }K_{0}(t^{\prime })]\exp
[i\int_{t_{0}}^{t}dt^{\prime }\Omega (t^{\prime })]
\end{equation*}%
\begin{equation*}
+ib_{-}(P,t)\Gamma _{01}^{-}(P,t)\exp [i\Delta P\int_{t_{0}}^{t}dt^{\prime
}K_{0}(t^{\prime })]\exp [-i\int_{t_{0}}^{t}dt^{\prime }\Omega (t^{\prime })]
\end{equation*}%
\begin{equation*}
+\int_{t_{0}}^{t}dt_{1}\{b_{+}(P,t_{1})\Theta _{01}^{+}(P,t_{1})\exp
[i\Delta P\int_{t_{0}}^{t_{1}}dt^{\prime }K_{0}(t^{\prime })]\exp
[i\int_{t_{0}}^{t_{1}}dt^{\prime }\Omega (t^{\prime })]\}
\end{equation*}%
\begin{equation}
+\int_{t_{0}}^{t}dt_{1}\{b_{-}(P,t_{1})\Theta _{01}^{-}(P,t_{1})\exp
[i\Delta P\int_{t_{0}}^{t_{1}}dt^{\prime }K_{0}(t^{\prime })]\exp
[-i\int_{t_{0}}^{t_{1}}dt^{\prime }\Omega (t^{\prime })]\}  \tag{163}
\end{equation}%
where the four amplitudes are given by 
\begin{equation*}
\Gamma _{01}^{+}(P,t)=-\Gamma _{01}^{-}(P,t)=-\frac{1}{\sqrt{2}}\frac{\Theta
(P,t)^{\ast }}{\Omega (t)},
\end{equation*}%
\begin{equation*}
\Theta _{01}^{+}(P,t)=-\Theta _{01}^{-}(P,t)=i\frac{\partial }{\partial t}(%
\frac{\Theta (P,t)^{\ast }}{\Omega (t)})-\frac{1}{2}\frac{\Gamma (P,t)\Theta
(P,t)^{\ast }}{\Omega (t)}-\frac{K_{0}(t)\Theta (P,t)^{\ast }}{\Omega (t)}%
\Delta P.
\end{equation*}%
A special point in the first-order exact solution (163) is that there is not
any term containing the solution $b_{0}(P,t)$ itself on the right-hand side
of (163). Thus, one may make directly the integration by parts on the last
two integrals on the right-hand side of (163). Here still denote that $%
\delta _{0}^{b}(P,t)=b_{0}(P,t)-b_{0}(P,t_{0})$ and $\delta _{\pm
}^{b}(P,t)=b_{\pm }(P,t)-b_{\pm }(P,t_{0}).$ Then the second-order exact
solution $\delta _{0}^{b2}(P,t)$ may be expressed as%
\begin{equation*}
\delta _{0}^{b2}(P,t)\equiv \delta _{0}^{b}(P,t)=ib_{+}(P,t)\Gamma
_{02}^{+}(P,t)\exp [i\Delta P\int_{t_{0}}^{t}dt^{\prime }K_{0}(t^{\prime
})]\exp [i\int_{t_{0}}^{t}dt^{\prime }\Omega (t^{\prime })]
\end{equation*}%
\begin{equation*}
+ib_{-}(P,t)\Gamma _{02}^{-}(P,t)\exp [i\Delta P\int_{t_{0}}^{t}dt^{\prime
}K_{0}(t^{\prime })]\exp [-i\int_{t_{0}}^{t}dt^{\prime }\Omega (t^{\prime })]
\end{equation*}%
\begin{equation*}
+ib_{0}(P,t_{0})\Gamma _{2}^{00}(P,t)+i\int_{t_{0}}^{t}dt_{1}\{\delta
_{0}^{b2}(P,t_{1})\Gamma _{2}^{0}(P,t_{1})\}
\end{equation*}%
\begin{equation*}
+\int_{t_{0}}^{t}dt_{1}\{b_{+}(P,t_{1})\Theta _{02}^{+}(P,t_{1})\exp
[i\Delta P\int_{t_{0}}^{t_{1}}dt^{\prime }K_{0}(t^{\prime })]\exp
[i\int_{t_{0}}^{t_{1}}dt^{\prime }\Omega (t^{\prime })]\}
\end{equation*}%
\begin{equation}
+\int_{t_{0}}^{t}dt_{1}\{b_{-}(P,t_{1})\Theta _{02}^{-}(P,t_{1})\exp
[i\Delta P\int_{t_{0}}^{t_{1}}dt^{\prime }K_{0}(t^{\prime })]\exp
[-i\int_{t_{0}}^{t_{1}}dt^{\prime }\Omega (t^{\prime })]\}  \tag{164}
\end{equation}%
where the amplitudes are given by%
\begin{equation*}
\Gamma _{02}^{+}(P,t)=\Gamma _{01}^{+}(P,t)-\frac{\Theta _{01}^{+}(P,t)}{%
\Omega (t)},\text{ }\Gamma _{02}^{-}(P,t)=\Gamma _{01}^{-}(P,t)+\frac{\Theta
_{01}^{-}(P,t)}{\Omega (t)},
\end{equation*}%
\begin{equation*}
\Gamma _{2}^{00}(P,t)=\int_{t_{0}}^{t}dt_{1}\Gamma _{2}^{0}(P,t_{1}),\text{ }%
\Gamma _{2}^{0}(P,t)=-\sqrt{2}\frac{\Theta (P,t)\Theta _{01}^{+}(P,t)}{%
\Omega (t)},
\end{equation*}%
\begin{equation*}
\Theta _{02}^{+}(P,t)=-\Theta _{02}^{-}(P,t)=i\frac{\partial }{\partial t}(%
\frac{\Theta _{01}^{+}(P,t)}{\Omega (t)})
\end{equation*}%
\begin{equation*}
-\frac{K_{0}(t)\Theta _{01}^{+}(P,t)}{\Omega (t)}\Delta P-\frac{1}{2}\frac{%
\Gamma (P,t)\Theta _{01}^{-}(P,t)}{\Omega (t)}.
\end{equation*}%
Now the first-order approximation solution may be obtained from the
first-order exact solution (163) by neglecting the last two integrals on the
right-hand side of (163). It is given by%
\begin{equation*}
\delta _{0}^{b}(P,t)=ib_{+}(P,t)\Gamma _{01}^{+}(P,t)\exp [i\Delta
P\int_{t_{0}}^{t}dt^{\prime }K_{0}(t^{\prime })]\exp
[i\int_{t_{0}}^{t}dt^{\prime }\Omega (t^{\prime })]
\end{equation*}%
\begin{equation}
+ib_{-}(P,t)\Gamma _{01}^{-}(P,t)\exp [i\Delta P\int_{t_{0}}^{t}dt^{\prime
}K_{0}(t^{\prime })]\exp [-i\int_{t_{0}}^{t}dt^{\prime }\Omega (t^{\prime })]
\tag{165}
\end{equation}%
and the truncation error is just the last two integrals on the right-hand
side of (163), and it can turn out by the integration by parts that the
truncation error is bounded by 
\begin{equation}
|E_{r1}^{0}(P,t)|\leq \frac{2|\Theta _{01}^{+}(P,t)|}{\Omega (t)}%
+\int_{t_{0}}^{t}dt_{1}\{|\Gamma _{2}^{0}(P,t_{1})|+2|\Theta
_{02}^{+}(P,t_{1})|\}.  \tag{166}
\end{equation}%
This error upper bound is clearly proportional to $\Omega (t)^{-2}.$ A
higher-order exact solution than the second-order one (164) may be obtained
by integrating by parts the last two integrals in (164), but it needs first
to eliminate the fourth term that contains the solution $\delta
_{0}^{b2}(P,t)$ itself on the right-hand side of (164). This can be done by
making an equivalent transformation on the solution $\delta _{0}^{b2}(P,t).$
In general, for the $k-$order exact solution $\delta _{0}^{bk}(P,t)$ ($%
k=2,3,...$) the equivalent transformation is written as%
\begin{equation}
\hat{\delta}_{0}^{bk}(P,t)=\delta _{0}^{bk}(P,t)\exp
[-i\int_{t_{0}}^{t}dt_{1}\Gamma _{k}^{0}(P,t_{1})].  \tag{167}
\end{equation}%
Then the $k-$order transformed solution $\hat{\delta}_{0}^{bk}(P,t)$ may be
expressed as%
\begin{equation*}
\hat{\delta}_{0}^{bk}(P,t)=ib_{+}(P,t)\hat{\Gamma}_{0k}^{+}(P,t)\exp
[i\Delta P\int_{t_{0}}^{t}dt^{\prime }K_{0}(t^{\prime })]\exp
[i\int_{t_{0}}^{t}dt^{\prime }\Omega (t^{\prime })]
\end{equation*}%
\begin{equation*}
+ib_{-}(P,t)\hat{\Gamma}_{0k}^{-}(P,t)\exp [i\Delta
P\int_{t_{0}}^{t}dt^{\prime }K_{0}(t^{\prime })]\exp
[-i\int_{t_{0}}^{t}dt^{\prime }\Omega (t^{\prime })]+ib_{0}(P,t_{0})\hat{%
\Gamma}_{k}^{00}(P,t)
\end{equation*}%
\begin{equation*}
+\int_{t_{0}}^{t}dt_{1}\{b_{+}(P,t_{1})\hat{\Theta}_{0k}^{+}(P,t_{1})\exp
[i\Delta P\int_{t_{0}}^{t_{1}}dt^{\prime }K_{0}(t^{\prime })]\exp
[i\int_{t_{0}}^{t_{1}}dt^{\prime }\Omega (t^{\prime })]\}
\end{equation*}%
\begin{equation}
+\int_{t_{0}}^{t}dt_{1}\{b_{-}(P,t_{1})\hat{\Theta}_{0k}^{-}(P,t_{1})\exp
[i\Delta P\int_{t_{0}}^{t_{1}}dt^{\prime }K_{0}(t^{\prime })]\exp
[-i\int_{t_{0}}^{t_{1}}dt^{\prime }\Omega (t^{\prime })]\}  \tag{168}
\end{equation}%
where the amplitudes satisfy the recursive relations: 
\begin{equation}
\hat{\Gamma}_{0k}^{+}(P,t)=\Gamma _{0k}^{+}(P,t)\exp
[-i\int_{t_{0}}^{t}dt_{1}\Gamma _{k}^{0}(P,t_{1})],  \tag{169a}
\end{equation}%
\begin{equation}
\hat{\Gamma}_{0k}^{-}(P,t)=\Gamma _{0k}^{-}(P,t)\exp
[-i\int_{t_{0}}^{t}dt_{1}\Gamma _{k}^{0}(P,t_{1})],  \tag{169b}
\end{equation}%
\begin{equation}
\hat{\Gamma}_{k}^{00}(P,t)=\int_{t_{0}}^{t}dt_{1}\{[\frac{\partial }{%
\partial t_{1}}\Gamma _{k}^{00}(P,t_{1})]\exp
[-i\int_{t_{0}}^{t_{1}}dt^{\prime }\Gamma _{k}^{0}(P,t^{\prime })]\}, 
\tag{169c}
\end{equation}%
\begin{equation}
\hat{\Theta}_{0k}^{+}(P,t)=[\Theta _{0k}^{+}(P,t)-\Gamma
_{0k}^{+}(P,t)\Gamma _{k}^{0}(P,t)]\exp [-i\int_{t_{0}}^{t}dt_{1}\Gamma
_{k}^{0}(P,t_{1})],  \tag{169d}
\end{equation}%
\begin{equation}
\hat{\Theta}_{0k}^{-}(P,t)=[\Theta _{0k}^{-}(P,t)-\Gamma
_{0k}^{-}(P,t)\Gamma _{k}^{0}(P,t)]\exp [-i\int_{t_{0}}^{t}dt_{1}\Gamma
_{k}^{0}(P,t_{1})].  \tag{169e}
\end{equation}%
From the $k-$order transformed solution $\hat{\delta}_{0}^{bk}(P,t)$ to the $%
(k+1)-$order solution $\delta _{0}^{bk+1}(P,t)$ there is an equivalent
transformation of the integration by parts. The $(k+1)-$order exact solution 
$\delta _{0}^{bk+1}(P,t)$ is still given by (164) as long as one makes the
replacement: $\delta _{0}^{b2}(P,t)\leftrightarrow \delta _{0}^{bk+1}(P,t)$
and the following replacements for the amplitudes: 
\begin{equation*}
\Gamma _{02}^{\alpha }(P,t)\leftrightarrow \Gamma _{0k+1}^{\alpha }(P,t),%
\text{ }\Theta _{02}^{\alpha }(P,t)\leftrightarrow \Theta _{0k+1}^{\alpha
}(P,t),\text{ }\alpha =+,-;
\end{equation*}%
\begin{equation*}
\Gamma _{2}^{00}(P,t)\leftrightarrow \Gamma _{k+1}^{00}(P,t),\text{ }\Gamma
_{2}^{0}(P,t)\leftrightarrow \Gamma _{k+1}^{0}(P,t).
\end{equation*}%
The recursive relations for the amplitudes between the two solutions $\hat{%
\delta}_{0}^{bk}(P,t)$ and $\delta _{0}^{bk+1}(P,t)$ ($k\geq 2$) are given by%
\begin{equation}
\Gamma _{0k+1}^{+}(P,t)=\hat{\Gamma}_{0k}^{+}(P,t)-\frac{\hat{\Theta}%
_{0k}^{+}(P,t)}{\Omega (t)},\text{ }\Gamma _{0k+1}^{-}(P,t)=\hat{\Gamma}%
_{0k}^{-}(P,t)+\frac{\hat{\Theta}_{0k}^{-}(P,t)}{\Omega (t)},  \tag{170a}
\end{equation}%
\begin{equation}
\Gamma _{k+1}^{00}(P,t)=\hat{\Gamma}_{k}^{00}(P,t)+\frac{1}{\sqrt{2}}%
\int_{t_{0}}^{t}dt_{1}\{-\frac{\Theta (P,t_{1})\hat{\Theta}_{0k}^{+}(P,t_{1})%
}{\Omega (t_{1})}+\frac{\Theta (P,t_{1})\hat{\Theta}_{0k}^{-}(P,t_{1})}{%
\Omega (t_{1})}\},  \tag{170b}
\end{equation}%
\begin{equation*}
\Gamma _{k+1}^{0}(P,t)=[-\frac{1}{\sqrt{2}}\frac{\Theta (P,t)\hat{\Theta}%
_{0k}^{+}(P,t)}{\Omega (t)}+\frac{1}{\sqrt{2}}\frac{\Theta (P,t)\hat{\Theta}%
_{0k}^{-}(P,t)}{\Omega (t)}]
\end{equation*}%
\begin{equation}
\times \exp [i\int_{t_{0}}^{t}dt_{1}\Gamma _{k}^{0}(P,t_{1})]\exp
[i\int_{t_{0}}^{t}dt_{1}\Gamma _{k-1}^{0}(P,t_{1})]...\exp
[i\int_{t_{0}}^{t}dt_{1}\Gamma _{2}^{0}(P,t_{1})],  \tag{170c}
\end{equation}%
\begin{equation}
\Theta _{0k+1}^{+}(P,t)=i\frac{\partial }{\partial t}(\frac{\hat{\Theta}%
_{0k}^{+}(P,t)}{\Omega (t)})-\frac{K_{0}(t)\hat{\Theta}_{0k}^{+}(P,t)}{%
\Omega (t)}\Delta P-\frac{1}{2}\frac{\Gamma (P,t)\hat{\Theta}_{0k}^{-}(P,t)}{%
\Omega (t)},  \tag{170d}
\end{equation}%
\begin{equation}
\Theta _{0k+1}^{-}(P,t)=-i\frac{\partial }{\partial t}(\frac{\hat{\Theta}%
_{0k}^{-}(P,t)}{\Omega (t)})+\frac{K_{0}(t)\hat{\Theta}_{0k}^{-}(P,t)}{%
\Omega (t)}\Delta P+\frac{1}{2}\frac{\Gamma (P,t)\hat{\Theta}_{0k}^{+}(P,t)}{%
\Omega (t)}.  \tag{170e}
\end{equation}%
It can turn out that the amplitudes $\Theta _{0k}^{+}(P,t)$ and $\Theta
_{0k}^{-}(P,t)$ of the $k-$order exact solution $\delta _{0}^{bk}(P,t)$ are
inversely proportional to $\Omega (t)^{-k}.$ Thus, the last two integrals of
the $k-$order exact solution $\delta _{0}^{bk}(P,t)$ have a negligible
contribution to the solution if the Rabi frequency $\Omega (t)$ is large.
Now the second-order approximation solution may be obtained from the
second-order exact solution $\hat{\delta}_{0}^{b2}(P,t)$ of (168), 
\begin{equation*}
\delta _{0}^{b2}(P,t)=ib_{+}(P,t)\Gamma _{02}^{+}(P,t)\exp [i\Delta
P\int_{t_{0}}^{t}dt^{\prime }K_{0}(t^{\prime })]\exp
[i\int_{t_{0}}^{t}dt^{\prime }\Omega (t^{\prime })]
\end{equation*}%
\begin{equation*}
+ib_{-}(P,t)\Gamma _{02}^{-}(P,t)\exp [i\Delta P\int_{t_{0}}^{t}dt^{\prime
}K_{0}(t^{\prime })]\exp [-i\int_{t_{0}}^{t}dt^{\prime }\Omega (t^{\prime })]
\end{equation*}%
\begin{equation}
-b_{0}(P,t_{0})\{1-\exp [i\int_{t_{0}}^{t}dt_{1}\Gamma _{2}^{0}(P,t_{1})]\},
\tag{171}
\end{equation}%
while the truncation error is given by the last two integrals on the
right-hand side of (168), and the truncation error upper bound is determined
from 
\begin{equation*}
|E_{r2}^{0}(P,t)|\leq \frac{|\hat{\Theta}_{02}^{+}(P,t)|}{\Omega (t)}+\frac{|%
\hat{\Theta}_{02}^{-}(P,t)|}{\Omega (t)}
\end{equation*}%
\begin{equation}
+\int_{t_{0}}^{t}dt_{1}\{|\Gamma _{3}^{0}(P,t)|+|\Theta
_{03}^{+}(P,t)|+|\Theta _{03}^{-}(P,t)|\}.  \tag{172}
\end{equation}%
This error upper bound is clearly proportional to $\Omega (t)^{-3}.$

Now by solving the three first-order approximation solutions (149), (159),
and (165), which are all linear algebra equations, one may obtain the
first-order approximation solution to the basic equations (141). At first
according to the first-order approximation solutions (149), (159), and (165)
and their truncation errors the first-order exact solution to the equations
(141) may be formally written as, (this is really the first-order
approximation solutions plus their truncation errors), 
\begin{equation}
b_{+}(P,t)=\alpha _{-}^{+}b_{-}(P,t)+\alpha _{0}^{+}\delta
_{0}^{b}(P,t)+\beta _{0}^{+}+E_{r1}^{+}(P,t),  \tag{173a}
\end{equation}%
\begin{equation}
b_{-}(P,t)=\alpha _{+}^{-}b_{+}(P,t)+\alpha _{0}^{-}\delta
_{0}^{b}(P,t)+\beta _{0}^{-}+E_{r1}^{-}(P,t),  \tag{173b}
\end{equation}%
\begin{equation}
\delta _{0}^{b}(P,t)=\alpha _{+}^{0}b_{+}(P,t)+\alpha
_{-}^{0}b_{-}(P,t)+E_{r1}^{0}(P,t),  \tag{173c}
\end{equation}%
where the truncation errors $E_{r1}^{\pm }(P,t)$ and $E_{r1}^{0}(P,t)$ may
be considered as small parameters and any other parameters such as $\alpha
_{-}^{+},$ $\alpha _{+}^{-},$ $\alpha _{+}^{0}$, $etc.,$ can be obtained
directly from the first-order approximation solutions (149), (159), and
(165). These three equations are linear algebra equations as $E_{r1}^{\pm
}(P,t)$ and $E_{r1}^{0}(P,t)$ are considered as the small parameters. The
exact solution to the three linear algebra equations (173) may be written as%
\begin{equation}
\delta _{\pm }^{b}(P,t)=\delta _{\pm m}(P,t)+E_{rt}^{\pm }(P,t),\text{ }%
\delta _{0}^{b}(P,t)=\delta _{0m}(P,t)+E_{rt}^{0}(P,t),  \tag{174}
\end{equation}%
where the main terms are written as%
\begin{equation*}
\delta _{\pm m}(P,t)=\mp i\frac{1}{\sqrt{2}}b_{0}(P,t_{0})F(P,t)
\end{equation*}%
\begin{equation*}
\times \{\frac{\Theta (P,t)}{\Omega (t)}\exp [-i\Delta
P\int_{t_{0}}^{t}dt^{\prime }K_{0}(t^{\prime })]\exp [\mp
i\int_{t_{0}}^{t}dt^{\prime }\Omega (t^{\prime })]
\end{equation*}%
\begin{equation}
-\frac{\Theta (P,t_{0})}{\Omega (t_{0})}\exp [\pm i\frac{1}{8}%
\int_{t_{0}}^{t}dt_{1}\frac{\Gamma (P,t_{1})^{2}+4|\Theta (P,t_{1})|^{2}}{%
\Omega (t_{1})}]\},  \tag{175a}
\end{equation}%
\begin{equation}
\delta _{0m}(P,t)=0.  \tag{175b}
\end{equation}%
Here the factors $F(P,t)$ and $G_{\pm }(P,t)$ are defined by%
\begin{equation*}
F(P,t)=\frac{1+\frac{1}{2}\frac{|\Theta (P,t)|^{2}}{\Omega (t)^{2}}}{1+\frac{%
1}{16}\frac{\Gamma (P,t)^{2}}{\Omega (t)^{2}}+\frac{|\Theta (P,t)|^{2}}{%
\Omega (t)^{2}}},\text{ }G_{\pm }(P,t)=\frac{\pm \frac{1}{4}\frac{\Gamma
(P,t)}{\Omega (t)}+\frac{1}{2}\frac{|\Theta (P,t)|^{2}}{\Omega (t)^{2}}}{1+%
\frac{1}{16}\frac{\Gamma (P,t)^{2}}{\Omega (t)^{2}}+\frac{|\Theta (P,t)|^{2}%
}{\Omega (t)^{2}}}.
\end{equation*}%
The factors $G_{\pm }(P,t)$ appear in the error terms $E_{rt}^{\pm }(P,t)$
and $E_{rt}^{0}(P,t),$ as can be seen below. Then the upper bound of the
error term that is generated by the main terms $\delta _{\pm m}(P,t)$ and $%
\delta _{0m}(P,t)$ is determined from%
\begin{equation*}
||E_{r}^{(1)}(P,t)||=\sqrt{|\delta _{+m}(P,t)|^{2}+|\delta
_{-m}(P,t)|^{2}+|\delta _{0m}(P,t)|^{2}}
\end{equation*}%
\begin{equation}
\leq F(P,t)[\frac{|\Theta (P,t)|}{\Omega (t)}+\frac{|\Theta (P,t_{0})|}{%
\Omega (t_{0})}].  \tag{176}
\end{equation}%
It is clear that this upper bound is proportional to $\Omega (t)^{-1}.$
Notice that the factor $F(P,t)$ is unity approximately if $|\Theta
(P,t)|^{2}/\Omega (t)^{2}<<1$ and $\Gamma (P,t)^{2}/\Omega (t)^{2}<<1.$ By
comparing the inequality (176) with the inequality (133) one can find that
if the factor $F(P,t)$ is unity, then the inequality (176) is really the
inequality (133) that leads to the first-order approximation adiabatic
condition (134).

Now investigate the secondary error terms $E_{rt}^{\pm }(P,t)$ and $%
E_{rt}^{0}(P,t)$ in the exact solution (174). These error terms contain the
first-order truncation errors $E_{r1}^{\pm }(P,t)$ and $E_{r1}^{0}(P,t)$
(See: (150), (160), and (166)). They may be written as%
\begin{equation}
E_{rt}^{\pm }(P,t)=E_{rt}^{\pm 2}(P,t)+E_{rt}^{\pm 3}(P,t)+E_{rt}^{\pm
4}(P,t),  \tag{177a}
\end{equation}%
\begin{equation}
E_{rt}^{0}(P,t)=E_{rt}^{02}(P,t)+E_{rt}^{03}(P,t)+E_{rt}^{04}(P,t)+E_{rt}^{05}(P,t),
\tag{177b}
\end{equation}%
where $E_{rt}^{\pm 2}(P,t)$, $E_{rt}^{02}(P,t)\varpropto \Omega (t)^{-2};$ $%
E_{rt}^{\pm 3}(P,t),$ $E_{rt}^{03}(P,t)\varpropto \Omega (t)^{-3};$ $%
E_{rt}^{\pm 4}(P,t),$ $E_{rt}^{04}(P,t)\varpropto \Omega (t)^{-4};$ $%
E_{rt}^{05}(P,t)\varpropto \Omega (t)^{-5}.$ It is not difficult to obtain
the upper bounds for all these error terms, but one needs only to consider
the dominating terms $E_{rt}^{\pm 2}(P,t)$ and $E_{rt}^{02}(P,t)$ that are
proportional to $\Omega (t)^{-2},$ since the truncation errors $E_{r1}^{\pm
}(P,t)$ and $E_{r1}^{0}(P,t)$ are proportional to $\Omega (t)^{-2}$ and the
other error terms $E_{rt}^{\pm k}(P,t)$ and $E_{rt}^{0l}(P,t)$ $(k=3,4;$ $%
l=3,4,5)$ are higher-order and can be neglected with respect to the
dominating terms $E_{rt}^{\pm 2}(P,t)$ and $E_{rt}^{02}(P,t)$. The
dominating error terms are given by%
\begin{equation*}
E_{rt}^{\pm 2}(P,t)=E_{r1}^{\pm }(P,t)F(P,t)\pm \frac{1}{\sqrt{2}}G_{\pm
}(P,t)\exp [\mp i\int_{t_{0}}^{t}dt^{\prime }2\Omega (t^{\prime })]
\end{equation*}%
\begin{equation*}
\times \{ib_{0}(P,t_{0})\frac{\Theta (P,t)}{\Omega (t)}\exp [-i\Delta
P\int_{t_{0}}^{t}dt^{\prime }K_{0}(t^{\prime })]\exp [\pm
i\int_{t_{0}}^{t}dt^{\prime }\Omega (t^{\prime })]
\end{equation*}%
\begin{equation}
-ib_{0}(P,t_{0})\frac{\Theta (P,t_{0})}{\Omega (t_{0})}\exp [\mp i\frac{1}{8}%
\int_{t_{0}}^{t}dt_{1}\frac{\Gamma (P,t_{1})^{2}+4|\Theta (P,t_{1})|^{2}}{%
\Omega (t_{1})}]\},  \tag{178a}
\end{equation}%
\begin{equation*}
E_{rt}^{02}(P,t)=E_{r1}^{0}(P,t)-b_{0}(P,t_{0})F(P,t)\frac{|\Theta (P,t)|^{2}%
}{\Omega (t)^{2}}
\end{equation*}%
\begin{equation*}
+b_{0}(P,t_{0})F(P,t)\frac{\Theta (P,t)^{\ast }}{\Omega (t)}\frac{\Theta
(P,t_{0})}{\Omega (t_{0})}\exp [i\Delta P\int_{t_{0}}^{t}dt^{\prime
}K_{0}(t^{\prime })]
\end{equation*}%
\begin{equation}
\times \cos \{\int_{t_{0}}^{t}dt_{1}[\Omega (t_{1})+\frac{\Gamma
(P,t_{1})^{2}+4|\Theta (P,t_{1})|^{2}}{8\Omega (t_{1})}]\}.  \tag{178b}
\end{equation}%
The last term on the right-hand sides of (178) for each one of the error
terms $E_{rt}^{\pm 2}(P,t)$ and $E_{rt}^{02}(P,t)$ is proportional to the
value $\Theta (P,t_{0})/\Omega (t_{0})$ at the initial time $t_{0}$ of the
basic STIRAP decelerating or accelerating process. As discussed before (See:
(133) and (134)), the initial value $|\Theta (P,t_{0})|/\Omega (t_{0})$ may
be controlled to be so small that it can be neglected. Then the norm for the
error vector $(E_{rt}^{02}(P,t),E_{rt}^{+2}(P,t),E_{rt}^{-2}(P,t))^{T}$ is
bounded by 
\begin{equation*}
||E_{r}^{(2)}(P,t)||=\sqrt{%
|E_{rt}^{+2}(P,t)|^{2}+|E_{rt}^{-2}(P,t)|^{2}+|E_{rt}^{02}(P,t)|^{2}}
\end{equation*}%
\begin{equation*}
\leq \sqrt{%
|E_{r1}^{0}(P,t)|^{2}+|F(P,t)|^{2}(|E_{r1}^{+}(P,t)|^{2}+|E_{r1}^{-}(P,t)|^{2})%
}
\end{equation*}%
\begin{equation}
+\frac{|\Theta (P,t)|}{\Omega (t)}\sqrt{|F(P,t)|^{2}\frac{|\Theta (P,t)|^{2}%
}{\Omega (t)^{2}}+\frac{1}{2}|G_{+}(P,t)|^{2}+\frac{1}{2}|G_{-}(P,t)|^{2}}. 
\tag{179}
\end{equation}%
This error upper bound is proportional to $\Omega (t)^{-2}.$ Now it follows
from (174) and (177) and then the inequalities (176) and (179) that the
total deviation of a real STIRAP adiabatic process from the ideal one at any
instant of time is bounded by%
\begin{equation*}
||E_{r}(P,t)||=\sqrt{|\delta _{+}^{b}(P,t)|^{2}+|\delta
_{-}^{b}(P,t)|^{2}+|\delta _{0}^{b}(P,t)|^{2}}
\end{equation*}%
\begin{equation*}
\leq ||E_{r}^{(1)}(P,t)||+||E_{r}^{(2)}(P,t)||
\end{equation*}%
\begin{equation*}
\leq F(P,t)\frac{|\Theta (P,t)|}{\Omega (t)}+\sqrt{%
|E_{r1}^{0}(P,t)|^{2}+|F(P,t)|^{2}(|E_{r1}^{+}(P,t)|^{2}+|E_{r1}^{-}(P,t)|^{2})%
}
\end{equation*}%
\begin{equation}
+\frac{|\Theta (P,t)|}{\Omega (t)}\sqrt{|F(P,t)|^{2}\frac{|\Theta (P,t)|^{2}%
}{\Omega (t)^{2}}+\frac{\frac{1}{16}\frac{\Gamma (P,t)^{2}}{\Omega (t)^{2}}+%
\frac{1}{4}\frac{|\Theta (P,t)|^{4}}{\Omega (t)^{4}}}{[1+\frac{1}{16}\frac{%
\Gamma (P,t)^{2}}{\Omega (t)^{2}}+\frac{|\Theta (P,t)|^{2}}{\Omega (t)^{2}}%
]^{2}}},  \tag{180}
\end{equation}%
where the upper bounds for the truncation errors $|E_{r1}^{+}(P,t)|,$ $%
|E_{r1}^{-}(P,t)|,$ and $|E_{r1}^{0}(P,t)|$ are obtained from (150), (160),
and (166), respectively. In order to obtain the global adiabatic condition
one needs to limit the maximum value on the rightest side of (180) not to be
more than a desired small value over the effective momentum region $[P]$ and
in the whole time period $t_{0}\leq t\leq t_{0}+T$ of the STIRAP
decelerating or accelerating process. Denote $A_{d}(P,t)$ as the function on
the rightest side of (180). Then the global adiabatic condition may be
expressed as%
\begin{equation}
||E_{r}(P,t)||\leq A_{d}(P,t)\leq \max_{P\in \lbrack P],\text{ }t_{0}\leq
t\leq t_{0}+T}\{A_{d}(P,t)\}\leq \varepsilon _{r},  \tag{181}
\end{equation}%
where $\varepsilon _{r}$ is the desired value and $\varepsilon _{r}<<1$.
Unlike the adiabatic condition (128) there is not an exponential correction
factor in the adiabatic condition (180) and (181). The adiabatic condition
(180) and (181) consists of the first-order term that is proportional to $%
\Omega (t)^{-1}$ and the second-order terms that are proportional to $\Omega
(t)^{-2},$ while the adiabatic condition (128) is the first-order term with
the factor $F(P,t)=1$ times the exponential correction factor. Thus, the
adiabatic condition (180) and (181) is much less severe than the adiabatic
condition (128), the latter is most severe for a quantum ensemble with a
broad momentum distribution. However, just like the adiabatic condition
(128) the adiabatic condition (180) and (181) is also strict and accurate
(It is not difficult to include the omitted secondary error terms of (177)
in the adiabatic condition (180)\ and (181)).\ Therefore, if any one of the
two general adiabatic conditions is met, then a better STIRAP pulse sequence
is obtained and the perfect state (or population) transfer may be realized
by the STIRAP pulse sequence. The adiabatic condition (180) and (181) may be
more useful in practice to realize the perfect STIRAP state (or population)
transfer in an atomic or molecular ensemble with a broad momentum
distribution. It may also be used to realize the STIRAP decelerating and
accelerating processes in the laser cooling and the quantum coherence
interference experiments of a cold atomic or molecular ensemble.

The adiabatic condition (180) and (181) is still slightly severe. This is
because it is based on the first-order approximation solution to the basic
equations (141), leading to that the upper bounds for the truncation errors $%
|E_{r1}^{+}(P,t)|,$ $|E_{r1}^{-}(P,t)|,$ and $|E_{r1}^{0}(P,t)|$ are not the
lowest ones. As known before, these truncation errors are proportional to $%
\Omega (t)^{-2}.$ A better adiabatic condition may be set up on the basis of
the second-order approximation solution to the basic equations (141), which
are given by (151), (161), and (171), and the corresponding truncation error
upper bounds $|E_{r2}^{+}(P,t)|,$ $|E_{r2}^{-}(P,t)|,$ and $%
|E_{r2}^{0}(P,t)| $ are given by (152), (162), and (172), respectively.
These upper bounds are proportional to $\Omega (t)^{-3}.$ Thus, they are the
lower ones with respect to those truncation error upper bounds of the
first-order solution, leading to that the adiabatic condition based on the
second-order approximation solution is better one.

It is known that in the adiabatic condition (180) and (181) the truncation
errors $|E_{r1}^{+}(P,t)|,$ $|E_{r1}^{-}(P,t)|,$ and $|E_{r1}^{0}(P,t)|$ are
proportional to $\Omega (t)^{-2}$ and the factor $F(P,t)$ is a function of
the ratio $\Gamma (P,t)^{2}/\Omega (t)^{2}.$ Then the parameter $\Gamma
(P,t) $ appears only in those terms that are proportional to $\Omega
(t)^{-2} $ on the rightest side of (180). This is different from the case
that the parameter $\Theta (P,t)$ may appear in the first term on the
rightest side of (180) that is proportional to $\Omega (t)^{-1}.$ Thus, the
parameter $\Gamma (P,t)$ has a smaller contribution to the adiabatic
condition (180) and (181) than the parameter $\Theta (P,t).$ As pointed out
before, the effect of the momentum distribution is dependent upon whether
the Raman laser light beams of a STIRAP\ experiment are copropagating or
counterpropagating. Consider that the Raman laser light beams are
copropagating. Then the momentum distribution could have a relatively small
effect, because in this case the parameter $\Theta (P,t)$ has a smaller
value. This is correct for the first-order approximation adiabatic condition
(134). However, it could not be so simple for the inequality (180) in a
quantum ensemble with a broad momentum distribution. When the first term on
the rightest side of (180) is smaller due to a small parameter $\Theta
(P,t), $ the second and third terms could become more important and have a
dominating contribution to the adiabatic condition (180) and (181). Then in
this case the parameters $\Gamma (P,t)$ and $K_{0}(t)$ become more important
in the adiabatic condition (180) and (181), resulting in that the momentum
distribution has a large effect on the adiabatic condition. Therefore, the
momentum distribution needs to be considered explicitly even for a
conventional STIRAP state (or population) transfer experiment in a quantum
ensemble with a broad momentum distribution, which uses the copropagating
Raman laser light beams. Obviously, for the STIRAP decelerating and
accelerating processes in a quantum ensemble with a broad momentum
distribution, which use the counterpropagating Raman laser light beams, one
needs to consider generally the effect of the momentum distribution on the
STIRAP\ state (or population) transfer.

The present adiabatic theoretical methods including the equivalent
transformation method to solve the basic differential equations (26) not
only can be used to set up a general adiabatic condition for the basic
STIRAP decelerating and accelerating processes of a single atom or molecule
or a quantum ensemble of the atoms or molecules, but also they will have an
extensive application in other research fields such as the NMR spectroscopy
(See: for example, Ref. [41]) and the magnetic resonance image (MRI). 
\newline
\newline
\newline
{\large 8. Discussion}

In the paper the standard three-state STIRAP population transfer theory in
the laser spectroscopy [4, 15, 16, 17, 18] has been developed to describe
theoretically the STIRAP-based\ unitary decelerating and accelerating
processes of a single freely moving atom by combining the superposition
principle in quantum mechanics [25] and the energy, momentum, and angular
momentum conservation laws for the atomic photon absorption and emission
processes [5, 19]. There are similar theoretical works or developments to
describe the atomic laser cooling process [5, 19, 20, 21] in a neutral atom
ensemble and the atomic quantum interference experiments [10, 12] in a cold
atomic ensemble. There are also a number of works to investigate the atomic
decelerating and accelerating processes by the laser light techniques [21,
23, 24]. However, the present work is focused on the analytical and
quantitative investigation how the momentum distribution of a superposition
of the momentum states of a pure-state quantum system such as a single
freely moving atom affects the state-transfer efficiency in these STIRAP
unitary decelerating and accelerating processes. It emphasizes the complete
STIRAP state transfer and the unitarity of these processes. This means that
in the present work any decoherence effect of the atomic system under study
is not considered. A main purpose to investigate the effect of the momentum
distribution on the STIRAP state transfer is to build up better STIRAP
unitary decelerating and accelerating sequences, so that the time- and
space-compressing processes of the quantum control process to simulate the
reversible and unitary state-insensitive halting protocol [22] can be
realized through these decelerating and accelerating processes. Thus, this
is involved in setting up a general adiabatic condition for the STIRAP
unitary decelerating and accelerating processes. In the paper two general
adiabatic conditions have been obtained analytically, one based on the Dyson
series solution to the basic differential equations to govern the STIRAP
processes, another based on the equivalent transformation method to solve
the basic differential equations. Both the general adiabatic conditions may
be used to set up a conventional STIRAP experiment and also the STIRAP-based
decelerating and accelerating processes. A complete STIRAP state transfer
could be achieved only in the ideal adiabatic condition. Generally, it is
hard to achieve a complete state transfer for the STIRAP processes when the
atomic momentum superposition state has a broad momentum distribution.
However, in the ideal or nearly ideal adiabatic condition an almost complete
STIRAP state transfer may be realized if the superposition of the momentum
states has a small effective wave-packet spreading or a narrow momentum
distribution. When the initial motional state of a freely moving atom is a
Gaussian wave-packet state, the final motional state of the atom is still a
perfect or almost perfect Gaussian wave-packet state after the atom
undergoes the STIRAP unitary decelerating (or accelerating) process in the
ideal or nearly ideal adiabatic condition. Therefore, in the paper it is
shown that the time- and space-compressing processes of the quantum control
process [22] can be realized almost perfectly through the STIRAP
decelerating and accelerating processes in the ideal or nearly ideal
adiabatic condition. This is one of the important results in the paper.

The standard STIRAP population transfer theory is generally based on the
semiclassical theory of electromagnetic radiation. In the semiclassical
theory the externally applied electromagnetic fields such as the Raman laser
light beams are considered as the classical electromagnetic fields, while
the atomic system itself and the interaction between the atomic system and
the external electromagnetic fields are treated quantum mechanically. It has
been shown that the semiclassical theory can describe almost perfectly the
three-state STIRAP population transfer experiments of the atomic and
molecular beams in the laser spectroscopy [4, 15, 16, 17, 18]. On the other
hand, the semiclassical theory is also successful to describe the
STIRAP-based laser cooling processes in a neutral atomic ensemble [20, 21]
and especially the atomic quantum interference experiments in a cold neutral
atom ensemble [10, 12, 13, 14]. In the paper the semiclassical theory also
is directly employed to describe the STIRAP-based\ unitary decelerating and
accelerating processes of a single atom. The semiclassical theory of
electromagnetic radiation generally can not explain reasonably the atomic
spontaneous emission [1, 25, 26]. However, it is suited to describe these
STIRAP unitary decelerating and accelerating processes due to that these
STIRAP processes can avoid the atomic spontaneous emission by setting
suitably the experimental parameters.

As far as the Hamiltonian of Eq. (4) to describe the three-state STIRAP\
experiment of an atom system is concerned, there are three requirements: $%
(i) $ the three-state subspace for the atomic internal states is closed
under the Hamiltonian; $(ii)$ the electric-dipole approximation is
satisfied; and $(iii)$ the rotating wave approximation ($RWA$) is
reasonable. The first requirement can be satisfied if one chooses suitably
the atom and its three internal states and the experimental parameters of
the Raman laser light beams. Since the size of an atom is generally much
less than the wave lengths of the Raman laser light beams at the optical
frequencies ($\thicksim \omega _{01}$ and $\omega _{02}$), the second
requirement may be met generally. If the Raman laser light beams are strong,
it can turn out that in the first approximation the strong laser light field
may generate a Bloch-Siegert shift to the transition frequency of the atomic
internal energy levels [1]. When the Rabi frequencies ($\Omega _{p}(t)$ and $%
\Omega _{s}(t)$) of the Raman laser light beams are much less than the
resonance frequencies ($\omega _{01}$ and $\omega _{02}$) of the atomic
internal energy levels and the detunings for the Raman laser light beams are
small, the magnitude of the Bloch-Siegert shifts generated by the Raman
laser light beams at the optical frequencies is very small and may be
negligible and hence the rotating wave approximation is reasonable [1].
However, it is also convenient to correct the Bloch-Siegert shifts in the
STIRAP experiment because one needs only to add the Bloch-Siegert shifts to
the resonance frequencies ($\omega _{01}$ and $\omega _{02}$). A simple
evaluation for the Bloch-Siegert shifts can be seen in Ref. [1] and a
general treatment may use the average Hamiltonian theory [34]. On the other
hand, it is also possible to apply an extra laser light field for each one
of the Raman laser light beams in the STIRAP\ experiment to compensate the
rotating-wave approximation. In fact, if each one of the two Raman laser
light beams in the STIRAP experiment is replaced with a pair of the laser
light beams with the orthogonal electric field vectors and the suitable
phases [38], one can eliminate the rotating-wave approximation. Similarly,
one may also use the circularly polarized lights to prepare the dipole
interaction Hamiltonian (10)\ without the rotating wave approximation [19,
38].

There is also another condition to be met that the electromagnetic field of
any Raman laser light beam is considered as an infinite plane-wave
electromagnetic field when calculating the time evolution process of a
freely moving atom under the STIRAP unitary decelerating and accelerating
processes. Here the infinite plane-wave electromagnetic field has spatially
uniform amplitude and phase. This condition can be met only when the
electromagnetic field can encompass sufficiently the whole wave-packet
motional state of the moving atom. Note that the electromagnetic field of
the Raman laser light beam propagates along a direction parallel to the
atomic moving direction in one-dimensional space. Since the atom is in a
Gaussian wave-packet motional state which has a finite wave-packet
spreading, then one can set suitably the experimental parameters for the
Raman laser light beam such that the electromagnetic field in space is much
wider than the effective wave-packet spreading of the atomic motional state
during the whole decelerating or accelerating process. Then in this case the
electromagnetic field in space can be reasonably considered as an infinite
and uniform plane-wave electromagnetic field for the atomic wave-packet
motional state. The condition may be satisfied more easily for a heavy atom
as the wave-packet motional state for such atom has a more narrow
wave-packet spreading. If the electromagnetic field in space has a finite
bandwidth less than or comparable to the wave-packet spreading of an atomic
motional state, then it can not be considered as an infinite and uniform
plane-wave electromagnetic field and the electric dipole interaction of Eq.
(10) with space-independent Rabi frequencies is not suited to describe the
STIRAP process, since in this case the Rabi frequencies are dependent upon
the spatial coordinate [35].

As an important result in the paper, it is shown that if a free atom is in a
Gaussian wave-packet motional state at the initial time, then it is still in
a Gaussian wave-packet motional state after it is decelerated (or
accelerated) by the STIRAP-based unitary decelerating (or accelerating)
sequence in the ideal or nearly ideal adiabatic condition. As far as a
Gaussian wave-packet state is concerned, there are two types of time- and
space-dependent unitary evolution processes that do not change the Gaussian
wave-packet shape of the atomic motional state. The first type is that the
unitary evolution processes may manipulate and control the center-of-mass
position and/or momentum of a Gaussian wave-packet state but can not
manipulate at will the complex linewidth of a Gaussian wave-packet state.
The second type is that the unitary evolution processes may manipulate and
control the complex linewidth of a Gaussian wave-packet state. The
STIRAP-based unitary decelerating and accelerating processes belong to the
first type. This type of the unitary evolution processes tend to have the
property that in the unitary evolution process the imaginary part of the
complex linewidth of a Gaussian wave-packet state increases linearly with
the time period of the unitary evolution process, while the real part
usually keeps unchanged, i.e., the wave-packet spreading of the Gaussian
wave-packet state becomes larger and larger early or late as the time period
increases. These unitary evolution processes which have the property also
include the free-particle motion and the atomic bouncing process off a hard
potential wall in the special case. Thus, the free-particle motion and the
atomic bouncing process may be assigned to the first type. The Hamiltonians\
of the quantum systems to create the first type of the time- and
space-dependent unitary propagators usually can not be singly used to
generate their inverse unitary propagators without any help of the
interactions from outside the quantum systems. Therefore, this type of
unitary evolution processes do not have their own inverse unitary
propagators in these quantum systems separated from the outside or their
environment. Obviously, these separate quantum systems also include the
isolated quantum systems in the quantum statistical physics and they are
described completely by the Hamiltonians of the quantum systems. These
unitary evolution processes could be considered to be self-irreversible in
the sense that there do not exist their own inverse unitary propagators in
the same separate quantum systems, although these processes obey the unitary
quantum dynamics and their inverse unitary propagators could be generated
with the help of the specific interactions from outside the quantum systems.
Take a free-particle motion as a typical example. A free-particle motion can
be described completely by the unitary propagator $U(t)=\exp
[-ip^{2}t/(2m\hslash )]$. Of course, one may also choose the unitary
propagator $U(t)^{+}=\exp [ip^{2}t/(2m\hslash )]$ to describe the
free-particle motion. However, once one chooses one of the two unitary
propagators to describe the free-particle motion, another is the inverse
propagator of the chosen unitary propagator. The free particle may move
toward the left or the right in the coordinate axis and both the motions can
be described by the same unitary propagator $U(t)$. Now the inverse
free-particle motion is described by the unitary propagator $U(t)^{+}$. The
Hamiltonian $H=p^{2}/2m$ of the free particle can really generate only the
unitary propagator $U(t)$ (or $U(t)^{+}$) but can not really generate both
the unitary propagator $U(t)$ and its inverse propagator $U(t)^{+}$
simultaneously by the Schr\"{o}dinger equation\ $i\hslash \partial
U(t)/\partial t=H(t)U(t).$ One may take a hermite conjugate on the Schr\"{o}%
dinger equation and then could obtain $U(t)^{+}$, but taking a Hermite
conjugate is not a real physical process and hence there does not exist the
inverse propagator $U(t)^{+}$ in the free-particle quantum system. Then the
inverse free-particle motion will never really take place if the free
particle is not acted on by any external interaction. In other words, one
may argue that the time reversal process (i.e., the inverse free-particle
motion) may be a real physical process [25, 30], but this process can not
really take place for the free particle without any help of the specific
external interactions. Then in this sense the free-particle motion is really
self-irreversible, although it is governed by the unitary propagator $U(t)$.
Of course, it is possible to construct the inverse unitary propagator of the
free particle if the specific external interactions are applied to the free
particle. How to construct these inverse propagators of the type of time-
and space-dependent unitary propagators mentioned above will be reported in
next paper. As pointed out in the previous papers [22, 36], such a situation
that a quantum system that obeys the unitary quantum dynamics can not really
have both the unitary propagator and its inverse propagator is often met in
the quantum systems which are used to implement a quantum computation. The
spontaneously irreversible processes of isolated quantum systems in the
quantum statistical physics [40] could be related to the situation. It has
been stressed in the previous papers [22, 36, 37] that these irreversible
processes should be understood through the unitary quantum dynamics instead
of the stochastic process and probability statistics [40], and they could be
related to the difference between the unitary evolution process and its
inverse process. On the other hand, there also exist other quantum systems
that have both the unitary propagators and their inverse propagators, where
both the types of the unitary propagators can be generated by the same
Hamiltonians of the quantum systems. A conventional harmonic oscillator is
typically one of such quantum systems. In the quantum statistical physics
such quantum systems obey completely the Poincar\'{e}$^{\prime }$s
recurrence theorem.

There are the second type of time- and space-dependent unitary propagators
that can manipulate and control the complex linewidth of a Gaussian
wave-packet motional state. A general quadratic Hamiltonian can generate a
time- and space-dependent unitary propagator that can keep the Gaussian
shape unchanged for a Gaussian wave-packet motional state under the action
of the unitary propagator. Obviously, the Hamiltonian is different from
those of the STIRAP unitary decelerating and accelerating processes. One
needs this type of unitary propagators to manipulate and control the complex
linewidth of a Gaussian wave-packet motional state to build up the quantum
circuit for the reversible and unitary state-insensitive halting protocol
and also needs their inverse unitary propagators to realize the efficient
quantum search process [22, 36]. How to construct these unitary propagators
will be reported in next paper (Arxiv: quant-ph/0708.2129).\newline
\newline
\newline
{\Large References}\newline
1. L. Allen and J. H. Eberly, \textit{Optical resonance and two-level atoms}%
, Dover, New York, 1987\newline
2. R. G. Brewer and E. L. Hahn, \textit{Coherent two-photon process:
Transient and steady-state cases}, Phys. Rev. A 11, 1641 (1975)\newline
3. U. Gaubatz, P. Rudecki, M. Becker, S. Schiemann, M. Kulz, and K.
Bergmann, \textit{Population switching between vibrational levels in
molecular beams}, Chem. Phys. Lett. 149, 463 (1988)\newline
4. K. Bergmann, H. Theuer, and B. W. Shore, \textit{Coherent population
transfer among quantum states of atoms and molecules}, Rev. Mod. Phys. 70,
1003 (1998)\newline
5. A. Aspect, \ E. Arimondo, R. Kaiser, N. Vansteenkiste, and C.
Cohen-Tannoudji, \textit{Laser cooling below the one-photon recoil energy by
velocity selective coherent population trapping}, Phys. Rev. Lett. 61, 826
(1988) \newline
6. M. Kasevich and S. Chu, \textit{Laser cooling below a photon recoil with
three-level atoms}, Phys. Rev. Lett. 69, 1741 (1992)\newline
7. (a) J. Reichel, O. Morice, G. M. Tino, and C. Salomon, \textit{Subrecoil
Raman cooling of Cesium atoms}, Europhys. Lett. 28, 477 (1994); (b) S.
Kulin, B. Saubamea, E. Peik, J. Lawall, T. W. Hijmans, M. Leduc, and C.
Cohen-Tannoudji, \textit{Coherent manipulation of atomic wave-packets by
adiabatic transfer}, Phys. Rev. Lett. 78, 4185 (1997)\newline
8. T. Esslinger, F. Sander, M. Weidemuller, A. Hemmerich, and T. W. Hansch, 
\textit{Subrecoil laser cooling with adiabatic transfer}, Phys. Rev. Lett.
76, 2432 (1996) \newline
9. V. Boyer, L. J. Lising, S. L. Rolston, and W. D. Phillips, \textit{Deeply
subrecoil two-dimensional Raman cooling}, Phys. Rev. A 70, 043405 (2004)%
\newline
10. P. Marte, P. Zoller, and J. L. Hall, \textit{Coherent atomic mirrors and
beam splitters by adiabatic passage in multilevel systems}, Phys. Rev. A 44,
R4118 (1991)\newline
11. (a)\ M. Kasevich and S. Chu, \textit{Measurement of the gravitational
acceleration of an atom with a light-pulse atom interferometer, }Appl. Phys.
B 54, 321 (1992); (b) D. S. Weiss, B. C. Young, and S. Chu, \textit{%
Precision measurement of }$\hslash /m_{C_{s}}$\textit{\ based on photon
recoil using laser-cooled atoms and atomic interferometry}, Appl. Phys. B
59, 217 (1994)\newline
12. M. Weitz, B. C. Young, and S. Chu, \textit{Atom manipulation based on
delayed laser pulses in three- and four-level systems: light shifts and
transfer efficiencies}, Phys. Rev. A 50, 2438 (1994) \newline
13. L. S. Goldner, C. Gerz, R. J. C. Spreeuw, S. L. Rolston, C. I.
Westbrook, W. D. Phillips, P. Marte, and P. Zoller, \textit{Momentum
transfer in laser-cooled Cesium by adiabatic passage in a light field},
Phys. Rev. Lett. 72, 997 (1994) \newline
14. J. Lawall and M. Prentiss, \textit{Demonstration of a novel atomic beam
splitter}, Phys. Rev. Lett. 72, 993 (1994) \newline
15. U. Gaubatz, P. Rudecki, S. Schiemann, and K. Bergmann, \textit{%
Population transfer between molecular vibrational levels by stimulated Raman
scattering with partially overlapping laserfields: a new concept and
experimental results,} J. Chem. Phys. 92, 5363 (1990) \newline
16. (a) T. F. Hioe, \textit{Theory of generalized adiabatic following in
multilevel systems}, Phys. Lett. A 99, 150 (1983); (b) J. Oreg, F. T. Hioe,
and J. H. Eberly, \textit{Adiabatic following in multilevel systems}, Phys.
Rev. A 29, 690 (1984) \newline
17. J. R. Kuklinski, U. Gaubatz, F. T. Hioe, and K. Bergmann, \textit{%
Adiabatic population transfer in a three-level system driven by delayed
laser pulses}, Phys. Rev. A 40, 6741 (1989) \newline
18. (a)\ G. W. Coulston and K. Bergmann, \textit{Population transfer by
stimulated Raman scattering with delayed pulses: analytical results for
multilevel systems}, J. Chem. Phys. 96, 3467 (1992); (b) A. Kuhn, G. W.
Coulston, G. Z. He, S. Schiemann, K. Bergmann, and W. S. Warren, \textit{%
Population transfer by stimulated Raman scattering with delayed pulses using
spectrally broad light}, J. Chem. Phys. 96, 4215 (1992)\newline
19. A. Aspect, E. Arimondo, R. Kaiser, N. Vansteenkiste, and C.
Cohen-Tannoudji, \textit{Laser cooling below the one-photon recoil energy by
velocity-selective coherent population trapping: theoretical analysis}, J.
Opt. Soc. Am. B 6, 2112 (1989) \newline
20. K. Moler, D. S. Weiss, M. Kasevich, and S. Chu, \textit{Theoretical
analysis of velocity-selective Raman transitions}, Phys. Rev. A 45, 342
(1992) \newline
21. S. Chu, \textit{Nobel Lecture: The manipulation of neutral particles},
Rev. Mod. Phys. 70, 685 (1998), and references therein\newline
22. X. Miao, \textit{The basic principles to construct a generalized
state-locking pulse field and simulate efficiently the reversible and
unitary halting protocol of a universal quantum computer},
http://arxiv.org/abs/quant-ph/0607144 (2006) \newline
23. C. N. Cohen-Tannoudji, \textit{Nobel Lecture: Manipulating atoms with
photons}, Rev. Mod. Phys. 70, 707 (1998), and references therein\newline
24.\ W. D. Phillips, \textit{Nobel Lecture: Laser cooling and trapping of
neutral atoms}, Rev. Mod. Phys. 70, 721 (1998), and references therein 
\newline
25. L. I. Schiff, \textit{Quantum mechanics}, 3rd, McGraw-Hill book company,
New York, 1968\newline
26. J. Dalibard and C. Cohen-Tannoudji, \textit{Atomic motion in laser
light: connection between semiclassical and quantum descriptions}, J. Phys.
B 18, 1661 (1985)\newline
27. (a) E. A. Power and S. Zienau, \textit{Coulomb gauge in non-relativistic
quantum electrodynamics and the shape of spectral lines}, Phil. Trans. Roy.
Soc. Lond. 251, 427 (1959); (b) Z. Fried, \textit{Vector potential versus
field intensity}, Phys. Rev. A 8, 2835 (1973)\newline
28. (a)\ G. Alzetta, A. Gozzini, L. Moi, and G. Orriols, \textit{An
experimental method for the observation of r.f. transitions and laser beat
resonances in oriented Na vapour}, Nuovo Cimento B 36, 5 (1976); (b) E.
Arimondo and G. Orriols, \textit{Non-absorbing atomic coherences by coherent
two-photon transition in a three-level optical pumping}, Lett. Nuovo Cimento
17, 333 (1976)\newline
29. M. Born and V. Fock, \textit{Beweis des Adiabatensatzes}, Z. Phys. 51,
165 (1928)\newline
30. A. Messiah, \textit{Quantum mechanics}, Vol. II, North-Holland,
Amsterdam, 1962 \newline
31 (a)\ M. L. Goldberger and K. M. Watson, \textit{Collision theory}, Chapt.
3, Wiley, New York, 1964; (b)\ R. G. Newton, \textit{Scattering theory of
waves and particles}, Chapt. 6, McGraw-Hill, New York, 1966; (c)\ E. J.
Heller, \textit{Time-dependent approach to semiclassical dynamics}, J. Chem.
Phys. 62, 1544 (1975); (d) R. G. Littlejohn, \textit{The semiclassical
evolution of wave packets}, Phys. Rep. 138, 193 (1986)\newline
32. See, for example, E. W. Weisstein, \textit{CRC concise encyclopeda of
mathematics}, P. 934 (Erfc(x)\ function) and P. 1162 (Gaussian integrals),
2nd, A CRC Press Company, New York, 2003\newline
33. See, for example, P. Hartman, \textit{Ordinary differential equations},
Chapt. 2, 2nd, Birkhauser, Boston, 1982; E. L. Ince, \textit{Ordinary
differential equations}, Chapt. III, Dover, New York, 1956\newline
34. M. Matti Maricq, \textit{Application of average Hamiltonian theory to
the NMR of solids}, Phys. Rev. B 25, 6622 (1982)\newline
35. K. Gibble, \textit{Difference between a photon}$^{\prime }$\textit{s
momentum and an atom}$^{\prime }$\textit{s recoil}, Phys. Rev. Lett. 97,
073002 (2006)\newline
36. X. Miao, \textit{Quantum search processes in the cyclic group state
spaces}, http:// arxiv.org/abs/quant-ph/0507236 (2005)\newline
37. X. Miao, \textit{Efficient multiple-quantum transition processes in an }$%
n-$\textit{qubit spin system}, http://arxiv.org/abs/quant-ph/0411046 (2004)%
\newline
38. S. L. McCall and E. L. Hahn, \textit{Self-induced transparency}, Phys.
Rev. 183, 457 (1969)\newline
39. C. E. Carroll and F. T. Hioe, \textit{Analytic solutions for three-state
systems with overlapping pulses}, Phys. Rev. A 42, 1522 (1990)\newline
40. L. D. Landau, E. M. Lifshitz, and L. P. Pitaevskii, \textit{Statistical
Physics (Part 1)}, 3rd edn, translated by J. B. Sykes and M. J. Kearsley,
Pergamon Press, New York, 1980\newline
41. X. Miao, \textit{Ultra-broadband heteronuclear Hartmann-Hahn
polarization transfer}, http://arxiv.org/abs/quant-ph/0203079 (2002)

\end{document}